\documentclass[aps, prd, amsmath, floats, floatfix, twocolumn,
superscriptaddress, nofootinbib, showpacs]{revtex4-1}
\usepackage[xetex]{graphicx}
\usepackage{color}
\usepackage{soul}
\usepackage{url}
\usepackage{bm}         
\usepackage{times}

\newcommand{\beq}{\begin{equation}}
\newcommand{\eeq}{\end{equation}}
\newcommand{\beqn}{\begin{eqnarray}}
\newcommand{\eeqn}{\end{eqnarray}}

\newcommand{\lo}{\mathrel{\raise.3ex\hbox{$<$}\mkern-14mu
    \lower0.6ex\hbox{$\sim$}}}
\newcommand{\go}{\mathrel{\raise.3ex\hbox{$>$}\mkern-14mu
    \lower0.6ex\hbox{$\sim$}}}

\usepackage{color}

\newcommand{\Caltech}{\affiliation{TAPIR, Walter Burke Institute for Theoretical Physics, MC 350-17,
    California Institute of Technology, Pasadena, California 91125, USA}}
\newcommand{\Einstein}{\affiliation{NASA Einstein Fellow}}
\newcommand{\Cornell}{\affiliation{Center for Radiophysics and Space
    Research, Cornell University, Ithaca, New York, 14853, USA}}
\newcommand{\WSU}{\affiliation{Department of Physics \& Astronomy,
	Washington State University, Pullman, Washington 99164, USA}}
\newcommand{\CITA}{\affiliation{Canadian Institute for Theoretical 
    Astrophysics, University of Toronto, Toronto, Ontario M5S 3H8, Canada}}

\newcommand{\CIFAR}{\affiliation{Canadian Institute for Advanced Research, 180 Dundas St.~West, Toronto, ON M5G 1Z8, Canada}} %
\newcommand{\LBL}{\affiliation{Lawrence Berkeley National Laboratory,
1 Cyclotron Rd, Berkeley, CA 94720, USA}}
\newcommand{\NCSU}{\affiliation{Department of Physics, North Carolina State University, Raleigh, North Carolina 27695, USA; Hubble Fellow}}
\newcommand{\AEI}{\affiliation{Max-Planck-Institut f?ur Gravitationsphysik, Albert-Einstein-Institut, D-14476 Golm, Germany}}
\usepackage{graphicx}
\usepackage{dcolumn}
\usepackage{bm}
\usepackage{epsf}

\begin{document}

\title{Post-merger evolution of a neutron star-black hole binary with neutrino transport}

\author{Francois Foucart}  \LBL \Einstein \CITA%
\author{Evan O'Connor}  \NCSU \CITA%
\author{Luke Roberts} \Einstein \Caltech %
\author{Matthew D. Duez} \WSU %
\author{Roland Haas} \Caltech \AEI %
\author{Lawrence E. Kidder} \Cornell %
\author{Christian D. Ott} \Caltech 
\author{Harald P. Pfeiffer} \CITA\CIFAR
\author{Mark A. Scheel} \Caltech %
\author{Bela Szilagyi} \Caltech %

\begin{abstract}
We present a first simulation of the post-merger evolution of a black hole-neutron star binary in full general relativity using an energy-integrated 
general relativistic truncated moment formalism for neutrino transport. We describe our implementation of the moment formalism
and important tests of our code, before studying the formation phase of an accretion disk after a black hole-neutron star merger. We use
as initial data an existing general relativistic simulation of the merger of a neutron star of mass $1.4M_\odot$ with a black hole of mass
$7M_\odot$ and dimensionless spin $\chi_{\rm BH}=0.8$.
Comparing with a simpler leakage scheme for the treatment of the neutrinos, we find noticeable differences in the neutron to proton ratio in and around the disk, 
and in the neutrino luminosity. We find that the electron neutrino luminosity is much lower in the transport simulations, and that both the disk and the disk
outflows are less neutron-rich. The spatial distribution of the neutrinos is significantly affected by relativistic effects, due to large velocities and curvature in the regions of strongest emission. Over the short timescale evolved, we do not observe purely neutrino-driven outflows. 
However, a small amount of material
($3\times 10^{-4}M_\odot$) is ejected in the polar region during the circularization of the disk. Most of that material
is ejected early in the formation of the disk, and is fairly neutron rich (electron fraction $Y_e\sim 0.15-0.25$). Through r-process nucleosynthesis, that material 
should produce high-opacity lanthanides in the polar region, and could thus affect the lightcurve of radioactively powered electromagnetic
transients. 
We also show that by the end of the simulation, while the bulk of the disk remains neutron-rich ($Y_e\sim 0.15-0.2$ and decreasing), 
its outer layers have a higher electron fraction: 10\% of the 
remaining mass has $Y_e>0.3$. As that material would be the first to be unbound by disk outflows on longer timescales, and as composition evolution is slower at later times, the changes
in $Y_e$ experienced during the formation phase of the disk could have an impact on nucleosynthesis outputs from neutrino-driven and viscously-driven outflows.
Finally, we find that the effective viscosity due to momentum transport by neutrinos is unlikely to have a strong effect on the growth of the magnetorotational instability in the post-merger accretion disk. 
\end{abstract}


\maketitle

\section{Introduction}
\label{sec:intro}

The likely detection of gravitational waves by the advanced LIGO/VIRGO detectors~\cite{Harry2010,AdV} in the coming years will open up an entirely new way
to observe the universe, complementing existing electromagnetic and neutrino observations. Mergers of black holes and neutron stars are expected to be among the first and most common sources of gravitational waves to be observed~\cite{Abadie:2010cfa}. Beyond the excitement associated with the first gravitational wave detection, binary mergers will provide us with a wealth of information which could constrain the formation and evolution of massive binaries~\cite{OShaughnessy2008,Gerosa2013}, the outcomes of core collapse supernovae, or the properties of the cold, dense neutron-rich matter in the core of neutron 
stars~\cite{Read2009b,ReadEtAl2013,Lackey:2013axa,DelPozzo:13}.

In the presence of at least one neutron star, the gravitational wave signal may be accompanied by electromagnetic and neutrino emissions. The joint detection
of a system by a gravitational wave observatory and an electromagnetic telescope would provide additional information about the location, environment, and parameters of the binary~\cite{metzger:11}. It could also help us understand some important astrophysical processes. Neutron star mergers are among the most likely
progenitors of short-hard gamma-ray bursts. They may also significantly contribute to the production of heavy elements in the universe, through r-process nucleosynthesis in the neutron-rich material unbound during some mergers~\cite{Roberts2011,korobkin:12,Wanajo2014}. This nucleosynthesis could be observed through radioactively powered transients, observable in the optical and/or infrared bands days after the merger (``kilonovae'')~\cite{Tanaka:2013ana,2013ApJ...775...18B}.

To understand which signals can be emitted by a merger, and how they depend on the initial parameters of the binary, we need numerical simulations
in full general relativity: the strong non-linearities in the evolution of the spacetime surrounding the binary make any approximate treatment of gravity
unreliable. Most general relativistic simulations focus on a very short
period around the time of merger ($\sim 100\,{\rm ms}$ for simulations involving neutron stars), when general relativistic effects are important. 
Over that time period and beyond gravity, the most important physical effects to take into account are the equation of state of neutron star matter, the effects of magnetic fields, and the cooling and composition evolution due to neutrino-matter interactions. Although numerical simulations of these effects
have significantly improved over the last few years, much work is needed to simulate them well-enough to reliably predict
the post-merger signals produced by compact binary mergers. The main object of this work is the development and testing of an improved method 
for the treatment of neutrinos in the SpEC code~\cite{SpECwebsite}, based on the moment formalism of Thorne~\cite{1981MNRAS.194..439T}, 
and its application to the post-merger evolution of a black hole-neutron star binary.

In terms of the neutron star equations of state, most general relativistic simulations use either a simple Gamma-law equation of state, 
or parametrized piecewise-polytropic equations of state with a thermal Gamma-law. 
When modeling the gravitational wave signal up to the point of merger,
this should be sufficient: the gravitational wave signal mostly probes a single parameter of the equation of state, its tidal deformability, while the detailed structure of the 
neutron star appears unimportant~\cite{ReadEtAl2013,Lackey:2013axa}. But the equation of state plays a crucial role when simulating the merger and post-merger evolution of the binary~\cite{Hotokezaka:2011dh,Takami:2014zpa}.
Only a few simulations have used cold~\cite{hotokezaka:13,2014ApJ...780...31T,Takami:2014zpa} or 
hot~\cite{Duez:2009yy,Deaton2013,Foucart:2014nda,Neilsen:2014hha,Wanajo2014} 
tabulated, nuclear-theory based equations of state, and of those only the equation of state used in~\cite{Wanajo2014} appears
consistent with the most recent models for dense neutron-rich matter~\cite{2013ApJ...771...51L,2013ApJ...774...17S}. 
In this work, we use a hot, composition dependent, nuclear-theory based equation of state by Lattimer \& Swesty~\cite{Lattimer:1991nc}, which leads
to an acceptable mass-radius relationship for neutron stars and allows us to include neutrino-matter interactions, but is not fully consistent with
the latest constraints from nuclear experiments.

The simulation of magnetic fields has seen some impressive recent improvements. General relativistic simulations of binary neutron star mergers assuming ideal magnetohydrodynamics have been performed with a number of codes (see e.g.~\cite{2014ARA&A..52..661L} for a recent review of the field), 
but resolving the growth of magnetic instabilities at an acceptable computational cost remains an open problem~\cite{Kiuchi2014}.
Simulations using force-free or resistive magnetohydrodynamics, needed to properly simulate magnetically dominated regions, 
have also been performed~\cite{2012arXiv1208.3487D,Palenzuela:2013hu,Ponce:2014sza}. So far they have, however, mostly studied the pre-merger evolution of the system. 
Here, we do not take the magnetic fields into account at all, and focus solely on neutrino effects.  

The third component needed to simulate compact binaries around the time of merger, neutrinos, is also the most computationally expensive to 
treat accurately.
In theory, for each species of neutrinos, one should evolve the distribution function of neutrinos in both physical and momentum space. The high dimensionality of the problem
places it out of reach of current numerical codes and computers. Accordingly, various approximations have been developed - most of them in Newtonian simulations and/or with the aim
of studying core collapse supernovae. In general relativistic simulations of binary mergers, a few simulations have used a simple cooling prescription (``leakage'')
to model the first-order impact of neutrinos on the cooling of the disk and the composition evolution of high-density 
regions~\cite{Deaton2013,Foucart:2014nda,Neilsen:2014hha}. The only published result
using a more advanced method~\cite{Wanajo2014} studied a binary neutron star merger using an energy-integrated (``gray'') version of
the moment formalism for radiation 
transport~\cite{1981MNRAS.194..439T,2011PThPh.125.1255S,Cardall2013}. The first
two moments of the neutrino distribution function (i.e. the energy density and flux of neutrinos) were evolved, and the analytical M1 closure was used 
to compute the third 
moment (pressure tensor). In this work, we discuss the implementation of a similar scheme within the SpEC code, and its application to the post-merger evolution of a black hole-neutron star merger. As in~\cite{Wanajo2014} , we use a gray M1 scheme.
We should note that there are significant differences between our implementation of the M1 formalism and the one used in~\cite{Wanajo2014}, in particular
in the treatment of optically thick regions\footnote{M. Shibata, Y. Sekiguchi, private communication}, which would make a comparison between the results of the two
codes an interesting test of the accuracy of the M1 formalism in the core of the disk or in the presence of a hypermassive neutron star. However, we will not discuss
this here, as we are considering a completely different physical setup for the evolution, and the details of the M1 implementation used in~\cite{Wanajo2014} have not yet been published.

The focus of this paper is to discuss three different questions. First, we describe the implementation and testing of the M1 formalism in SpEC. 
An overview of the methods is offered in Sec.~\ref{sec:M1}, while the interested reader will find the details of the algorithm and important tests of the code in the Appendices. Second, we simulate the evolution of a black hole-neutron star binary in the critical phase in which the accretion disk is formed, from the moment at which neutrino emission increases due to the heating of the forming disk, to the time at which the disk reaches a quasi-equilibrium state (about $20\,{\rm ms}$ after merger). We pay particular attention to the temperature and composition evolution, the ejection of matter along the spin axis of the black hole, the geometry of the neutrino radiation, and the potential astrophysical consequences of our results. This is the main focus of Secs.~\ref{sec:disk}-\ref{sec:concl}. And third, we study the impact of using the M1 formalism instead of the simpler leakage scheme, and the impact of the various approximations which have to be made when using a gray scheme. This is the focus of Sec.~\ref{sec:nutreat}.

\section{Moment Formalism}
\label{sec:M1}

We will start here with an overview of the M1 formalism, and of its numerical implementation in SpEC. 
A more detailed description of many aspects of the algorithm, as well as tests of our implementation, are
available in the appendices.

\subsection{Evolution equations}

For each neutrino species $\nu_i$, we can describe
the neutrinos by their distribution function
$f_{(\nu)}(x^\alpha,p^\alpha)$, where $x^\alpha=(t,x^i)$
gives the time and the position of the neutrinos,
and $p^\alpha$ is the 4-momentum of the neutrinos.
The distribution function $f_{(\nu)}$ evolves according to the 
Boltzmann equation
\beq
p^\alpha \left[\frac{\partial f_{(\nu)}}{\partial x^\alpha} - \Gamma^\beta_{\alpha \gamma}
p^\gamma \frac{\partial f_{(\nu)}}{\partial p^\beta} \right] = \left[\frac{d f_{(\nu)}}{d\tau}\right]_{\rm coll}\,\,,
\eeq
where the $\Gamma^\alpha_{\beta \gamma}$ are the Christoffel symbols and the right-hand side includes all collisional
processes (emissions, absorptions, scatterings). In general, this is a 7-dimensional problem
which is extremely expensive to solve numerically. Approximations to the Boltzmann equation
have thus been developed for numerical applications. In this work, we consider the moment
formalism developed by Thorne~\cite{1981MNRAS.194..439T}, in which only the lowest moments
of the distribution function in momentum space are evolved. Our code is largely inspired by the implementation
of Thorne's formalism into general relativistic hydrodynamics simulations proposed by Shibata {\it et al.}~\cite{2011PThPh.125.1255S}
and Cardall {\it et al.}~\cite{Cardall2013}.
We limit ourselves to the use of this formalism in the ``gray'' approximation, that
is we only consider energy-integrated moments. Although the moment
formalism can in theory be used with a discretization in neutrino energies, this makes the
simulations significantly more expensive and involves additional technical difficulties in the treatment
of the gravitational and velocity redshifts. We will also only consider three independent neutrino
species: the electron neutrinos $\nu_e$, the electron antineutrinos $\bar \nu_e$, and the heavy-lepton
neutrinos $\nu_x$. The latter is the combination of 4 species
($\nu_\mu, \bar\nu_\mu,\nu_\tau,\bar\nu_\tau$).  This merging is
justified because the temperatures and neutrino energies reached in
our merger calculations are low enough to suppress the formation of
the corresponding heavy leptons whose presence would require including
the charged current neutrino interactions that differentiate between
these individual species.

In the gray approximation, and considering only the first two moments of the distribution function,
we evolve for each species projections of the
stress-energy tensor of the neutrino radiation $T^{\mu \nu}_{\rm rad}$. 
One possible decomposition of $T^{\mu \nu}_{\rm rad}$ is~\cite{2011PThPh.125.1255S}
\beq
T^{\mu \nu}_{\rm rad} = J u^\mu u^\nu + H^\mu u^\nu + H^\nu u^\mu + S^{\mu \nu}\,\,,
\eeq
with $H^\mu u_\mu = S^{\mu \nu}u_\mu = 0$ and $u^\mu$ the 4-velocity of
the fluid. The energy $J$, flux $H^\mu$
and stress tensor $S^{\mu \nu}$ of the neutrino radiation as observed by an
observer comoving with the fluid are related to the neutrino distribution function by
\beqn
J &=& \int_0^\infty d\nu\,\nu^3 \int d\Omega\,f_{(\nu)}(x^\alpha, \nu,\Omega)\,\,,\\
H^\mu &=& \int_0^\infty d\nu\,\nu^3 \int d\Omega\,f_{(\nu)}(x^\alpha,\nu,\Omega) l^\mu\,\,,\\
S^{\mu\nu} &=& \int_0^\infty d\nu\,\nu^3 \int d\Omega\,f_{(\nu)}(x^\alpha,\nu,\Omega) l^\mu l^\nu\,\,,
\eeqn
where $\nu$ is the neutrino energy in the fluid frame, $\int d\Omega$ denotes integrals
over solid angle on a unit sphere in momentum space, and
\beq
p^\alpha = \nu (u^\alpha + l^\alpha)\,\,,
\eeq
with $l^\alpha u_\alpha = 0$ and $l^\alpha l_\alpha =1$.
We also consider the decomposition of $T^{\mu \nu}_{\rm rad}$ in
terms of the energy, flux and stress tensor observed by an
inertial observer,
\beq
T^{\mu \nu}_{\rm rad} = E n^\mu n^\nu + F^\mu n^\nu + F^\nu n^\mu + P^{\mu \nu}\,\,,
\eeq
with $F^\mu n_\mu=P^{\mu \nu} n_\mu = F^t = P^{t\nu}=0$, 
and $n^\alpha$ the unit normal to a $t={\rm constant}$ slice.

We use the 3+1 decomposition of the metric,
\beqn
ds^2 &=& g_{\alpha \beta}dx^\alpha dx^\beta\\
&=& -\alpha^2 dt^2 + \gamma_{ij} (dx^i + \beta^i)(dx^j+\beta^j)\,\,,
\eeqn
where $\alpha$ is the lapse, $\beta^i$ the shift, and $\gamma_{ij}$
the 3-metric on a slice of constant coordinate $t$. The extension
of $\gamma_{ij}$ to the full 4-dimensional space is the projection
operator 
\beq
\gamma_{\alpha \beta} = g_{\alpha \beta} + n_\alpha n_\beta\,\,.
\eeq
We similarly define a projection operator onto the reference frame 
of an observer comoving with the fluid, 
\beq
h_{\alpha \beta}=g_{\alpha \beta}+u_\alpha u_\beta\,\,.
\eeq
We can then
write equations relating the fluid-frame variables to the inertial frame variables~\cite{Cardall2013}:
\beqn
E &=& W^2 J + 2W v_\mu H^\mu + v_\mu v_\nu S^{\mu \nu}\,\,,\\
F_\mu &=& W^2 v_\mu J + W (g_{\mu\nu}-n_\mu v_\nu)H^\nu \nonumber\\
&& +Wv_\mu v_\nu H^\nu + (g_{\mu\nu}-n_\mu v_\nu)v_\rho S^{\nu
\rho} \label{flux_equation} \,\,,\\
P_{\mu \nu} &=& W^2 v_\mu v_\nu J + W (g_{\mu\rho}-n_\mu v_\rho)v_\nu H^\rho \nonumber\\
&&+W(g_{\rho\nu}-n_\rho v_\nu)v_\mu H^\rho \nonumber\\
&&+(g_{\mu\rho}-n_\mu v_\rho)(g_{\nu\kappa}-n_\nu v_\kappa)S^{\rho \kappa}\label{eq:Pij}\,\,,
\eeqn
using the decomposition of the 4-velocity
\beq
u^\mu = W (n^\mu + v^\mu)\,\,,
\eeq
with $v^\mu n_\mu=0$ and $W=\sqrt{1+\gamma^{ij}u_i u_j}$.

Evolution equations
for $\tilde E = \sqrt{\gamma}E$ and $\tilde F = \sqrt{\gamma}F^i$ can then be written in conservative form:
\beqn
\partial_t \tilde E &+& \partial_j(\alpha \tilde F^j -\beta^j \tilde E)\label{eq:Enu}\\
&=&\alpha (\tilde P^{ij}K_{ij} -\tilde F^j \partial_j \ln{\alpha} - \tilde S^\alpha n_\alpha)\nonumber\,\,,\\
\partial_t \tilde F_i &+& \partial_j(\alpha \tilde P^j_{i} -\beta^j \tilde F_i) \label{eq:Fnu}\\
&=&(-\tilde E\partial_i\alpha+\tilde F_k\partial_i\beta^k+\frac{\alpha}{2} \tilde P^{jk}\partial_i \gamma_{jk}+\alpha \tilde S^\alpha \gamma_{i\alpha})\nonumber\,\,,
\eeqn
where $\gamma$ is the determinant of $\gamma_{ij}$, $\tilde P_{ij}=\sqrt{\gamma} P_{ij}$,
and $\tilde S^\alpha$ includes all collisional source terms.

To close this system of equations, we need two additional ingredients: a prescription for the 
computation of $P^{ij}(E,F_i)$ (`closure relation', which we choose following Minerbo 1978~\cite{Minerbo1978}), 
and the collisional source terms $\tilde S^\alpha$. 
In the M1 formalism, the neutrino pressure tensor $P^{ij}$ is recovered as an interpolation between
its known limits for an optically thick medium and an optically thin medium with a unique direction
of propagation for the neutrinos. We provide details on its computation in Appendix~\ref{sec:closure}.
For the source terms $\tilde S^\alpha$, we will consider that the fluid has an emissivity $\eta$ due
to the charged-current reactions
\beqn
p + e^- &\rightarrow& n + \nu_e\,\,,\\
n + e^+ &\rightarrow& p + \bar \nu_e\,\,,
\eeqn
as well as electron-positron pair annihilation
\beq
e^+ + e^- \rightarrow \nu_i \bar\nu_i\,\,,
\eeq 
plasmon decay
\beq
\gamma \rightarrow \nu_i \bar\nu_i\,\,,
\eeq
and nucleon-nucleon Bremsstrahlung
\beq
N + N \rightarrow N + N + \nu_i + \bar \nu_i\,\,.
\eeq
The inverse reactions are responsible for an absorption opacity
$\kappa_a$. We also consider a scattering opacity $\kappa_s$ due to elastic scattering
of neutrinos on nucleons and heavy nuclei. The source terms are then
\beq
\tilde S^\alpha = \sqrt{\gamma} \left(\eta u^\alpha - \kappa_a J u^\alpha - (\kappa_a+\kappa_s) H^\alpha\right)\,\,.
\eeq
Details of the choices made for the computation of the gray $\eta$, $\kappa_a$ and $\kappa_s$
are provided in Appendix~\ref{sec:sources}.

\subsection{Numerical scheme}

We add the evolution of neutrinos with the moment scheme to the SpEC code~\cite{SpECwebsite}, which
already includes a general relativistic hydrodynamics module~\cite{Duez:2008rb}. The latest methods
used for evolving in SpEC the coupled system formed by Einstein's equation and the general relativistic equations of hydrodynamics 
are described in~\cite{Foucart:2013a}, Appendix A.

In the M1 formalism, we evolve the variables $\tilde E$ and $\tilde F_i$ according
to Eqs.~(\ref{eq:Enu}-\ref{eq:Fnu}), and couple the neutrinos to the fluid evolution. The coupling takes the form
\beqn
\partial_t \tau &=& ...\, + \alpha \tilde S^\alpha n_\alpha \label{eq:sourcetau}\,\,,\\
\partial_t S_i &=& ...\, - \alpha \tilde S^\alpha \gamma_{i\alpha}\label{eq:sourceS}\,\,,\\
\partial_t (\rho_* Y_e) &=& ...\, - {\rm sign}(\nu_i) \alpha \sqrt{\gamma} \frac{\eta - \kappa_a J}{\langle\epsilon_\nu\rangle}\label{eq:sourceRhoYe}\,\,,
\eeqn
where $\rho_*$, $\rho_*Y_e$, $\tau$ and $S_i$ are the conservative hydrodynamics variables which are evolved in SpEC,
\beqn
\rho_* &=& \rho_0 W \sqrt{\gamma} \,\,,\\
\tau &=& \rho_* (hW-1)-\sqrt{\gamma}P\,\,,\\
S_i &=& \rho_* h u_i\,\,,
\eeqn 
$\rho_0$ is the baryon density of the fluid, $P$ its pressure, $Y_e$ its electron fraction, 
and $h$ its specific enthalpy.
$\langle\epsilon_\nu\rangle$ is the weighted average energy of neutrinos, which should be
\beq
\langle\epsilon_\nu\rangle = \frac{\int_0^\infty \left[\eta(\epsilon_\nu) - \kappa_a(\epsilon_\nu) J(\epsilon_\nu)\right] \,d\epsilon_\nu}{\int_0^\infty \frac{\eta(\epsilon_\nu) - \kappa_a(\epsilon_\nu) J(\epsilon_\nu)}{\epsilon_\nu} \,d\epsilon_\nu}\,\,,
\eeq 
and ${\rm sign}(\nu_i)$ is $1$ for $\nu_e$,
$-1$ for $\bar \nu_e$, and $0$ for heavy-lepton neutrinos. Lacking the knowledge of the neutrino
spectrum, we use for $\langle\epsilon_\nu\rangle$ the approximation
\beq
\langle\epsilon_\nu\rangle \approx \frac{F_5(\eta_\nu)}{F_4(\eta_\nu)} T \max{\left(1,\sqrt{\frac{\langle\epsilon^2_{\nu,leak}\rangle}{\langle\epsilon^2_{\nu,fluid}\rangle}}\right)}\,\,,
\eeq
where
\beq
F_k(\eta_\nu) = \int_0^{\infty} \frac{x^k}{1+\exp{(x-\eta_\nu)}}\,dx
\eeq
is the Fermi integral, $\eta_\nu=\mu_\nu/T$ is the degeneracy parameter, and $\mu_\nu$ is the chemical potential
of neutrinos in equilibrium with the fluid.
$\langle\epsilon^2_{\nu,leak}\rangle$ is the global estimate of the average square energy of neutrinos
obtained from the simpler leakage scheme~\cite{Foucart:2014nda}, and $\langle\epsilon^2_{\nu,fluid}\rangle$ would be the
average square energy of the neutrinos if they obeyed a Fermi-Dirac distribution at the local temperature 
$T$ of the fluid and with neutrino chemical potential $\mu_\nu$ (see Appendix~\ref{sec:sources}).
This form for $\langle\epsilon_\nu\rangle$
is exact when $\kappa_a \propto \epsilon_\nu^2$, the neutrinos are in equilibrium
with the fluid, and $\langle\epsilon^2_{\nu,leak}\rangle \leq \langle\epsilon^2_{\nu,fluid}\rangle$ 
-- conditions which, for $\nu_e$ and $\bar\nu_e$, are
approximately satisfied in optically thick regions.
In particular, the last condition is satisfied in post-merger accretion disks because $\langle\epsilon^2_{\nu,leak}\rangle$ is set by the temperature
on the neutrinosphere, which is generally colder than points of higher optical depth in the disk. 
When one of these conditions is not satisfied, on the other hand , it is one of the 
approximations which we have to make to accommodate the use of a gray scheme, 
one that is well motivated by the $\nu^2$ dependence of the dominant absorption
processes, and the relative homogeneity of the fluid temperature around the neutrinospheres.

For the applications that we are considering here, the back-reaction of the neutrinos
onto the fluid evolution is weak (except for transients when we turn on neutrino emission).
Accordingly, we separate the hydrodynamics and neutrino evolution. Our evolution scheme
thus proceeds as follow:
\begin{itemize}
\item Evolve the Einstein equations and the general relativistic hydrodynamics equations, without
taking neutrinos into account, over a time step $\Delta t_{H}$ chosen as in~\cite{Foucart:2013a}, Appendix A.3.
\item Evolve the neutrino radiation, potentially taking multiple time steps, so that the neutrinos
are also evolved by $\Delta t_H$. The time step used to evolve the neutrinos is chosen
as described in Appendix~\ref{sec:timestepping}. Each time step proceeds as follow:
\begin{itemize}
\item Reconstruct the fields $E,F_i/E$ at cell faces, using the $minmod$ reconstruction method.
\item Compute the closure relation at cell faces to get the fluxes $(\alpha \tilde F^i - \beta^i \tilde E)$
and $(\alpha \tilde P^i_j - \beta^i \tilde F_j)$.
\item Use those fluxes to compute the divergence terms in Eqs.~(\ref{eq:Enu}-\ref{eq:Fnu}),
using the shock-capturing methods and corrections in high optical depth regions described in Appendix~\ref{sec:flux}.
\item Compute the closure relation at cell centers.
\item Compute the gravitational source terms on the right-hand side of Eqs.~(\ref{eq:Enu}-\ref{eq:Fnu})
[everything but the terms proportional to $\tilde S^\alpha$] from $E$ and $F_i$ at the beginning of the neutrino step.
\item Solve Eqs.~(\ref{eq:Enu}-\ref{eq:Fnu}) by treating implicitly the collisional source terms proportional
to $\tilde S^\alpha$, following the method described in Appendix~\ref{sec:timestepping}.
\item Compute the coupling to the hydrodynamics variables, and update $(\tau,S_i,\rho_*Y_e)$
according to Eqs.~(\ref{eq:sourcetau}-\ref{eq:sourceRhoYe}).
\item If we have evolved the neutrinos by $\Delta t_H$, go back to the GR-Hydro evolution. Otherwise, take
the next neutrino time step.
\end{itemize}
\end{itemize}

A more complete description of the different steps of this algorithm is provided in Appendices~\ref{sec:closure}-\ref{sec:timestepping}.

\section{Initial Conditions and Numerical Setup}
\label{sec:ID}

As a first astrophysical application of our code, we consider the disk formation phase of a black hole-neutron star merger
from Foucart {\it et al.} 2014~\cite{Foucart:2014nda}. This phase is particularly interesting to study with a general relativistic code
and neutrino transport because general relativistic effects and the evolution of the metric remain important at this point. Furthermore
due to the high temperatures experienced by the fluid during disk formation, the neutrino luminosity is higher and the composition 
evolution is faster than at any other time. Finally the fluid is initially very close to the black hole, where relativistic effects cannot be neglected,
and it is far from equilibrium. This phase of the evolution would thus not be properly captured by simulations which start from an equilibrium torus 
configuration, or which model general relativistic effects through the use of pseudo-Newtonian potentials. 

In the specific merger that we are considering, the masses of the compact objects before merger are $M_{\rm NS}^i=1.4M_\odot$
and $M_{\rm BH}^i=7M_{\odot}$. The initial dimensionless spin of the black hole is $\chi_{\rm BH}^i=0.8$, and it is aligned with the orbital
angular momentum of the binary. The neutron star is initially
non-spinning. We showed in~\cite{Foucart:2014nda} that the disruption of the neutron star results in the ejection of about $0.06M_\odot$ 
of material, and the formation of an accretion disk of mass $M_{\rm disk}\sim 0.1M_\odot$. The final properties of the black hole
are $M_{\rm BH}^f=8M_\odot$ and $\chi_{\rm BH}^f=0.87$. We use the equation of state of Lattimer \& Swesty~\cite{Lattimer:1991nc} with
nuclear incompressibility parameter $K_0=220\,{\rm MeV}$ and symmetry energy $S_\nu = 29.3\,{\rm MeV}$, using a table
available on http://www.stellarcollapse.org and described in O'Connor \& Ott 2010~\cite{OConnor2010}. For a neutron star of mass $M_{\rm NS}=1.4M_\odot$,
this results in a neutron star radius $R_{\rm NS}=12.7\,{\rm km}$.

The post-merger configuration obtained as a result of the merger is expected to be fairly typical for black hole-neutron star mergers in
which the neutron star is disrupted by the black hole. In~\cite{Foucart:2014nda}, we studied the range of initial black hole masses 
$M^i_{\rm BH}=(7-10)\, M_\odot$ and neutron stars masses $M^i_{\rm NS}= (1.2-1.4)\,M_\odot$ currently deemed most likely
from the observation of galactic stellar mass black holes~\cite{Ozel:2010,Kreidberg:2012} and of neutron stars in compact binary 
systems~\cite{Stairs2004,Lorimer2008,2010arXiv1012.3208L}.
The distribution of neutron star and black hole masses may be different for black hole-neutron star binaries, but no such binary
has been observed so far.
For those parameters, a moderate to high initial black hole spin, $\chi^i_{\rm BH}\geq (0.5-0.9)$, is a requirement for the neutron star to be disrupted by the black hole and thus allow the formation
of an accretion disk~\cite{Foucart2012}. Once that condition is satisfied, however, we showed that the local properties of the disk are 
largely independent of the binary parameters.  $10\,{\rm ms}$ after merger, we observe $T\sim 5-10\,{\rm MeV}$, $Y_e\sim 0.15$ and a typical
density $\rho_0^{\rm max}\sim 10^{10-11}\,\rm{g/cm^3}$. Global properties, such as the neutrino luminosity, mostly scale with the mass of the disk.
The accretion disks have masses $M_{\rm disk}\sim (0.05-0.15)\,M_\odot$, optical depths of a few ($\tau_{\nu_e}\sim 5$, $\tau_{\bar\nu_e} \sim \tau_{\nu_x} \sim 1$), 
and neutrino luminosities $L_{\nu_e}\sim L_{\bar\nu_e}\sim 10^{53}\,{\rm erg/s}$ and $L_{\nu_x}\sim 10^{52}\,{\rm erg/s}$ for all 4 species combined.
The only exceptions appear to be the accretion disks formed in mergers with rapidly rotating, relatively low mass black holes 
(see also~\cite{Deaton2013,Lovelace:2013vma}), which can be much more massive [$M_{\rm disk}\sim (0.2-0.5)\,M_\odot$], 
more optically thick, and more luminous (particularly in heavy lepton neutrinos $\nu_x$). They may also develop additional 
hydrodynamics instabilities causing long-term heating in the disk~\cite{Deaton2013}. 

In this paper, we want to assess the effects of a better treatment of the neutrinos on the evolution of the disk and of its immediate
neighborhood. To do so, we compare simulations using the same leakage scheme as in~\cite{Deaton2013,Foucart:2014nda} with simulations using the M1 formalism presented here.
The leakage scheme should give us a rough estimate of the cooling of the disk due to the neutrinos and of the evolution of the composition of the high 
density regions of the disk, but does not include heating or composition evolution due to neutrino absorption in the low-density regions. 
The M1 formalism should give us a better estimate of both the total luminosity and the effects of neutrino-matter
interactions. 

We evolve the post-merger accretion disk with the following treatments of the neutrinos:
\begin{itemize}
\item A ``Leakage'' simulation using the same algorithm as in~\cite{Deaton2013,Foucart:2014nda} .
\item A ``Leakage'' simulation in which the neutrino opacities have been corrected as in our M1 code (to guarantee that the energy
density of neutrinos in the optically thick regions is correct, see Appendix~\ref{sec:sources}).
\item A ``M1'' simulation using the standard methods described in this paper.
\item A ``M1'' simulation with a simpler treatment of the emissivity in optically thin regions (i.e. without the correction described in Eq.~\ref{eq:etacorr}),
to get a first estimate of the errors due to the use of a gray scheme: we find in our test of the neutrino-fluid interactions in a post-bounce supernova profile 
(see Appendix~\ref{sec:testcollapse}) that variations in the results with
and without that correction are comparable to the difference between the gray results and the results of an energy-dependent code.
\item Three ``M1'' simulations with a numerical grid covering a larger volume and improved boundary conditions (see below), with and without the correction 
to the emissivities described in Eq.~\ref{eq:etacorr}, and with and without the correction to the opacities described in Eq.~\ref{eq:kcorr}.
\end{itemize}

\begin{figure}
\flushleft
\includegraphics[width=1.\columnwidth]{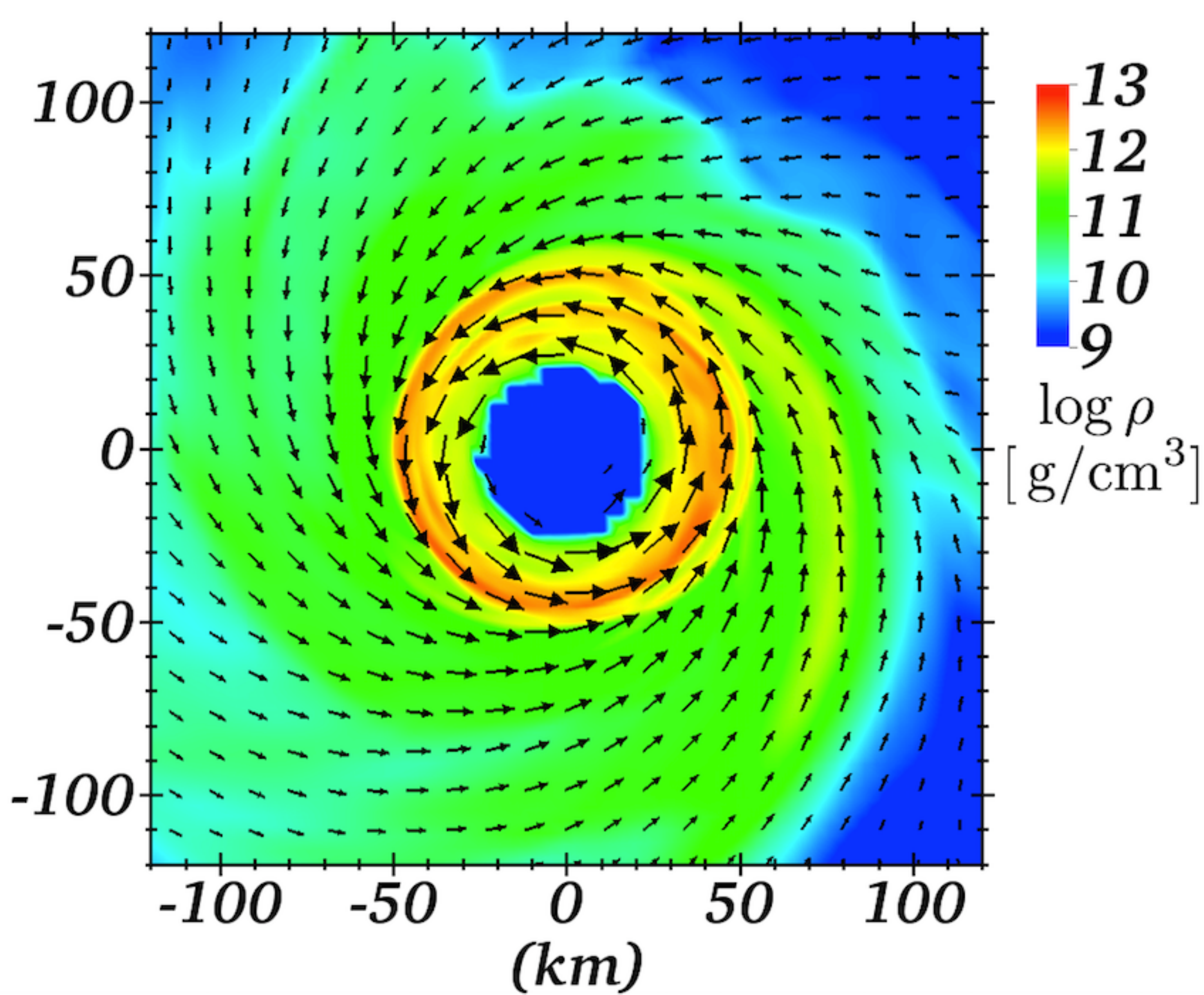}
\caption{Density and velocity field in the equatorial plane of the black hole at the initial time $t_0 \sim t_{\rm merge}+6.1\,{\rm ms}$.
For reference, the velocity at the peak of the density distribution is about $0.6c$.}
\label{fig:rhoInit}
\end{figure}

As initial configuration, we consider a snapshot of simulation M14-7-S8 of~\cite{Foucart:2014nda} at $t_0=t_{\rm merge}+6.1\,{\rm ms}$,
where $t_{\rm merge}$ is the time at which half of the neutron star material has been accreted by the black hole. This is around the time
when an accretion disk is forming and neutrino effects begin to significantly affect its evolution. The initial configuration is shown in Fig.~\ref{fig:rhoInit}.
From the disruption of the neutron star to $t_0$, the 
hydrodynamics equations were evolved using two levels of refinement, each with $100^2 \times 50$ grid points and with grid spacings 
$\Delta x\approx (1.5\,{\rm km},3\,{\rm km})$
in the equatorial plane and $\Delta z\approx (0.5\,{\rm km},1\,{\rm km})$ in the vertical direction. 
For comparison, the horizon of the black hole has a radius $r_{\rm BH}\sim 25\,{\rm km}$ in the coordinates of our simulation,
the peak of the matter density in the accretion disk is initially at $r_{\rm disk}\sim 50\,{\rm km}$, and the scale height of the disk is
$H\sim (0.2-0.3)r$. We chose the grid structure in order
for the finest grid level to cover the forming accretion disk. We also used the equatorial symmetry of the system to only evolve the $z\geq 0$ region.
Note that, in SpEC, Einstein's equations are evolved on a separate pseudospectral grid extending $3700\,{\rm km}$ away from the center of the black hole.
This is why we can use such a small finite difference grid if we only want to study the evolution of the post-merger accretion disk. A smaller grid
means that we can maintain a reasonable resolution in the disk at a fairly low computational cost\footnote{Although such a resolution is only reasonable
because we do not include the effects of magnetic fields. It would be much too coarse to capture the effects of MHD instabilities.}, 
but also that the unbound material, and some of the bound
material in the tidal tail formed when the neutron star is disrupted, is allowed to leave the grid. 
By $t_0$, we still have a baryonic mass $M_b=0.12M_\odot$ on the grid, while we have allowed 
$0.06M_\odot$ of unbound material and $0.06M_\odot$ of bound material to escape. In~\cite{Foucart:2014nda}, we also performed lower resolution
simulations following the material in the tidal tail farther out. From these simulations, we can deduce that the material that we allowed to escape
would start to fall back onto the disk at $t\sim t_{\rm merge}+20\,{\rm ms}$. At later times, material neglected in our simulation may affect the evolution
of the disk. 

Our simulations do not include the effects of magnetic fields, and the only viscosity is due to the finite resolution of our numerical grid.
Evolving for much longer would thus not be very informative. The growth of magnetic fields due to the magnetorotational instability is expected to act on a timescale comparable to the orbital timescale of the disk, i.e. a few milliseconds, and angular momentum transport due to magnetic turbulence is 
expected to act on timescales $t_{\rm visc}\sim(30-700)\,{\rm ms}$ (see Sec.~\ref{sec:diskEv}). 
Accordingly, we only study the evolution of the post-merger disk from $t_0$ to $t_f = t_{\rm merge}+20\,{\rm ms}$. We will see that this is long enough for the fluid and the neutrinos
to reach a quasi-equilibrium state. This also allows us to study the main differences between simulations using the leakage scheme and simulations using the M1 formalism. Finally, we can get a first estimate of the outflows emitted in the region above the disk and of the effects of interactions between the disk and the tidal tail.

For the evolution between $t_0$ and $t_f$, we use a different finite difference grid. Indeed, although the system still respects the equatorial symmetry after merger, 
small perturbations which are not equatorially symmetric might grow in the disk due to
hydrodynamical instabilities. To make sure that our grid structure does not artificially suppress such perturbations, we remove the assumption
of equatorial symmetry.
We also want to capture the radial growth of the disk, potential matter outflows, and the evolution of the bound material in the tidal tail remaining on the grid (which is still expanding at $t=t_0$). 
For the post-merger evolutions, we thus use 
a third level of refinement (with $\Delta x\sim 6\,{\rm km}$ in the equatorial plane and $\Delta z \sim 2\,{\rm km}$ in the vertical direction), keeping as before $\Delta x\approx (1.5\,{\rm km},3\,{\rm km})$
in the equatorial plane and $\Delta z\approx (0.5\,{\rm km},1\,{\rm km})$ in the vertical direction for the finer levels. Each refinement level has $100^3$ grid points. In three of the M1 simulations, 
a fourth level is added with $\Delta x=12\,{\rm km}$ and $\Delta z=4{\rm km}$. 
This was done because, with only 3 levels of refinement, we cannot follow the evolution of low-density
material far enough along the direction of the black hole spin axis to extract information about potential disk winds.
We note that, due to our choice of map between the grid and the laboratory frame, the resolution in the grid is not uniform. We quote here the resolution close to the black hole, but the grid spacing far from the black hole is actually $\sim 30\%$ coarser.

We also performed simulations with different numerical resolution, to assess our numerical errors. A detailed discussion of these errors is provided
in Appendix~\ref{sec:accuracy}. In summary, we find that the errors due to finite resolution are at most comparable to the errors due to the use of a
gray scheme, and much smaller than the errors in the leakage scheme.

As opposed to previous SpEC simulations, we want here to study the behavior of low-density regions.
Accordingly, for these last M1 simulations we also made a few modifications to the handling of low-density matter.
In general relativistic hydrodynamics simulations, the region surrounding the neutron star or accretion disk is generally
filled with lower-density material in order to avoid evolving towards either negative densities or unphysical values of
the evolved variables. Additionally, corrections are applied to all regions of baryon density $\rho_0$ lower than an ``atmosphere'' threshold
$\rho_0^{\rm atm}$. In our case, those corrections set the temperature to $T=0.5\,{\rm MeV}$ and the spatial components of the 4-velocity to $u_i=0$, 
although different prescriptions also work in practice. This generally results in large portions of the grid being filled with 
material at $\rho_0 \sim \rho_0^{\rm atm}$. In previous SpEC simulations with the LS220 equation of state~\cite{Deaton2013,Foucart:2014nda},
that threshold was set at $\rho_0^{\rm atm}=10^{-5}\rho_0^{\rm max}$, with $\rho_0^{\rm max}$ the maximum value of the baryon
density on the grid at the current time. In accretion disks, this leads to $\rho_0^{\rm atm}\sim 10^{6-7}\,{\rm g/cm^3}$. Due to larger numerical
errors close to the black hole, $\rho_0^{\rm atm}$ is also higher in a region about one apparent horizon radius away from the black hole horizon, 
by as much as a factor of $100$ on the horizon itself. 
In this work, we are interested in disk outflows with density $\rho_0\sim 10^{7-8}\,{\rm g/cm^3}$ in the regions
covered by our grid. Such outflows cannot be launched with that high an atmosphere threshold. This is in part because of the pressure
exerted by the atmosphere material, and in part because any material decompressing to $\rho_0^{\rm atm}$ is immediately slowed down
to $u_i=0$, and falls back onto the disk. We thus modify our choice of atmosphere threshold to
\beqn
\rho_0^{\rm atm} &=& \rho_0^{\rm floor} +\rho_0^{\rm max} \bigg[10^{-5} \left(\frac{2r_{\rm AH}}{r_{\rm AH}+r}\right)^2 \\
&+& 10^{-3} \exp{\left(-\left(\frac{r-r_{\rm AH}}{0.5r_{\rm AH}}\right)^2\right)}\bigg]\,\,,
\nonumber
\eeqn
with $r_{\rm AH}$ the grid-coodinate radius of the black hole apparent horizon, which is by construction a sphere in the coordinates of our numerical grid. 
$r$ is the distance to the center of the black hole in the same coordinates, and $\rho_0^{\rm floor} \sim 10^{5}\,{\rm g/cm^3}$. 
With this choice, the atmosphere threshold remains as before close to the black hole,
but now drops to $\rho_0^{\rm atm} \sim \rho_0^{\rm floor}$ at larger distances, which is sufficient for our current purpose.
 A lower threshold would be required if we wanted to follow disk
outflows farther from the black hole.

Finally, in previous simulations the table containing the equation of state information only covered the range 
$10^8\,{\rm g/cm^3}<\rho_0<10^{16}\,{\rm g/cm^3}$, and was artificially extended to lower densities using a simple ideal gas equation of state.
In this last simulation, we instead use a table going down to $10^5\,{\rm g/cm^3}$. The new table uses 345 logarithmically spaced points to cover that
range of densities, instead of 250 for the smaller table. Both tables are also discretized in temperature ($0.01\,{\rm MeV}<T<251\,{\rm MeV}$ with 136
logarithmically spaced points) and composition ($0.035<Y_e<0.53$ with 50 linearly spaced points). Below $10^8\,{\rm g/cm^3}$, the compositional
information of the equation of state is approximated by the nuclear statistical equilibrium of matter at $10^8\,{\rm g/cm^3}$, for the same temperature
and electron fraction (see O'Connor \& Ott 2010~\cite{OConnor2010} for details).
 
\section{Disk Evolution}
\label{sec:disk}

In the following, we describe the physical results of the simulations. We mainly focus on the M1 simulation using our standard
algorithm, including the improved treatment of the low-density regions for the hydrodynamical variables, as it offers the most accurate predictions. 
The effects of different neutrino
treatments are discussed in Sec.~\ref{sec:nutreat}.

\subsection{Global properties of the disk}
\label{sec:diskEv}

At the beginning of the simulation ($t_0=t_{\rm merge}+6.1\,{\rm ms}$), we have about $0.08M_\odot$ of material in a forming accretion disk extending 
$\sim 60\,{\rm km}$ away from the black hole. As shown in Fig.~\ref{fig:rhoInit}, this material is still far from being circularized or axisymmetric.
Another $0.04M_\odot$ of material is on the grid in an extended tidal tail with a relatively flat density profile. As discussed in the previous section, an 
additional $0.06\,M_\odot$ of unbound material and $0.06\,M_\odot$ of bound material with fallback time longer than $\sim 20\,{\rm ms}$ were allowed to escape the grid before $t_0$.
The initial configuration has a sharp density peak around $r=45\,{\rm km}$. Most of the tail material is cold ($T<1.5\,{\rm MeV}$), while the disk material has a broad temperature distribution,
with most of the matter at temperatures $2\,{\rm MeV}<T<12\,{\rm MeV}$. The composition of the disk and tail is sharply peaked at $0.05<Y_e<0.07$, and the small amount of material at $Y_e>0.1$ is in the hot regions close to the black hole and rapidly accreted.

Although neutrinos have an important effect on the evolution of the system, to first order, purely hydrodynamical effects dictate the evolution. 
During the time period considered in this simulation, $5\,{\rm ms}-20\,{\rm ms}$ after merger, the circularization of the disk material is the most important effect. At the initial time, most of the disk material is at or close to periastron. Shock heating
and the contraction of the disrupted material cause the fluid to heat to its maximum average temperature, $\langle T\rangle=6.4\,{\rm MeV}$ at $t=t_0+1.5\,{\rm ms}$
\footnote{Here and in the rest of the text, the average fluid properties refer to density-weighted averages}.
Afterwards, the forming disk goes through a damped cycle of expansions and contractions, with a period of about $6\,{\rm ms}$. During each expansion
period, the temperature of the fluid and the accretion rate decrease. During contractions, shock heating causes the temperature to rise, and the accretion rate increases. Neutrino emissions and absorptions, although not critical to the dynamics, contribute to a smoothing of the temperature distribution and
determine the composition of the fluid. Their total luminosity mostly follows the oscillations of the fluid temperature. Energy lost to neutrino emissions and to the accretion of hot material
onto the black hole also causes a slower global cooling of the fluid, by about $1\,{\rm MeV}$ over the $14\,{\rm ms}$ of evolution.

\begin{figure}
\flushleft
\includegraphics[width=1.\columnwidth]{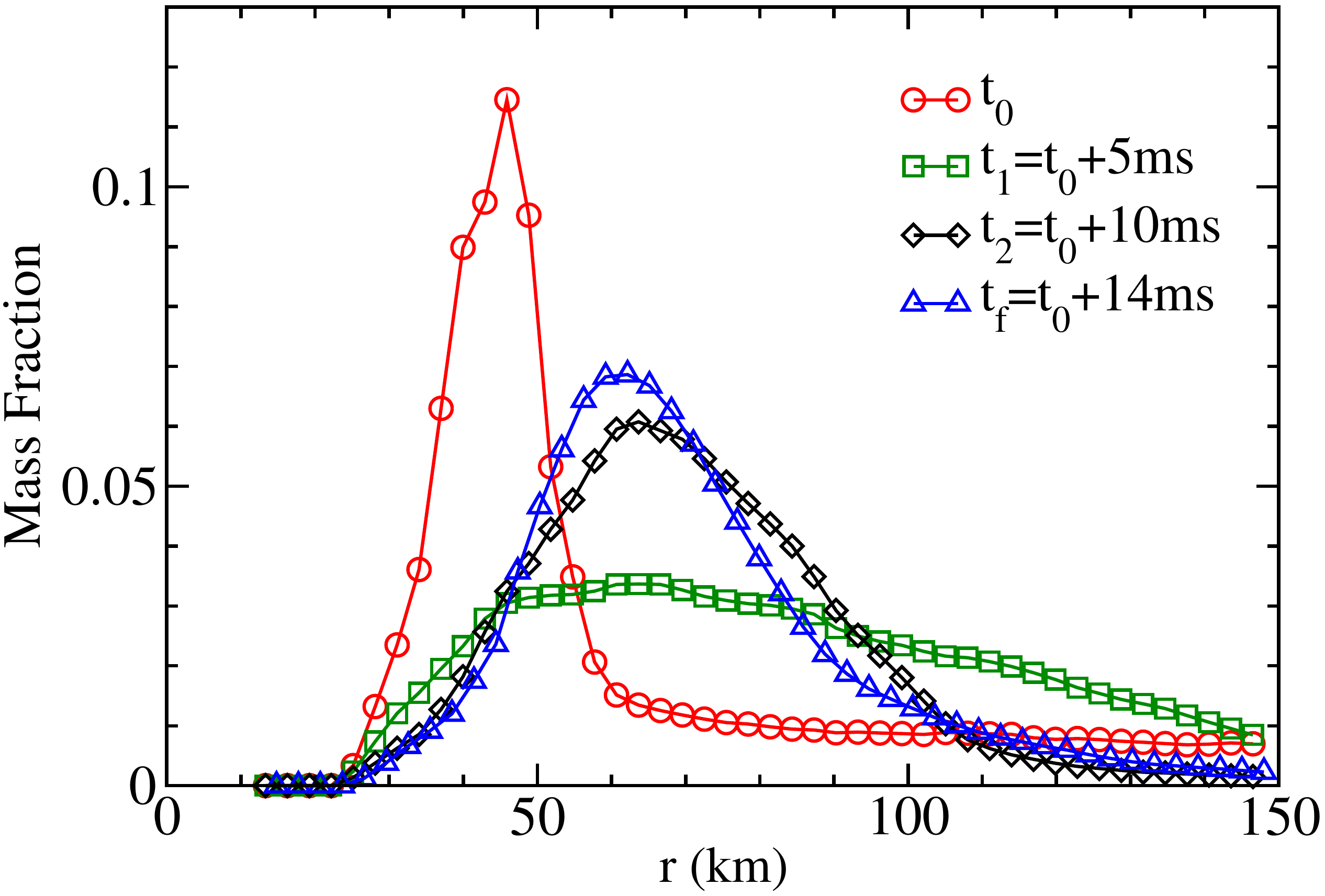}
\caption{Radial distribution of the material outside of the black hole. We plot the fraction of the total mass within bins of width 
$\Delta r \approx 3\,{\rm km}$ at 4 representative times of the simulation. Most of the matter is initially near the black hole
($r\lesssim 60\,{\rm km}$) and close to periastron. The disk then expands, and later circularizes.}
\label{fig:RhoEv}
\end{figure}

\begin{figure}
\flushleft
\includegraphics[width=1.\columnwidth]{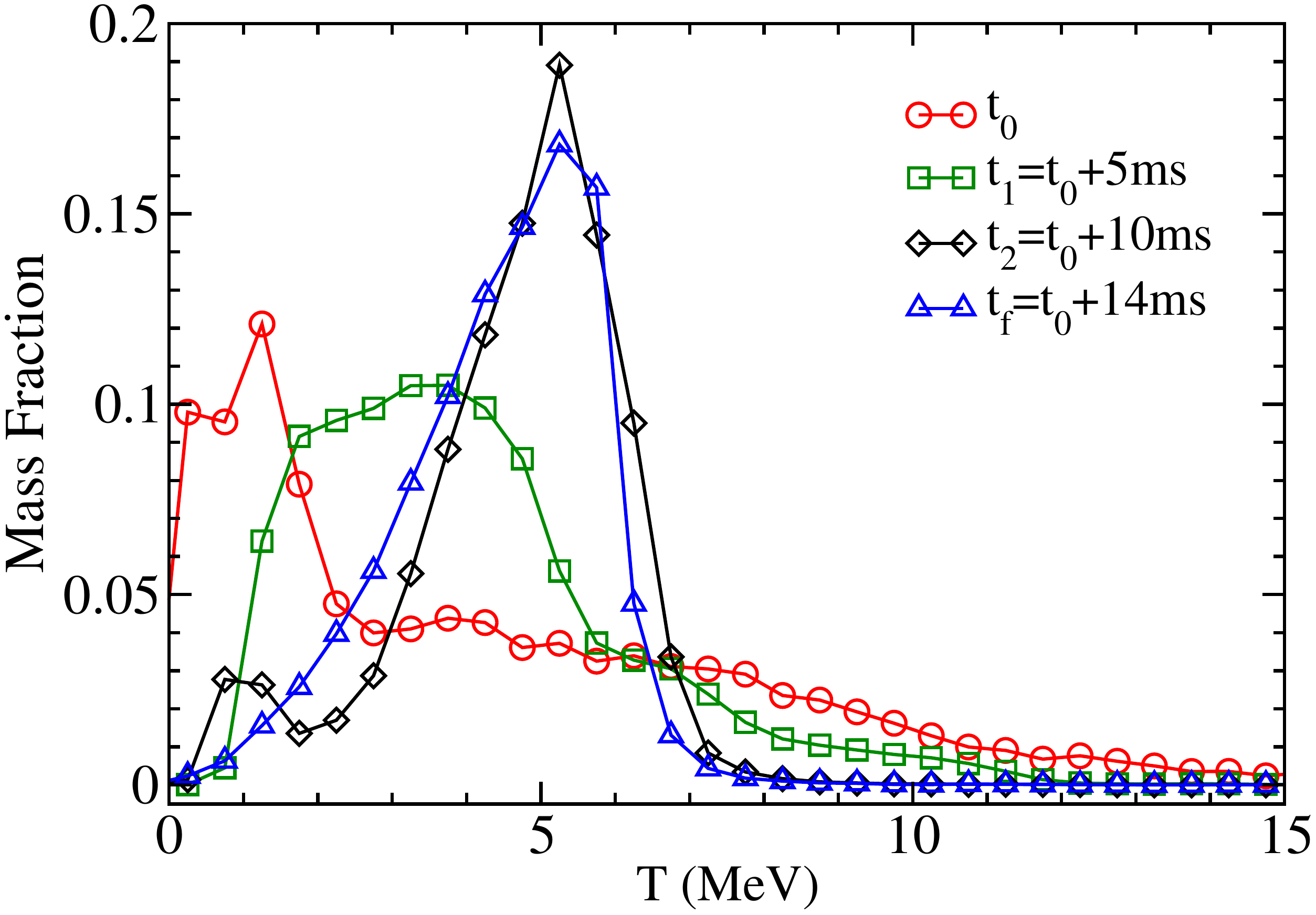}
\caption{Fraction of the total mass within temperature bins of width $\Delta T = 0.5\,{\rm MeV}$, plotted at 4 representative times of the simulation.
Within $10\,{\rm ms}$, the temperature of the disk becomes very homogeneous.}
\label{fig:TEv}
\end{figure}

\begin{figure}
\flushleft
\includegraphics[width=1.\columnwidth]{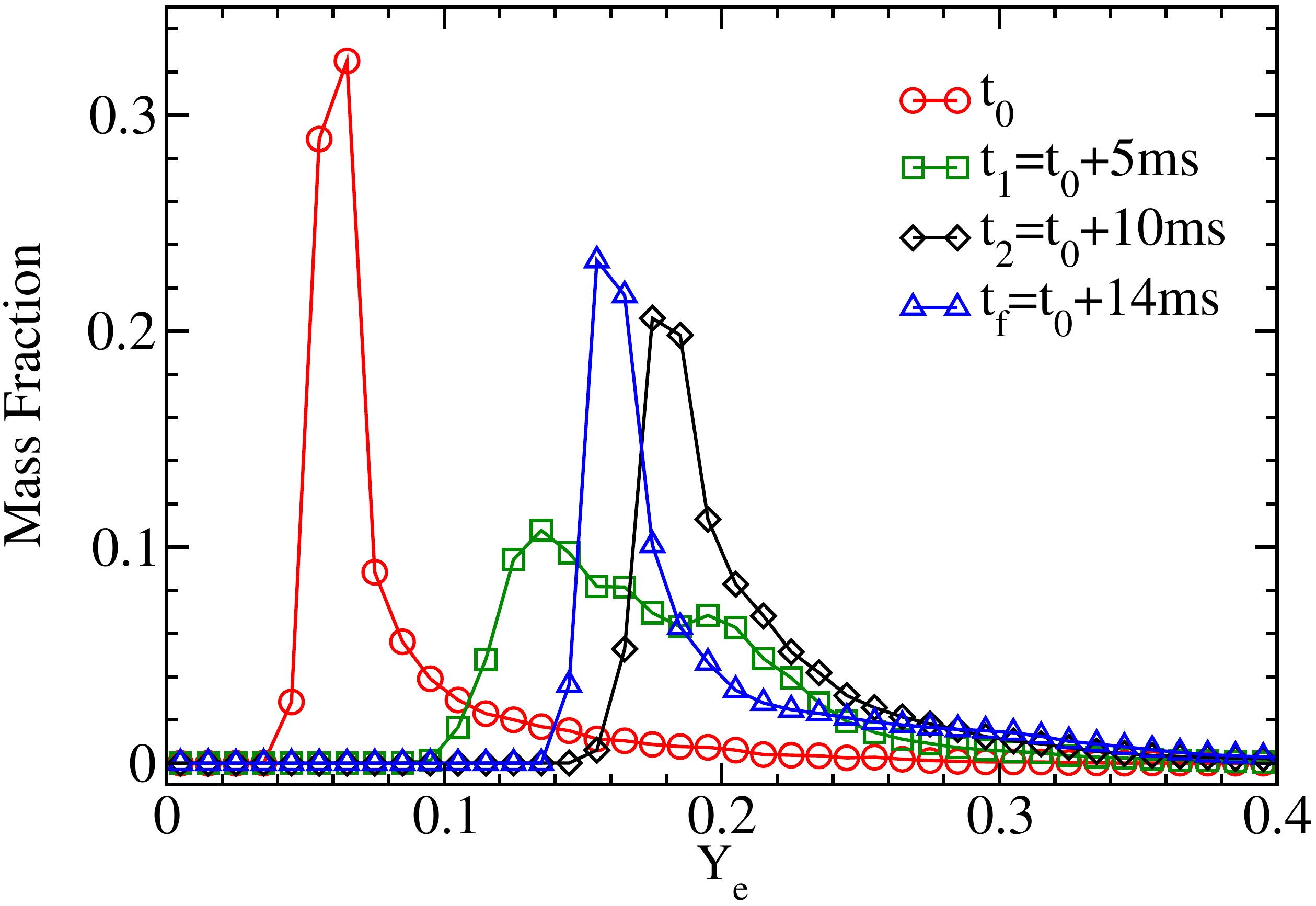}
\caption{Fraction of the total mass within electron fraction bins of width $\Delta Y_e = 0.01$, plotted at 4 representative times of the simulation.
Within $10\,{\rm ms}$, the electron fraction of the disk becomes very homogeneous. The tail of higher-$Y_e$ material is due to lower
density regions around the disk.}
\label{fig:YeEv}
\end{figure}

We look at the evolution of the disk through snapshots of the simulation taken at $t_0$, $t_1=t_0+5\,{\rm ms}$, $t_2=t_0+10\,{\rm ms}$, and $t_f=t_0+14\,{\rm ms}$. 
We plot the fraction of the local mass observed at a given radius in Fig.~\ref{fig:RhoEv}, at a given temperature in Fig.~\ref{fig:TEv}, and at a given electron fraction in Fig.~\ref{fig:YeEv}. All plots are normalized to the total mass on the grid. Due to accretion onto the black hole, the total mass decreases from $0.12\,M_\odot$ at $t_0$ to
$0.097\,M_\odot$ at $t_1$, $0.075\,M_\odot$ at $t_2$, and $0.069\,M_\odot$ at $t_f$. By the end of the simulation, nearly all of the remaining material is in a circularized accretion disk. We stop the simulations at $t_f=t_{\rm merge}+20\,{\rm ms}$, when the material which was allowed to leave the numerical grid
should begin to affect the evolution of the disk.

The density evolution (Fig.~\ref{fig:RhoEv}) is mostly a consequence of the accretion of material at the inner edge of the disk and the circularization of 
disk and tail material. 
At first, the large eccentricity of most of the disrupted material causes the matter distribution to broaden. This is most visible in the difference
between $t_0$ and $t_1$, which correspond to times close to the moment of maximum contraction and maximum expansion of the forming disk.
Afterwards, a more slowly evolving disk forms with a density peak  $r\sim 60\,{\rm km}-70\,{\rm km}$, and some milder oscillations.  
Turbulent angular momentum transport due to the magnetorotational instability, which is not modeled here, should eventually take
over and cause a viscous spreading of the disk. This should happen over the viscous timescale $t_{\rm visc}\sim (\alpha \delta^2 \Omega)^{-1}$, 
where $\alpha$ is the standard viscosity parameter,  $\delta=H/r$ and $H$ is the scale height of the disk. 
For the expected $\alpha\sim 0.01-0.1$ and the simulation $\delta\sim 0.2-0.3$, we get $t_{\rm visc}\sim (30-700)\,{\rm ms} > t_f-t_0 = 14\,{\rm ms}$.
We thus expect angular momentum transport due to magnetically-driven turbulence to be fairly unimportant on the timescales considered here.

The temperature evolution (Fig.~\ref{fig:TEv}) is largely a result of mixing in the fluid, shock heating during the circularization of the disk, 
and energy transport by neutrinos. 
Over the first $10\,{\rm ms}$ of evolution, the temperature goes from a wide, nearly flat distribution with $2\,{\rm MeV}<T<12\,{\rm MeV}$ to a more uniform temperature
$T\sim 4\,{\rm MeV}-6\,{\rm MeV}$. Some mixing of fluid elements with different temperatures does occur in purely hydrodynamical simulations,
or in simulations using a simple cooling prescription for the neutrinos (leakage runs). But the process is significantly more efficient when using a transport scheme, as now neutrinos
can transport energy from hot parts of the disks to cool parts of the disk. The homogenization of the temperature is helped, in both transport and leakage simulations, by
the steep dependence of the neutrino emissivities on the temperature, which causes the hottest points in the disk to cool much more rapidly than their neighbors. But Fig.~\ref{fig:TEv}
shows that, in addition to the rapid cooling of the hottest points, neutrino transport also causes a heating of the cooler points in the disk.
Variations in the average temperature of the fluid are due mostly to the oscillations observed as the disk circularizes.

As for the density, magnetically-driven turbulence is expected to affect the evolution of the temperature over long time scales. Newtonian simulations have
shown that an isotropic viscosity can heat the disk to maximum temperatures of 
$T_{\rm max} \sim 3\,{\rm MeV}-10\,{\rm MeV}$ for $\alpha \sim 0.001-0.1$~\cite{Setiawan2006}. For the largest viscosities, viscous heating would be relevant to the thermal evolution of the disk towards the end of our simulations. For lower viscosities, it would remain largely irrelevant until the disk cools down, but
magnetically-driven turbulence might still hasten the homogenization of the disk temperature.
As viscous heating is the main process stopping the cooling of the disk, it is of course always relevant to the long term evolution of the disk.

Finally, Fig.~\ref{fig:YeEv} shows that the disk also reaches its equilibrium composition within about $10\,{\rm ms}$. By that point the disk is in 
a quasi-equilibrium state in which charged-current reactions no longer cause net changes of the electron fraction of the disk.
The composition then evolves more slowly as the properties of the disk change, adapting nearly instantaneously to the disk evolution. At the temperatures and electron fractions observed in most of the disk, matter becomes less neutron rich when the disk heats up, and more neutron rich as it cools down.
The electron fraction also decreases for denser material. At times beyond $20\,{\rm ms}$ past merger, the disk is expected to cool and
the core of the disk will reneutronize, at least until viscous heating balances neutrino cooling. However, this is not necessarily true for the high-$Y_e$ tail of 
the distribution which corresponds to cooler, lower density points whose composition has been significantly affected by neutrino
irradiation and which evolve fairly slowly at late times. 

\subsection{Final disk configuration}

From the discussion in Sec.~\ref{sec:diskEv}, 
we expect that at $t_f$ the simulated hydrodynamical properties of the accretion disk (density, temperature, composition) are probably a
good representation of the state of the system about $20\,{\rm ms}$ after merger. Magnetically-driven turbulence and magnetically-driven winds could affect the results.
But so far simulations have found that for reasonable initial values of the magnetic fields, magnetic outflows only appear more than $100\,{\rm ms}$ after the merger~\cite{Paschalidis2014},
or are not observed at all~\cite{2012PhRvD..85f4029E,Etienne:2012te} \footnote{The main difference between the two sets of simulations is
the inclusion of an initial dipole magnetic field outside of the neutron star in~\cite{Paschalidis2014}.}.
As discussed in the previous section, MRI-driven angular momentum transport is also expected to occur on timescales
longer than the duration of our simulation. In this section, we thus offer a more detailed description of the disk at the end of the simulation,
assuming that it is a fairly good description of a post-merger accretion disk $\sim 20\,{\rm ms}$ after a black hole-neutron star merger.

\begin{figure}
\flushleft
\includegraphics[width=1.0\columnwidth]{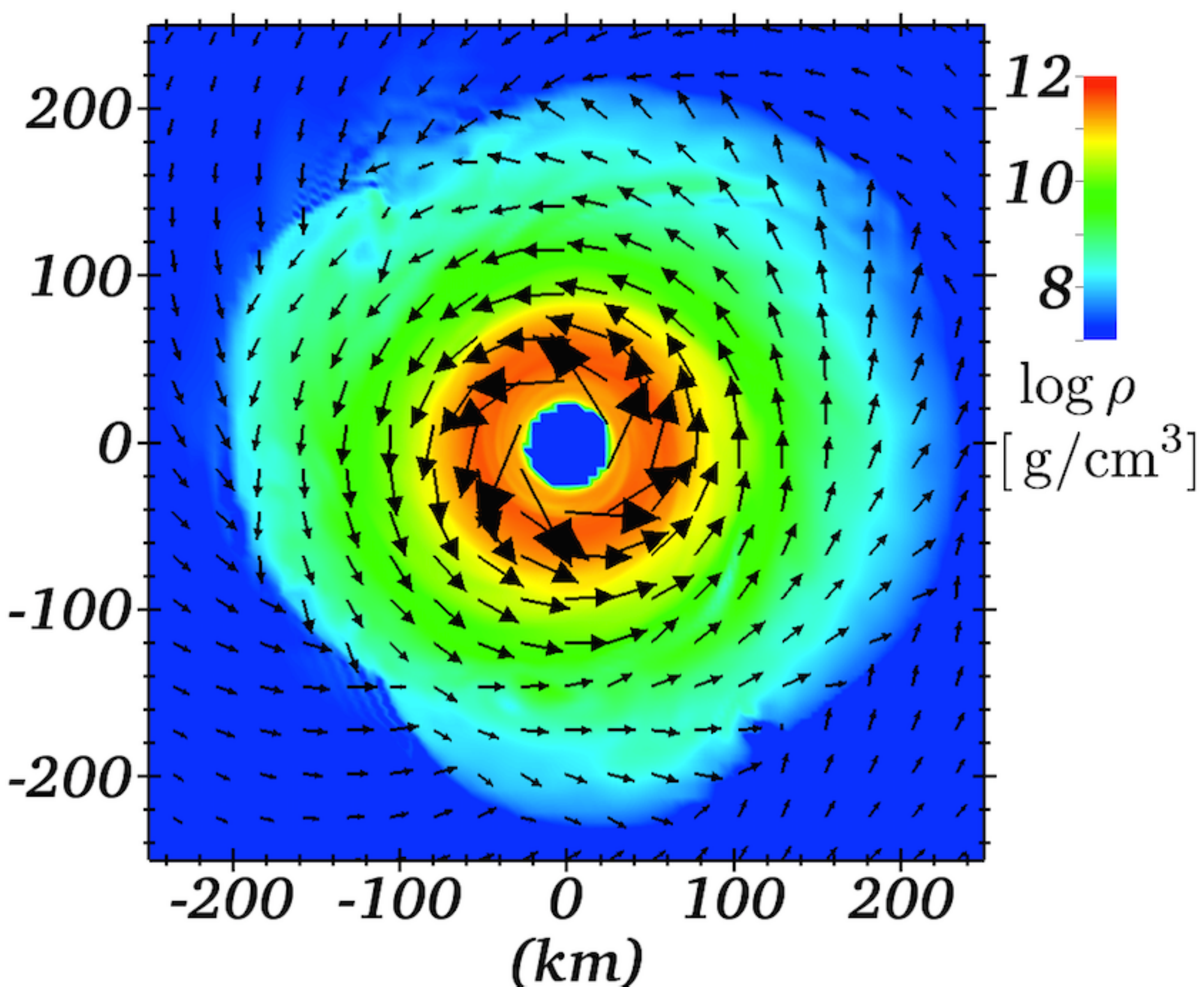}
\caption{Density and velocity field in the equatorial plane of the black hole at the end of the simulation, $20\,{\rm ms}$ after merger. 
For scale, the velocities at the peak of the
density distribution are $v\sim 0.5c$.}
\label{fig:rhoEq}
\end{figure}

\begin{figure}
\flushleft
\includegraphics[width=1.\columnwidth]{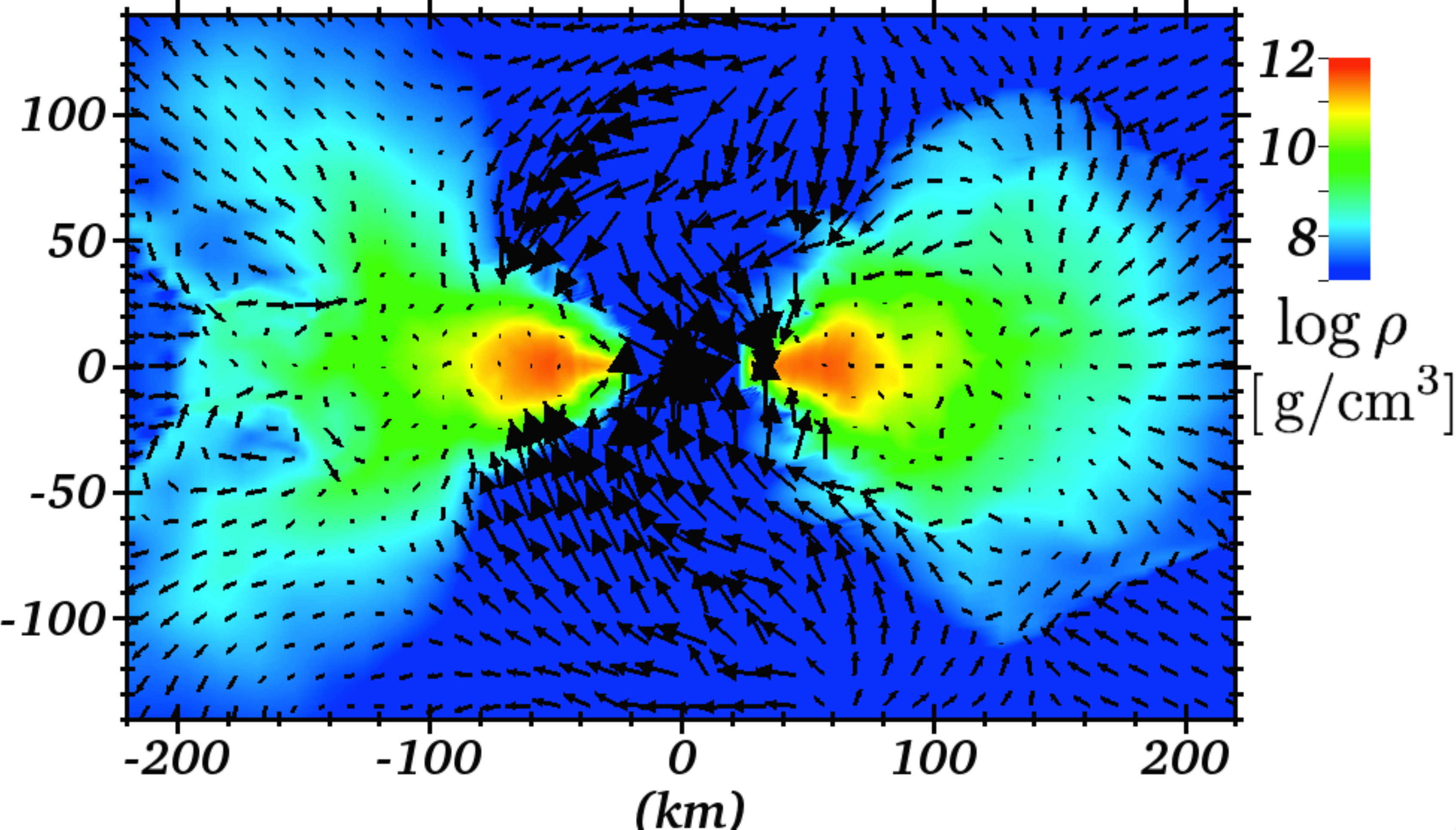}
\caption{Density (in ${\rm g/cm^3}$) and velocity field in the $y=y_{\rm BH}$ plane at the end of the simulation, $20\,{\rm ms}$ after merger.
For reference, the largest poloidal velocities are about $0.6c$. In the core of the disk, the velocities are mostly directed out of the 
plane of the visualization.}
\label{fig:rho}
\end{figure}

We first show the density and velocity field in the equatorial plane of the black hole (Fig.~\ref{fig:rhoEq}), and in the vertical plane 
$y=y_{\rm BH}$ (Fig.~\ref{fig:rho}). Here, we define the velocity as the ``transport velocity'' $v^i_T = \frac{\alpha}{W} u^i$,
which satisfies $\partial_t \rho_* + \partial_i (\rho_* v^i_T) =0$. This is a convenient definition of the velocity for visualization purposes, 
as it directly shows the motion of fluid elements on the grid with all general relativistic coordinate effects taken into account.
From Fig.~\ref{fig:rhoEq} we infer that the disk is well circularized up to $r\sim 100\,{\rm km}$. Beyond that radius, we have either the last remnants of the tidal tail
falling back onto the disk (upper left quadrant), or low density material expanding into the neighboring medium. 
Most of the mass is in the inner region of the disk,
at densities $\rho_0 \sim 10^{11-12} {\rm g/cm^3}$. But beyond $r\sim 70{\rm km}$, the density drops rapidly to $\rho_0\sim 10^{8-9}{\rm g/cm^3}$ at 
the edge of the disk. The velocity in the inner disk is mildly relativistic ($v_T\sim 0.5c$ at $r\sim 60{\rm km}$) which, as we shall see, has important effects on the geometry of the neutrino radiation.

Fig.~\ref{fig:rho} shows some interesting additional features. The core of the disk at the end of our simulation 
is of moderate geometrical thickness ($H/r\sim 0.2$), 
and still noticeably asymmetric (note that the density is plotted on a logarithmic scale). The region close to the spin axis of the black hole
is free of any material, except for the accretion of low density material from the disk at height $z \leq 80\,{\rm km}$, with 
$\rho_0 \sim 10^{7-8}\,{\rm g/cm^3}$. There are however resolved outflows coming from the contact regions between the disk and the tail (left side of the plot),
at densities $\rho_0\sim 10^8\,{\rm g/cm^3}$. At the grid boundary, this material has not reached the escape velocity (it has $u_t>-1$).
But most of the outflowing fluid has enough energy to be unbound (i.e.\ it satisfies the condition $h u_t <-1$, thanks to temperatures $T\sim 2\,{\rm MeV}$).
We discuss the disk/tail interactions in more detail in Sec.~\ref{sec:disktail}, and the unbound material in Sec.~\ref{sec:unbound}. 
For now, we simply note that the outflows, although helped by neutrino heating and radiation pressure, are mostly a consequence of purely hydrodynamical interactions at the disk/tail interface, as can be verified by their existence in the simulations using a leakage scheme. Earlier in the evolution, shocks during the circularization power similar outflows, albeit more axisymmetric and closer to the black hole spin axis.

On the opposite side of the disk, we observe a more homogeneous expansion of low-density material. This expansion should in practice be stopped once 
the outflowing disk material begins interacting with the tail material neglected in this simulation, i.e.\ the material falling back onto the disk on a timescale longer
than the duration of the simulation. 

\begin{figure}
\flushleft
\includegraphics[width=1.\columnwidth]{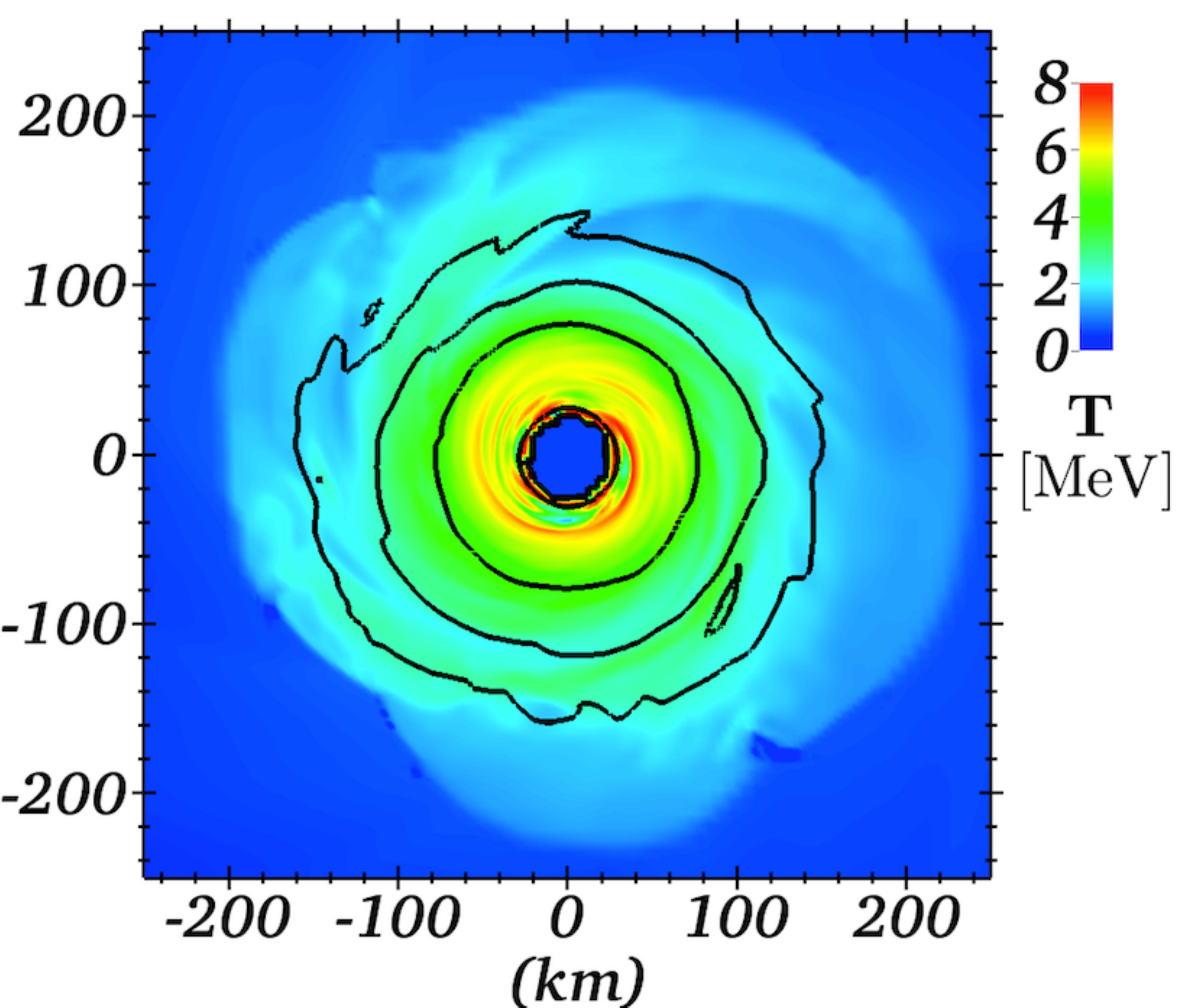}
\caption{Temperature  in the equatorial plane of the black hole at the end of the simulation, $20\,{\rm ms}$ after merger. Density contours are shown as thick black lines
for $\rho_0=10^{9,10,11}\,{\rm g/cm^3}$.}
\label{fig:TempEq}
\end{figure}

\begin{figure}
\flushleft
\includegraphics[width=1.\columnwidth]{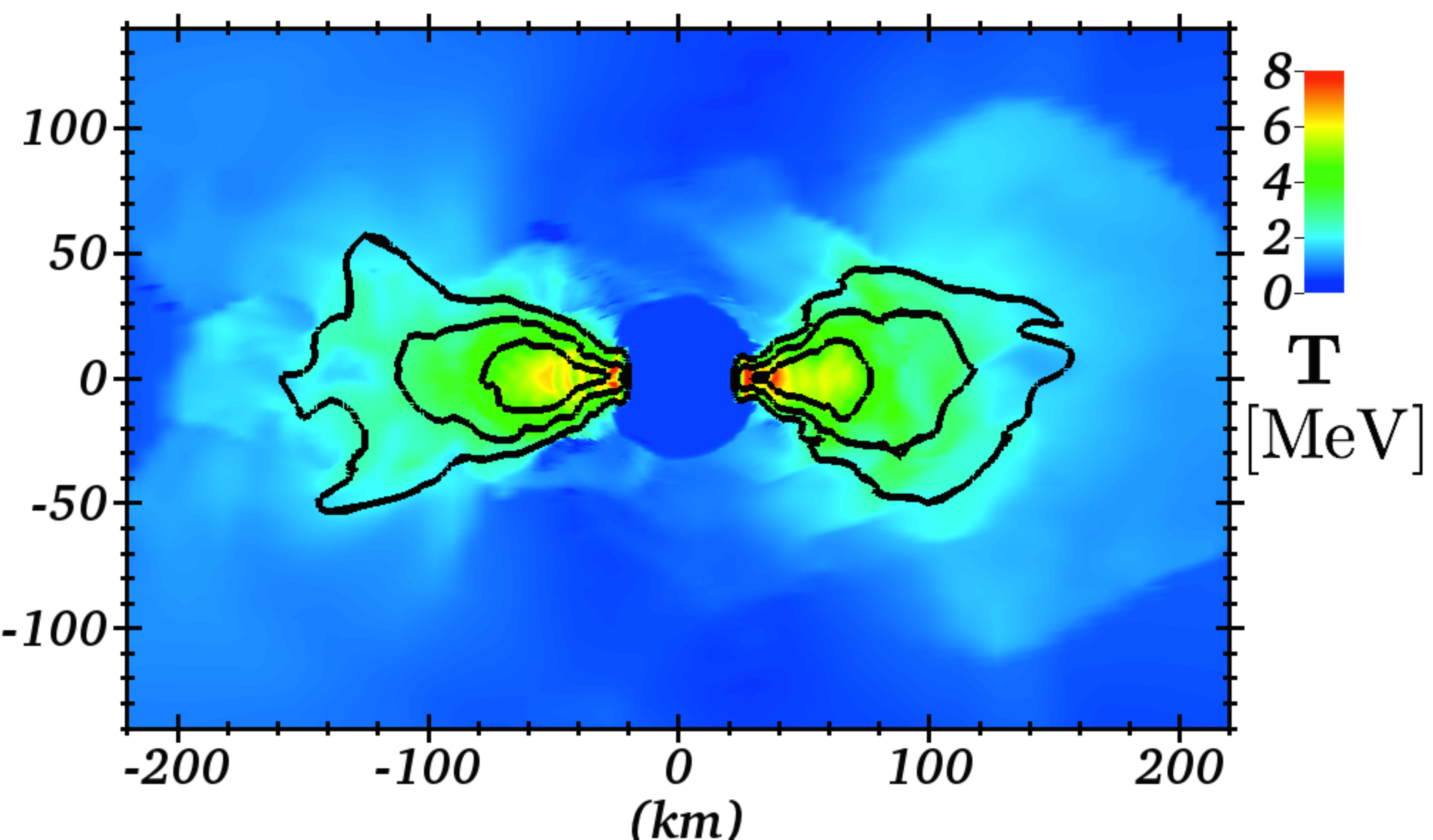}
\caption{Temperature  in a plane orthogonal to the equatorial plane of the disk the at the end of the simulation, $20\,{\rm ms}$ after merger. Density contours are shown as thick black lines for $\rho_0=10^{9,10,11}\,{\rm g/cm^3}$.}
\label{fig:Temp}
\end{figure}

Similar visualizations for the temperature are shown in Fig.~\ref{fig:TempEq} and Fig.~\ref{fig:Temp}. Asymmetric features in the temperature distribution
remain clearly visible despite the smoothing effect of the neutrino cooling and heating. Nearly all of the material, including the tidal tail, has been heated to 
$T>1\,{\rm MeV}$, by a combination of shocks and neutrino absorption. Denser material with $\rho_0 \gtrsim 10^{11}\,{\rm g/cm^3}$ 
is heated to $T\gtrsim 4\,{\rm MeV}$. The relatively small variations of the temperature are one of the main reasons a gray scheme for the neutrino
radiation is more reliable for merger simulations than in core collapse supernovae, where large temperature gradients exist and small changes
in the interactions between the neutrinos and the fluid can significantly modify the dynamics of the system.

\begin{figure}
\flushleft
\includegraphics[width=1.\columnwidth]{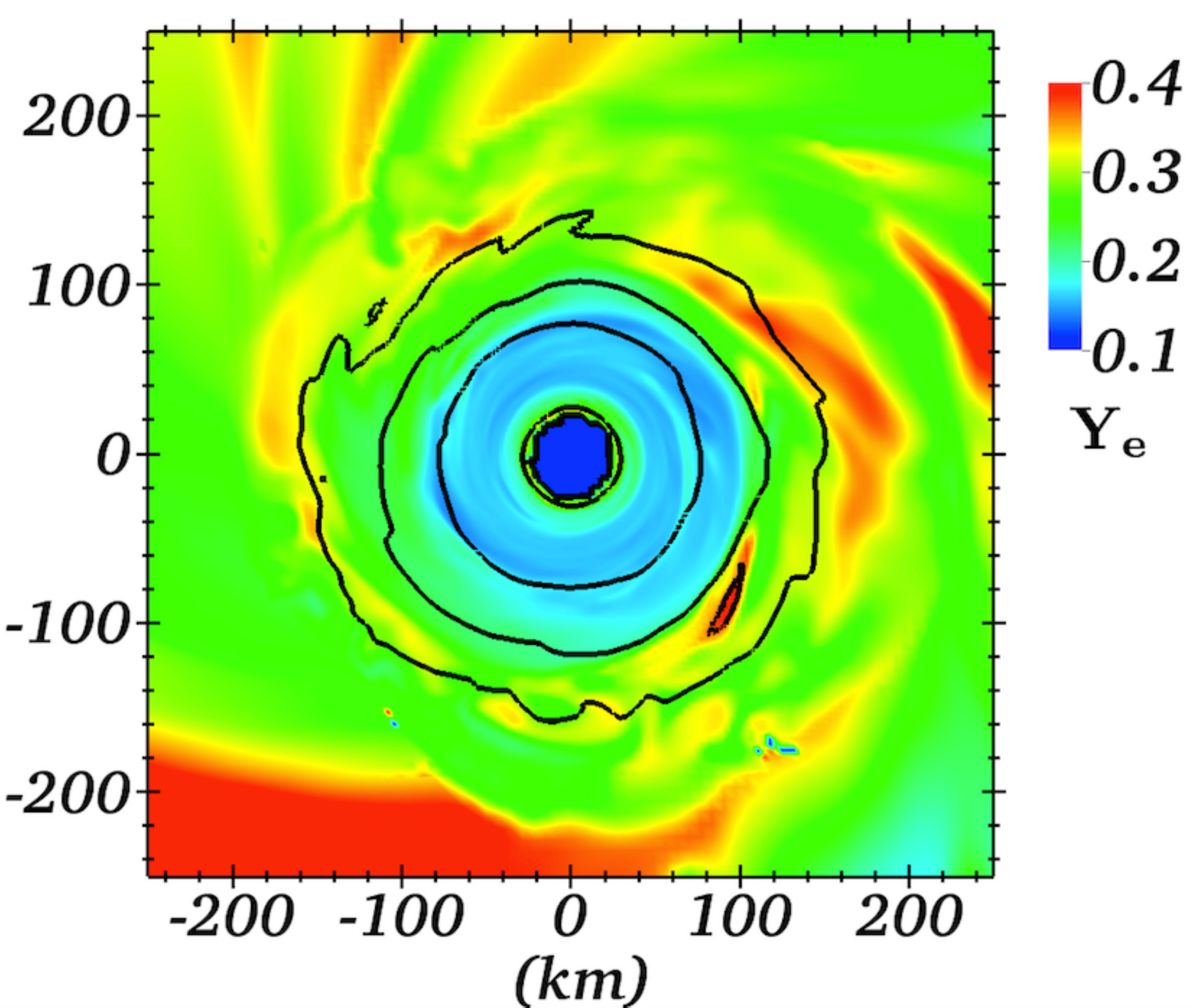}
\caption{Electron fraction $Y_e$  in the equatorial plane of the black hole at the end of the simulation, $20\,{\rm ms}$ after merger. 
Density contours are shown as thick black lines for $\rho_0=10^{9,10,11}\,{\rm g/cm^3}$}
\label{fig:YeEq}
\end{figure}

\begin{figure}
\flushleft
\includegraphics[width=1.\columnwidth]{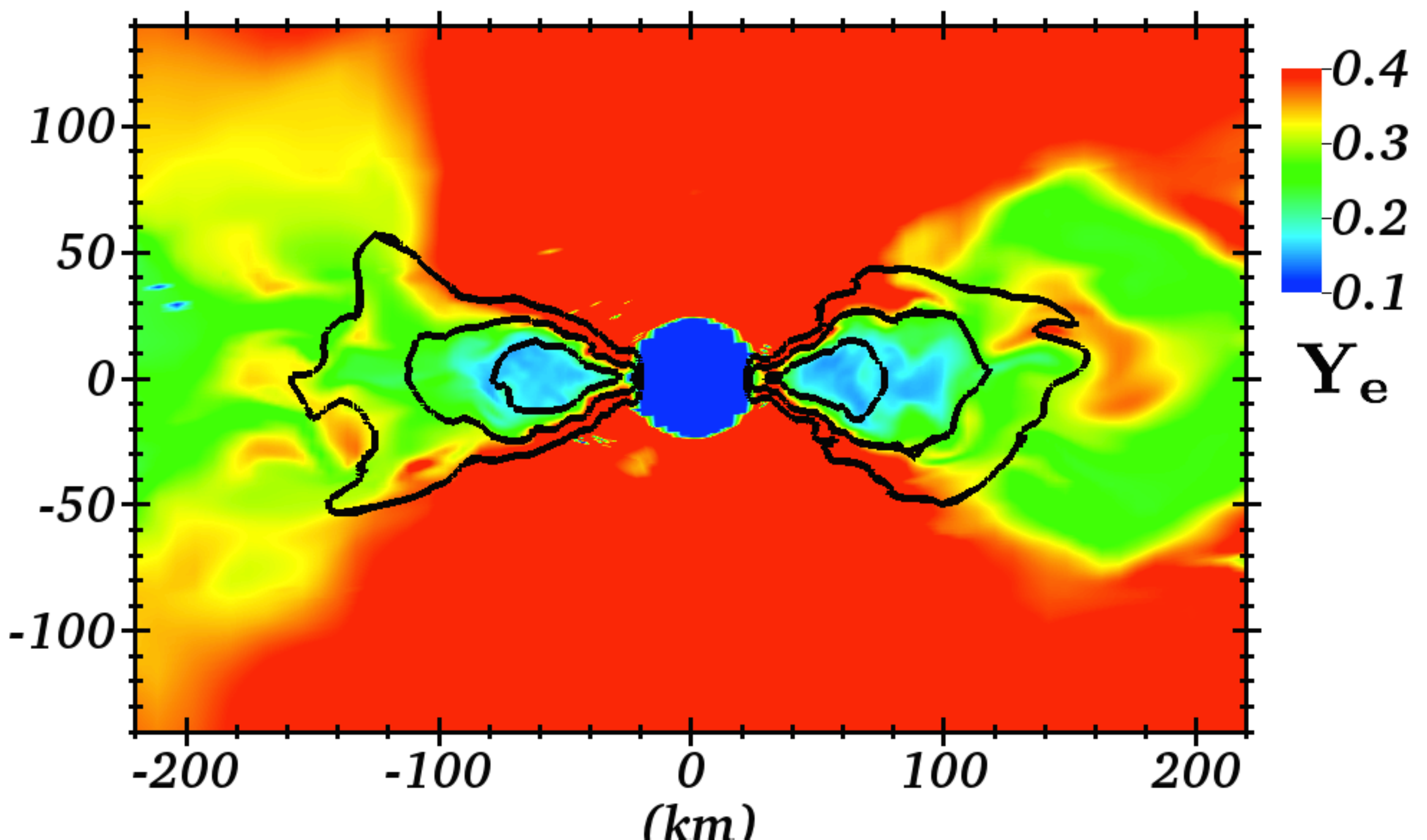}
\caption{Electron fraction $Y_e$ in a plane orthogonal to the equatorial plane of the disk the at the end of the simulation, $20\,{\rm ms}$ after merger. Density contours are shown as thick black lines for $\rho_0=10^{9,10,11}\,{\rm g/cm^3}$.}
\label{fig:Ye}
\end{figure}

Finally, we consider the composition of the fluid, through its electron fraction $Y_e$, in Fig.~\ref{fig:YeEq} and Fig.~\ref{fig:Ye}.
Being able to reliably predict the electron fraction in low-density regions is one of the main advantage of the use
of the moment formalism over the leakage scheme. We see that the core of the disk remains relatively neutron rich 
($0.15\lesssim Y_e \lesssim 0.20$), but material below $\rho_0 \sim 10^{10}\,{\rm g/cm^3}$ is significantly protonized.
About $30\%$ of the total mass is at $Y_e>0.2$, and about $10\%$ at $Y_e>0.3$. More importantly, the less neutron-rich
material, which surrounds the neutron-rich core of the disk, is more likely to be unbound by disk winds. 
Lee {\it et al}.~\cite{Lee:2005se}, Fernandez \& Metzger~\cite{Fernandez2013,Fernandez:2014} and Just {\it et al}~\cite{Just2014}
have shown that $5\%-25\%$ of the matter in accretion disks formed in binary neutron star or black hole-neutron star mergers
is eventually unbound, mostly due to viscous heating in the disk. 
In those 2D simulations, and in the absence of a long-lived hypermassive neutron star, 
these outflows remained too
neutron-rich to significantly affect the electromagnetic emission resulting from r-process nucleosynthesis in the
ejecta (i.e.\ their nucleosynthesis output does not significantly differ from the output of the more massive dynamical ejecta)~\cite{Fernandez2013}. 
However, the initial conditions for these simulations were taken to be $Y_e\sim 0.1$ everywhere, at a time
at which the disk is already wider and the composition evolution due to neutrino emission slower than in the simulations presented here. 
It is possible that a higher initial electron fraction in the outer regions of the disk can have a measurable impact on the properties of 
kilonovae. We discuss this in more detail in Sec.~\ref{sec:unbound}.  
Finally, we consider the material in the disk winds. By the end of the simulation, these outflows have a high electron fraction, 
with $Y_e\sim 0.4$. However, as discussed in Sec.~\ref{sec:unbound}, most of the material unbound in that region is ejected at earlier
times, and is more neutron-rich.

\subsection{Stability and disk/tail interactions}
\label{sec:disktail}

From the beginning of the simulation, strong asymmetries and shocks are present in the disk, and the system is well out of equilibrium. 
However, as the disk circularizes, it reaches a more stable configuration. By the end of the simulation, although asymmetries remain, fluid
elements follow nearly circular trajectories at the expected orbital frequency, up to radii $r\sim 100\,{\rm km}$. The core of the disk
is convectively stable, and close to equilibrium. Around the disk/tail interface, however, this is not the case. First, the tail material
is not circularized, which creates a shear layer at the outer edge of the disk. Additionally, according to the Soldberg-Hoiland 
criterion~\cite{1975ApJ...197..745S}, the disk is convectively unstable in both the vertical and radial directions. 
We should note that as this is also the region in
which the disk begins to deviate from hydrostatic equilibrium, the Soldberg-Hoiland criterion is not strictly applicable. But the presence
of convective regions is clearly observable at all times in the simulation. These instabilities cause the creation of large scale eddies 
close to the outer edge of the disk, and are strongly correlated with outflows launched above the disk. 
Given how out of equilibrium the system is, picking a single instability responsible for the creation
of the eddies or causally associating these eddies with the outflows is difficult. However, we note that 
regions in which there no longer are any interactions between the accretion disk and the tidal tail are
devoid of both eddies and outflows. 
We can infer that the ejection of disk material is probably helped by the convection of hotter fluid from the core of the disk to the top of the 
disk/tail interface. The high neutrino fluxes in that same region probably play a role as well, especially in directing the outflows at late times
and setting their composition. Comparing Fig.~\ref{fig:rho} and Fig.~\ref{fig:Enue}, it is clear that the disk outflows are, at the end of the simulation,
aligned with the neutrino radiation. This is not the case at earlier times, when the outflows are stronger and closer to the spin axis of the black hole.

\subsection{Unbound material}
\label{sec:unbound}

The ejection of unbound material by black hole-neutron star and neutron star-neutron star mergers is, from an astrophysical perspective,
one of the most important consequences of these mergers. This is
because these ejecta are one of the most likely locations for r-process nucleosynthesis to occur~\cite{1976ApJ...210..549L,Roberts2011,korobkin:12}.
 Neutron star mergers were long thought to happen too late in the evolution of a galaxy to explain observations of r-process 
 elements~\cite{2004A&A...416..997A}, but recent studies incorporating updated population synthesis predictions for neutron star binaries are 
 more favorable to the merger process~\cite{2014MNRAS.438.2177M,vandevoort2015}. 
 Additionally, the radioactive decay
 of nuclei in the neutron-rich ejecta during r-process nucleosynthesis powers electromagnetic transients potentially detectable in the 
 optical or, more likely, in the infrared~\cite{Li:1998bw,Kasen:2013xka,Tanaka:2013ana}. The luminosity, duration and peak frequency
 of these electromagnetic signals can significantly vary with the
 mass, composition, entropy and velocity of the 
 ejecta~\cite{2011ApJ...743..155S,Kasen:2013xka,2013ApJ...775...18B}. 
 In particular, the composition will significantly affect the results of the nucleosynthesis: low $Y_e$ material will produce
 mostly heavy elements (strong r-process), while high $Y_e$ material will produce more iron-peak elements (weak r-process).
The transition occurs at $Y_e\sim 0.25-0.3$ for conditions typical of a viscously-driven wind in a post-merger accretion 
 disk~\cite{Fernandez:2014b}.  However, nucleosynthesis results also depend on the entropy of the ejected material, and the exact dividing point 
 for the lower entropy ejecta observed in our simulations is not, to
 our knowledge, known at this point. This difference in the products of r-process nucleosynthesis impacts the lightcurve of
 radioactively powered electromagnetic transients. Indeed, high-opacity lanthanides are produces in the case of a strong r-process, 
 causing the emission to be fainter, redder and longer lived than in the case of a weak r-process~\cite{Kasen:2013xka,Tanaka:2013ana}.
 
 In black hole-neutron star binaries, matter can be ejected through different processes during and after the merger. First, if the neutron star
 is disrupted, a mass $M_{\rm ej}\sim 0.01M_\odot-0.1M_\odot$ (depending on the parameters of the binary) 
 is typically unbound during the tidal disruption of the neutron
 star~\cite{Foucart:2013a,Kyutoku:2013,Foucart:2014nda}. The amount of ejected material can even be larger for rapidly spinning, low mass
black holes~\cite{Lovelace:2013vma,Deaton2013}. The tidally
ejected material is very neutron rich ($Y_e<0.1$), cold ($T<1\,{\rm MeV}$), confined close
to the equatorial plane, and strongly asymmetric.

After that, material can be ejected during the formation of the accretion disk and at the interface between
the disk and the tidal tail. This is the main source of outflows observed in this simulation, and is discussed in more detail below.

Third, disk winds may be triggered by magnetic effects and/or neutrino absorption.  
Newtonian simulations of binary neutron star mergers using an energy-dependent leakage 
scheme (with ad-hoc neutrino absorption)~\cite{Perego2014} or energy-dependent flux-limited diffusion~\cite{Dessart2009}
observed the formation of a neutrino-driven wind within $\sim 100\,{\rm ms}$
of the merger, and neutrino driven winds were also
observed in two-dimensional simulations of an accretion disk, with initial condition taken from a black hole-neutron star merger~\cite{Just2014}. 
In~\cite{Perego2014}, the observed outflows were more neutron-rich in the equatorial region, with conditions favorable to a strong r-process, 
and less neutron-rich in the polar region, with the expectation of a weaker
r-process (and bluer kilonovae). In~\cite{Just2014}, a wide range of electron fraction is observed in all directions, although the polar outflows remain less
neutron rich. The disk generated in our black hole-neutron star mergers are likely, over longer timescales than those simulated here, 
to create winds similar to those observed in~\cite{Perego2014,Just2014}. In those studies, about $1\%$ of the disk mass was eventually ejected in 
the neutrino-driven wind. In black hole-neutron star mergers, this is generally much less material than in the dynamical ejecta. Accordingly, that ejecta
is only interesting if its different composition has observable consequences.
 
Finally, over longer timescales (seconds), and using an idealized initial disk profile, 2D Newtonian simulations
have also shown that viscous heating can drive strong outflows in the disk~\cite{Lee:2009uc,Fernandez2013}.
The total ejected mass is about $5\%-25\%$ of the mass of the disk, with some significant dependance on the spin of the black 
hole~\cite{Fernandez:2014}. 
Starting from idealized initial conditions (equilibrium torus with $Y_e\sim 0.1$), these simulations find that most of the unbound material
remains at electron fractions too low to avoid the production of lanthanides -- and they thus lead to an electromagnetic signal peaking
in the infrared, and to strong r-process nucleosynthesis. For an initial black hole spin similar to our simulation ($\chi_{\rm BH}=0.8$), however,
$0.1\%-1\%$ of the mass of the disk is ejected in a less neutron-rich outflow, which does not produce any lanthanides and could lead
to a `blue bump' in the kilonova lightcurve~\cite{Fernandez:2014}.
More recent results using more accurate neutrino methods also find large ejected masses,
of the order of $20\%-25\%$ of the mass of the disk~\cite{Just2014}.

In our simulation, we observe nearly-polar outflows during the circularization of the disk, mostly coming from the region in which the accretion 
disk and the tidal tail interact.
The total mass ejected is only $3\times 10^{-4}M_\odot \sim 0.4\% M_{\rm disk}$, a negligible amount compared to the mass ejected
during the disruption of the neutron star ($0.06M_\odot$). But because it is ejected early, and in a direction in which no material has been
unbound so far, its effects cannot be neglected. First, material in the polar region can affect the formation and collimation of a jet, if the merger
leads to a short gamma-ray burst. Additionally, because this material is ejected before any disk wind, it could obscure the blue component
of a kilonova if enough high-opacity lanthanides are formed during r-process nucleosynthesis and the ejected material
obscures a significant fraction of the polar regions. 
The properties of the outflows observed in our simulation vary significantly in time. 
The front of the outflow has a relatively low electron fraction ($Y_e\lesssim 0.2$) and specific entropy $s\sim 30k_B$ per baryon.
At later times, the outflows become less neutron rich and colder. By the end of the simulation, we measure $Y_e\gtrsim 0.3$ and $s\sim 15 k_B$.
Overall, about $15\%$ of the ejected material has $Y_e \lesssim 0.2$, and about $15\%$ has $Y_e \gtrsim 0.3$. Accordingly, we would expect
most of the material unbound during disk formation to undergo strong r-process nucleosynthesis, and produce lanthanides. By the end of the simulation,
the mass loss to these outflows has stabilized to about $0.005M_\odot/{\rm s}$. It is thus conceivable that after the end of our simulation, another
$\sim 10^{-4}M_\odot$ of less neutron-rich material would be unbound through the same process. 

Putting all potential components of the 
ejecta together, we can get estimates of the various components of the ejecta for the specific binary parameters studied here.
The quantitative results will of course vary with the initial parameters of the binary, but we expect the main qualitative features
to be fairly robust within the most likely range of black hole and neutron star masses, at least as long as the black hole spin is large enough
for the neutron star to be disrupted before merger. Here, we find:
\begin{itemize}
\item About $0.06M_\odot$ ejected asymmetrically in the equatorial plane at the time of tidal disruption. This material is cold, neutron rich, and certainly leads
to strong r-process nucleosynthesis and electromagnetic emission peaking in the infrared.
\item About $3\times 10^{-4}M_\odot$ ejected at higher latitudes on a timescale of about $10\,{\rm ms}$ after merger, due
to the circularization of the disk. This material is
less neutron rich and hotter, with more uncertain nucleosynthesis output / lightcurves (possibly depending on the initial parameters of the binary).
For the configuration considered here, however, we expect most of that material to undergo strong r-process nucleosynthesis.
\item Probably about $10^{-3}M_\odot$ ejected in neutrino-driven winds on a timescale of about $100\,{\rm ms}$ after the merger, if our disk behaves
as for the lower mass system studied in~\cite{Perego2014}.
About half of that material should be ejected in the polar direction, with a higher $Y_e$ / entropy than for the material ejected in our simulation. The rest is closer
to the equatorial plane, and more neutron rich.
\item Probably about $0.01 M_\odot$ of material unbound due to viscous heating, on timescales of the order of
$1\,{\rm s}$~\cite{Lee:2009uc,Fernandez2013,Just2014}. Most of that material is too neutron-rich to avoid strong r-process nucleosynthesis, but a small fraction
($\sim 10^{-3}M_\odot$) could be unbound in a hotter, less neutron rich front~\cite{Fernandez:2014b}.
\end{itemize}
We would thus end up with about $0.07M_\odot$ of neutron-rich material ejected either in the equatorial regions or late in the polar regions.
In the polar region, that material is preceded by ejecta with a much more complex composition. On the order of $10^{-3}M_\odot$ of 
less neutron-rich material is ejected, with the potential to produce different nucleosynthesis. If it is not obscured by neutron-rich
material, it can also produce different kilonovae lightcurves (i.e.\ a blue bump). However, electromagnetic
radiation from that material could be significantly obscured by the lower $Y_e$ outflows observed in our simulation. Accordingly, the spatial distribution
and exact mass of the outflows produced during disk formation probably should not be neglected when studying kilonova lightcurves from black hole-neutron
star mergers. Considering that we only studied one configuration,
and that the ejected mass observed here in the polar regions ($3\times 10^{-4}M_\odot$) is barely large enough to matter, it would be
dangerous to draw overly generic conclusions from our results. The initial parameters of the binary, which significantly affect the distribution of matter
between the disk and tidal tail and the conditions under which the disk form, are very likely to cause variations in the properties of these early disk outlows. 
But in the configuration studied here, the outflows
cover a fairly wide part of the high-latitude regions, roughly $\theta \lesssim 45^\circ$ with $\theta$ the angle with respect to the spin axis of the black hole.

\subsection{Neutrino Emission}
\label{sec:nurad}

\begin{figure}
\flushleft
\includegraphics[width=1.\columnwidth]{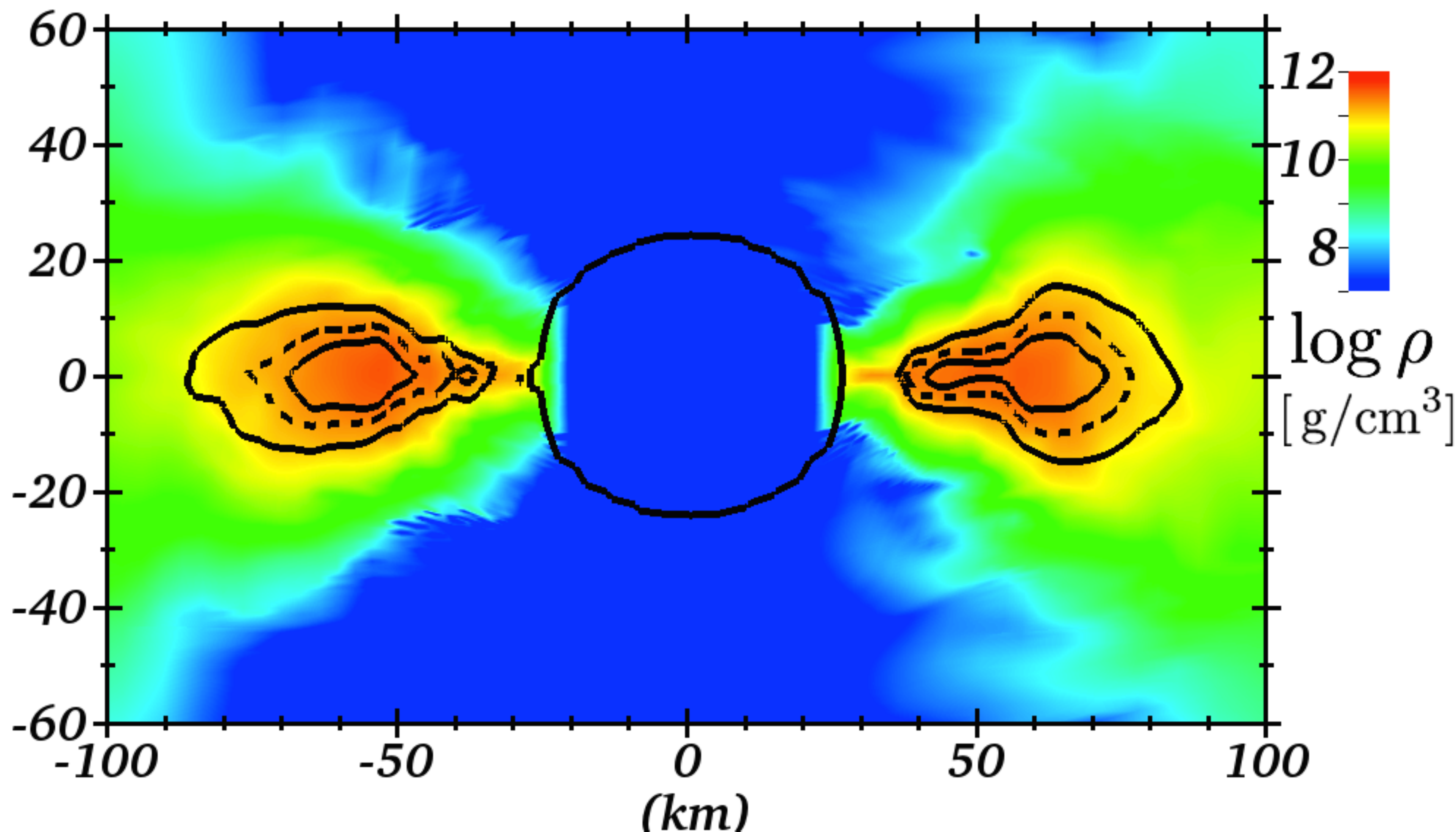}
\caption{Neutrinospheres $\tau=2/3$ at the end of the simulation, $20\,{\rm ms}$ after merger. We show the vertical plane $y=y_{\rm BH}$. The three contours are, 
from the outside to the inside, the neutrinospheres for $\nu_e$ (outer solid line), $\bar \nu_e$ (dashed line) and $\nu_x$ (inner solid line).
The disk is colored according to its baryon density $\rho_0$.}
\label{fig:nuspheres}
\end{figure}

With our M1 code, we can for the first time examine the spatial distribution of neutrinos in the first $20\,{\rm ms}$ following a black hole-neutron
star merger. To understand the main properties of the neutrino radiation, it is however useful to take a step back and look at a few of the quantities
which could already be predicted using our leakage scheme. Indeed, the leakage scheme gives us a good estimate of the optical depth in the disk, a useful quantity to understand
where the neutrinos effectively decouple from the fluid. For the electron neutrinos, although there are always a few hotter regions in which the optical depth
is $\tau_{\nu_e} \sim 10$, the density averaged optical depth is only $\langle\tau_{\nu_e}\rangle \sim 2$. It is even lower for the electron antineutrinos, with
$\langle\tau_{\bar \nu_e}\rangle\sim 1$, and heavy lepton neutrinos, with $\langle\tau_{\nu_x}\rangle\sim 0.5$. The neutrinospheres, defined as $\tau=2/3$, are shown
for a vertical cut at the end of the simulation in Fig.~\ref{fig:nuspheres}. We see that neutrinos are only trapped in the core of the disk, and mostly
free streaming everywhere else.

\begin{figure}
\flushleft
\includegraphics[width=1.\columnwidth]{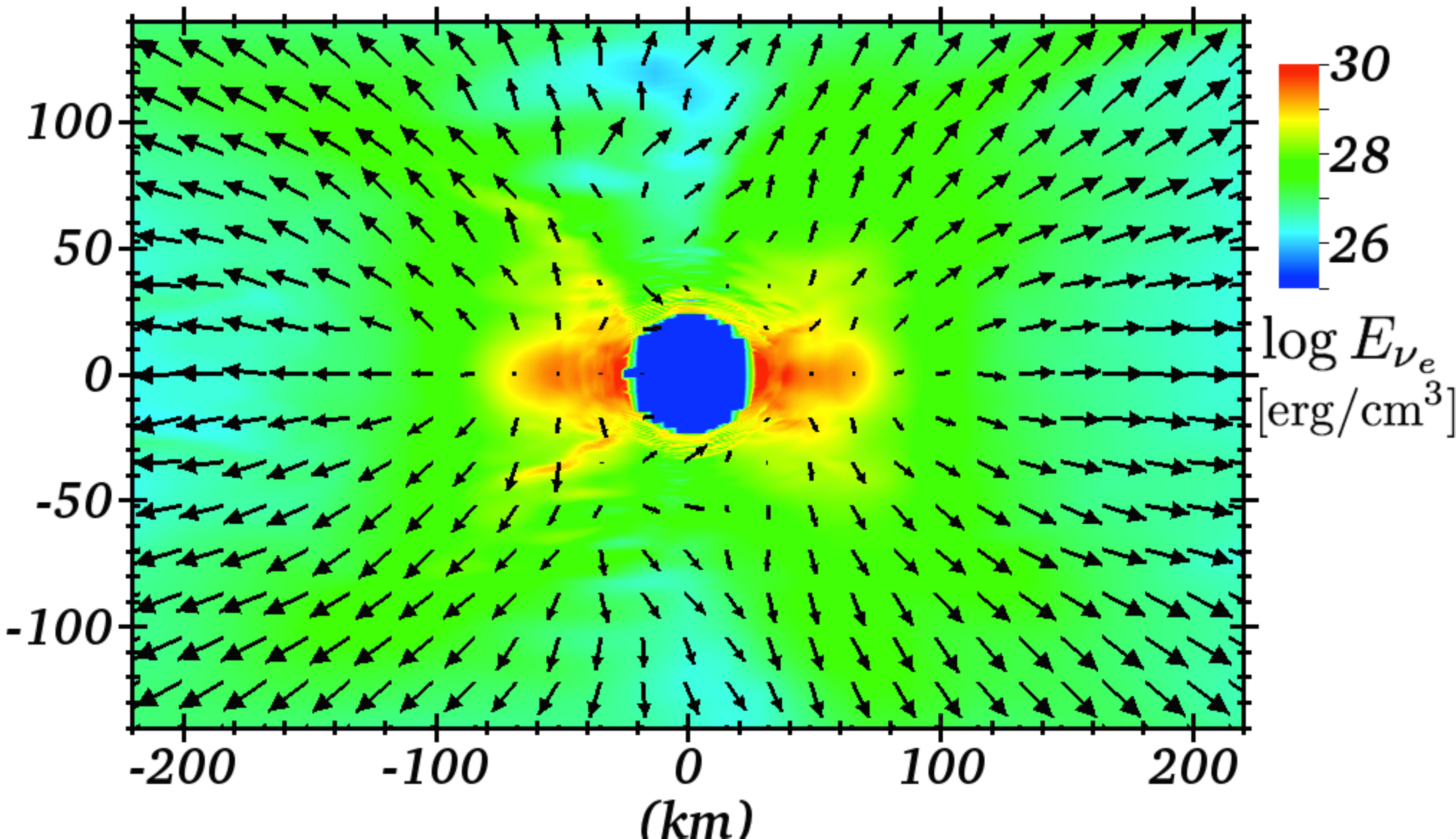}
\caption{Energy density and normalized flux ($\alpha F^i/E-\beta^i$) of the electron neutrinos at the end of the simulation, $20\,{\rm ms}$ after merger.
We show the vertical plane $y=y_{\rm BH}$.}
\label{fig:Enue}
\end{figure}

\begin{figure}
\flushleft
\includegraphics[width=1.\columnwidth]{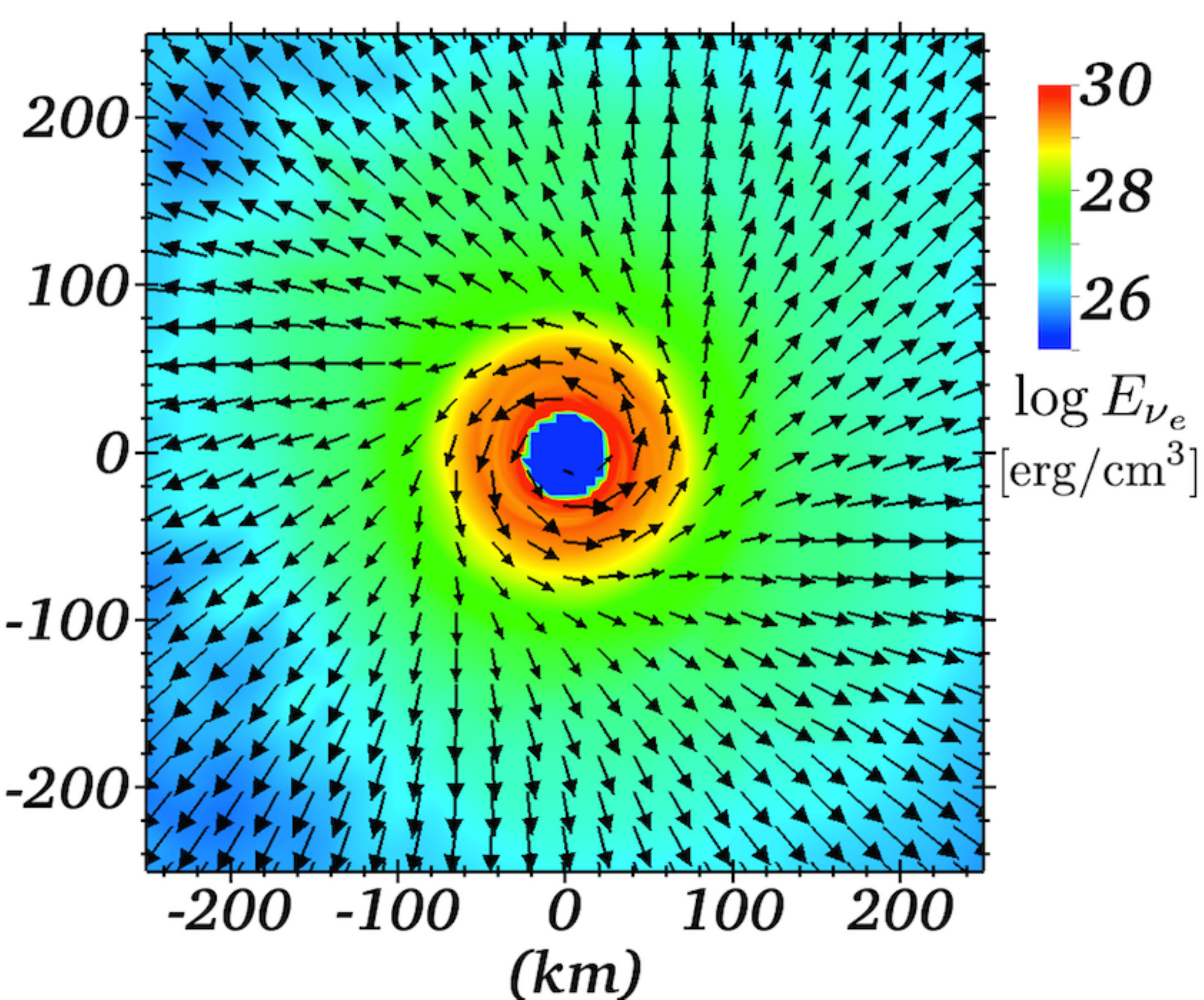}
\caption{Energy density and normalized flux ($\alpha F^i/E-\beta^i$) of the electron neutrinos at the end of the simulation ($20\,{\rm ms}$ after merger),
in the equatorial plane of the disk.}
\label{fig:EnueEq}
\end{figure}

To illustrate the main properties of the neutrino radiation, we plot in Figs.~\ref{fig:Enue}-\ref{fig:EnueEq} the energy density and ``normalized flux''
$F^i_{\rm norm} = \alpha F^i/E-\beta^i$ in vertical and horizontal slices of the disk, for the electron neutrinos. 
The normalized flux is chosen so that it represents an effective transport velocity for the neutrino energy. In the core of the disk, Fig.~\ref{fig:EnueEq}
shows that the neutrino energy is mostly transported with the fluid, as befits an optically thick region. Outside of the expected neutrinosphere, i.e.
for $r \gtrsim 90\,{\rm km}$, the neutrinos transition to free streaming away from the disk. The energy density is maximal
close to the inner edge of the disk, in part due to higher temperatures and in part due to gravitational redshifting.

The vertical slice on Fig.~\ref{fig:Enue} shows a few additional features of interest. First, we note the random orientation of the fluxes
along the speed axis of the black hole, up to a height $|z| \lesssim 80M_\odot$. 
This is a known problematic feature of the M1 approximation, due to the convergence of neutrinos from all around the disk. 
Beyond this issue, we can also see that the emission at large radii is clearly asymmetric. We find significantly lower energy densities 
within about $20^\circ$ of the equatorial plane as well as in the (less reliable) polar regions. This is a purely geometric effect due to
the projected shadow of the disk and the relativistic beaming of the neutrinos.

\begin{figure}
\flushleft
\includegraphics[width=1.\columnwidth]{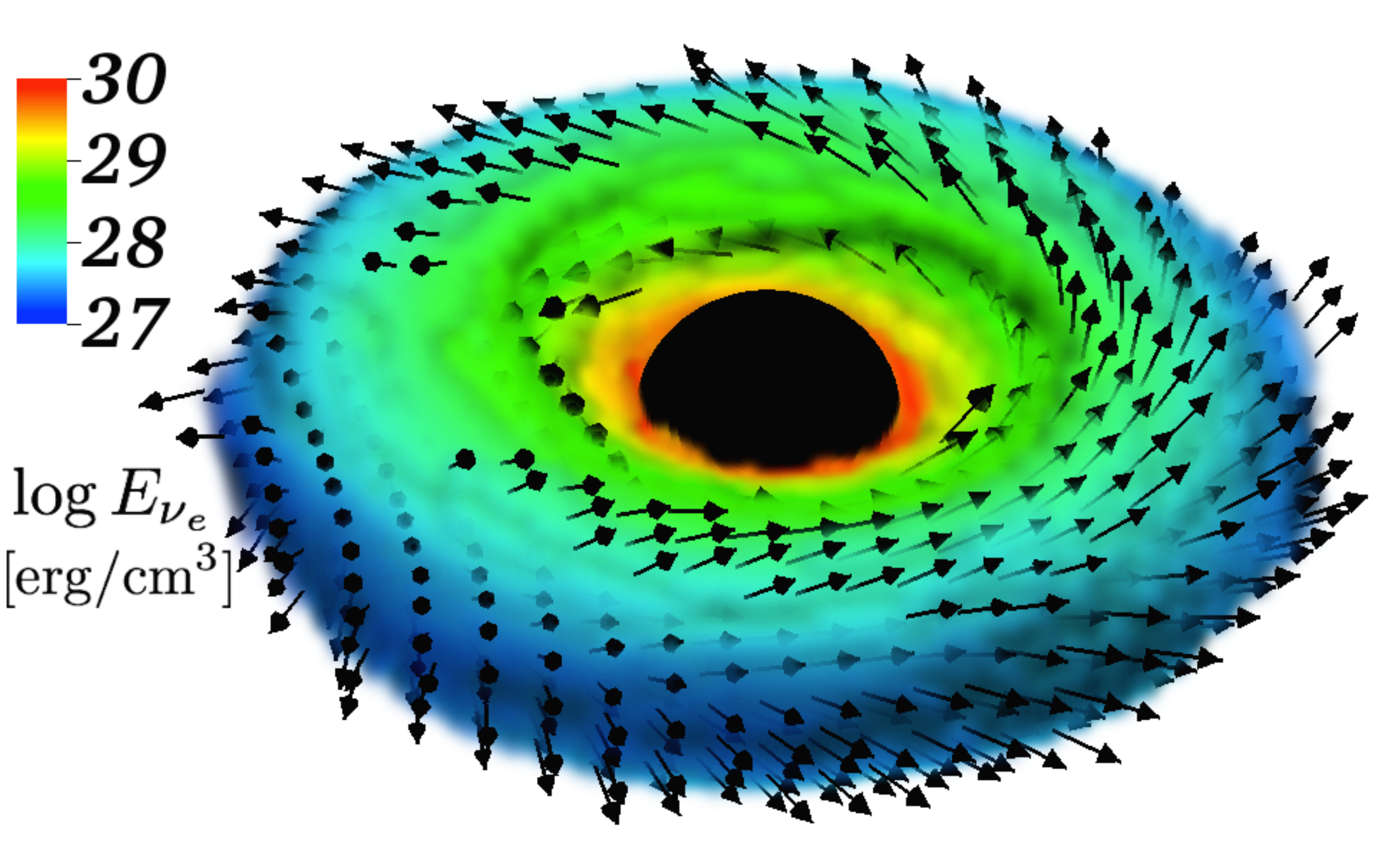}
\caption{Energy density and normalized fluxes of the electron neutrinos on the $\tau_{\nu_e}=0.1$ surface.}
\label{fig:TauNue}
\end{figure}

To understand this beaming, we first have to remember that the velocities in the disk are mildly relativistic, with $v\sim 0.5c$.
This means that the emission of neutrinos will be focused within a relatively large beam centered on the direction of motion
of the fluid, which nearly follows a circular orbit around the black hole. Only a small fraction of the neutrinos are emitted in the vertical direction.
Additionally, most of the neutrinos decouple from the matter around the location of the neutrinosphere, and then
free stream away from the disk. Accordingly, if the inner regions of the disk are brighter than its outer regions, as is the case here,
the disk casts a shadow along the equatorial plane. This is more clearly illustrated by Fig.~\ref{fig:TauNue}, which shows the energy density and fluxes
for the electron neutrinos on the surface $\tau_{\nu_e}=0.1$. Most of the energy comes from the inner disk, and neutrinos cannot easily escape along
the equatorial plane. The shadow along the equatorial plane observed in Fig.~\ref{fig:Enue} exactly matches the thickness of the disk. 
Relativistic beaming, on the other hand, causes
neutrinos to be preferentially emitted nearly tangent to the disk. Hence the neutrino flux is larger just outside of the shadow than at high latitudes.
The equatorial shadow is of course not perfect, as some neutrinos are emitted from the outer edge of the neutrinosphere. But this is a relatively
small fraction of the total neutrino luminosity, as can be seen in Fig.~\ref{fig:TauNue}. 

The radiation fields of the other neutrino species are qualitatively similar, except for the fact that the neutrinosphere is located deeper
inside the disk.
We find that most of the energy emitted in neutrinos goes into electron antineutrinos, at least at early times.
The luminosity in $\nu_e$ is fairly constant during the evolution, at 
$L_{\nu_e}\sim 5-8\times 10^{52}\,{\rm ergs/s}$.
But the luminosity in $\bar \nu_e$ peaks at $L_{\bar \nu_e}\sim 5 \times 10^{53}\,{\rm ergs/s}$ during the rapid evolution of the electron
fraction from its very neutron-rich initial value to the equilibrium value in the disk. Within $5\,{\rm ms}$, it decreases to 
$L_{\bar \nu_e}\sim 10^{53}\,{\rm ergs/s}$, and by the end of the simulation it becomes lower than the electron neutrino luminosity,
with  $L^f_{\nu_e}=7\times 10^{52}\,{\rm ergs/s}$ and $L^f_{\bar \nu_e}=5\times 10^{52}\,{\rm ergs/s}$. Finally, there is an equally brief burst of heavy-lepton neutrinos, with $L_{\nu_x}\sim 3\times 10^{52}\,{\rm ergs/s}$
for each species, due to the existence of hot spots early in the disk formation. That luminosity rapidly decreases as the temperature becomes more homogeneous. Within $3\,{\rm ms}$, we measure $L_{\nu_x}\sim 5\times 10^{51}\,{\rm ergs/s}$ for each species, and by the end of the simulation,
$L^f_{\nu_x} = 10^{51}\,{\rm ergs/s}$ per species.

The properties of the neutrino spectrum are not directly measurable within our gray formalism. Nonetheless, it is possible to get reasonable estimates
from either the predictions of the leakage scheme or the properties of the fluid on the neutrinosphere. The leakage scheme predicts
average energies of $\langle\epsilon_{\nu_e}\rangle=11\,{\rm MeV}-13\,{\rm MeV}$,  $\langle\epsilon_{\bar \nu_e}\rangle=14\,{\rm MeV}-15\,{\rm MeV}$, and
$\langle\epsilon_{\nu_x}\rangle=16\,{\rm MeV}-18\,{\rm MeV}$ during the last $10\,{\rm ms}$ of evolution. All of the energies peak $\sim 4\,{\rm MeV}$ higher
at the beginning of the simulation, when hot spots are present. Neglecting corrections due to the finite chemical potential of neutrinos in the emitting
regions and assuming a redshift factor of $2$ between the emitting region and the observer, 
this would correspond to fluid temperatures of roughly $6\,{\rm MeV}$, $7\,{\rm MeV}$, and $8\,{\rm MeV}$. Considering the temperature distribution
observed in the disk, the increasing temperature as the neutrinosphere recedes deeper into the disk, and the higher emissivity of high temperature
points, this appears fairly reasonable. 

\subsection{Neutrino viscosity and the growth of magnetic instabilities}
\label{sec:nuvis}

The magnetorotational instability (MRI) is expected to play a crucial role in the growth of magnetic fields in post-merger accretion disk,
and may also be important for the generation of jets after a compact binary merger. However, recent work has shown that the transport of
momentum by neutrinos can significantly affect the growth timescale and wavelength of the MRI~\cite{Masada2008,Guilet2014}. 
In the context of protoneutron stars, Guilet {\it et al.}~\cite{Guilet2014} showed that if the wavelength of the fastest growing mode of the MRI in the absence of
neutrinos $\lambda_{\rm MRI}$
is larger than the neutrino mean free path $\lambda_\nu$, an effective viscosity from the transport of neutrinos causes the MRI 
to grow slower than expected for
small magnetic fields, but at a fastest-growing wavelength independent of the magnetic field strength 
(instead of the wavelength decreasing with the magnetic field strength). If instead $\lambda_{\rm MRI}<\lambda_\nu$,
the neutrinos can act as a drag force on the magnetized fluid. For large neutrino energy densities, this slows the growth timescale of the fastest
growing mode of the MRI, and slightly increases its wavelength. 
In accretion disks, and when neutrinos can be modeled through an effective viscosity, a similar increase of the growth timescale of the MRI has been
measured~\cite{Masada2008}.
By applying the model of Guilet {\it et al.}~\cite{Guilet2014} to our accretion disk, we can obtain a first estimate
of the expected effect of neutrinos on the growth of the MRI. 

First, we need to determine the critical magnetic field $B_c$ at which $\lambda_{\rm MRI}=\lambda_\nu$, using $\lambda_{\rm MRI} \sim 2\pi b /\sqrt{b^2+\rho_0 h}$ and $\lambda_\nu \sim 1/(\kappa_a+\kappa_s)$ (where $b$ is the strength of the magnetic field observed by an observer comoving with the fluid). At the highest density points in the disk, we get $B_c \sim 10^{13}\,{\rm G}$, while $B_c$ smoothly increases with decreasing density, to $B_c\sim 10^{15}\,{\rm G}$ for $\rho_0 \sim 10^{10}\,{\rm g/cm^3}$. These values are larger than the initial 
magnetic field in most merging neutron stars, but smaller than the expected saturation amplitude of the MRI, 
thus indicating that both the viscous and neutrino drag regimes might be relevant during the evolution of a post-merger accretion disk.

As far as the neutrino drag is concerned (e.g. for $B<B_c$), its effects on the growth of the MRI should however be negligible. 
Guilet {\it et al.}~\cite{Guilet2014}
find that the importance of neutrinos on the growth of the MRI in this regime is determined by the value of the dimensionless parameter
\beq
\frac{\Gamma}{\Omega} = \frac{2 (\kappa_a+\kappa_s) E_\nu}{15\rho_0 \Omega}\,\,.
\eeq
For $\Gamma/\Omega \gtrsim 1$, neutrinos affect the growth of the MRI. However, in our disk, we find $\Gamma/\Omega \lesssim 0.01$ everywhere. Thus
the growth of the MRI should remain unaffected at low magnetic field strengths.

In the viscous regime ($B>B_c$), the importance of neutrino effects is determined by the value of the Elasser number
$\epsilon = v_A^2 / (\nu \Omega)$ where $v_A$ is the Alfven speed and
\beq
\nu = \frac{2E_\nu }{15\rho_0 (\kappa_a+\kappa_s)}
\eeq
is the effective viscosity due to neutrinos. Viscosity affects the growth of the MRI for $E_\nu \lesssim 1$. For $B\sim B_c$, and the
conditions observed in our simulation, we would get $\epsilon \sim 1$ (and the Elasser number then grows as $B^2$).
Accordingly, within the simple model used here, neutrinos could plausibly affect the growth
of the MRI for $B\sim B_c$. But for $\epsilon\sim 1$, these effects are mild, and neutrinos are not expected to have any effect
for $B \ll B_c$ or $B \gg B_c$.

\subsection{Impact of the neutrino treatment}
\label{sec:nutreat}

Having performed simulations with both a leakage scheme and the M1 formalism,
we can now obtain better estimates of the impact that an approximate treatment of the neutrinos has on the evolution
of a post-merger accretion disk. 

\subsubsection{Comparison with a leakage scheme}

We can first look at differences between the leakage simulations and the M1 simulations. Not surprisingly, the two methods provide reasonable
qualitative agreement for the global properties of the high density regions of the disk: the simulations with neutrino leakage capture the formation of the disk, its early protonization and later re-neutronization, and the global temperature evolution due to the initial expansion of the disk, neutrino cooling and shock heating. Lacking neutrino absorption, the leakage simulation however tends to maintain larger temperature differences between neighboring regions 
of the disk. It underestimates the timescale necessary for the disk to cool down, and the magnitude of its protonization. More quantitatively, the average
electron fraction in the leakage simulation is lower by $\Delta Y_e \sim 0.05$, and the average temperature is lower by $\Delta T = (0.5-0.8)\,{\rm MeV}$.
Additionally, without neutrino absorption the evolution of the fluid composition in low-density regions is entirely unreliable. The composition of the tidal tail material falling back onto the black hole and of the outflows are thus radically different.

\begin{figure}
\flushleft
\includegraphics[width=1.\columnwidth]{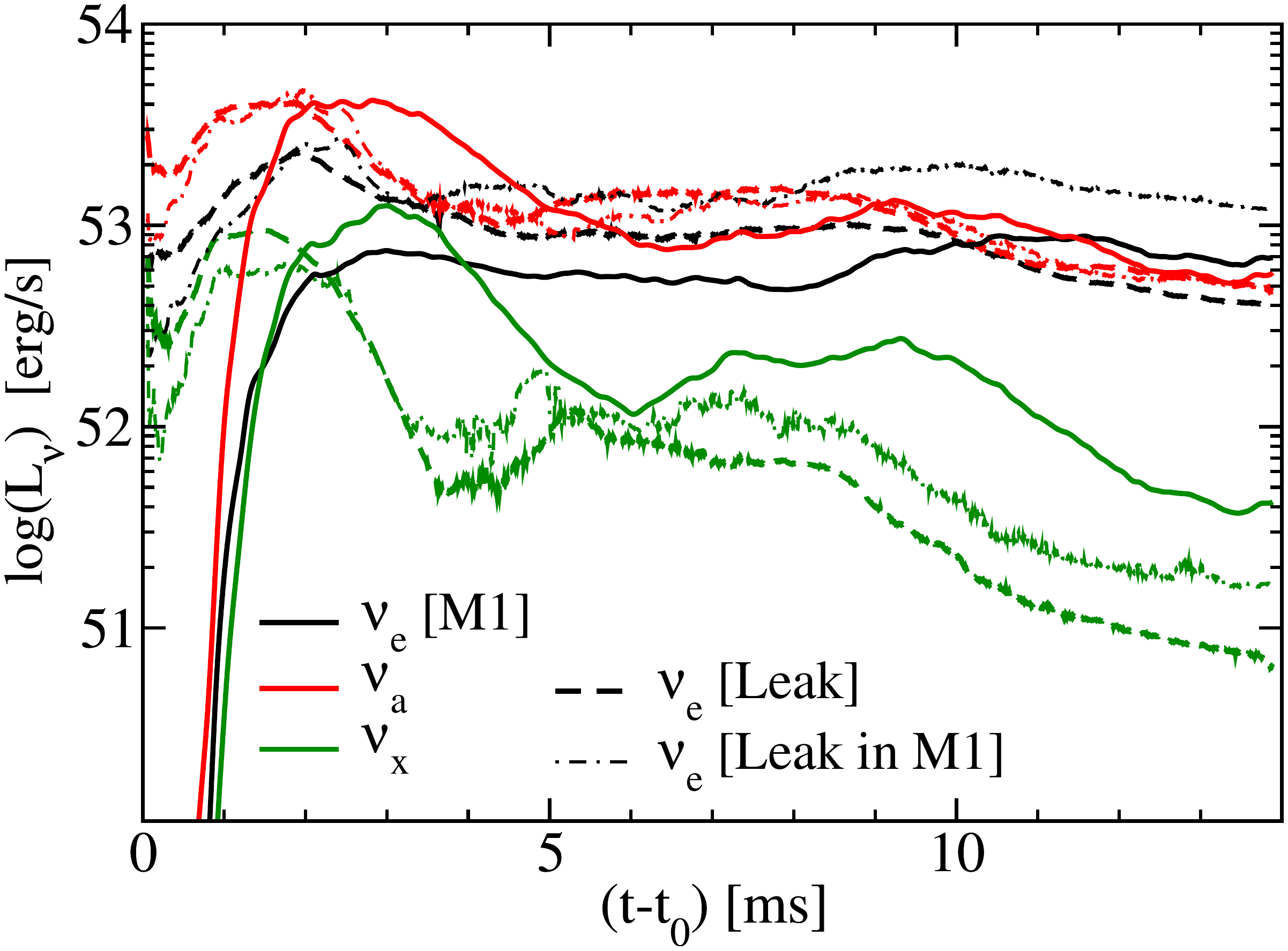}
\caption{Luminosities for the different neutrino species as a function of time. Solid curves show the M1 results. Dashed curves show results
using a leakage scheme. Thin dot-dashed curves show the prediction of the leakage scheme, but for the value of the hydrodynamic variables obtained
by evolving the disk with the M1 scheme. Heavy-lepton neutrino luminosities $\nu_x$ are for all 4 species combined.}
\label{fig:lum}
\end{figure}

There are also significant differences in the neutrino luminosities, for the electron neutrinos by a factor of two about $10\,{\rm ms}$ after merger 
($5\,{\rm ms}$ after the beginning of the simulation), 
and by about $30\%$ by the end of the simulation. For heavy-type neutrinos, the difference is typically a factor of $2-4$, presumably because 
of the steeper dependence of their emissivities on the temperature of the disk. We show the luminosities for the leakage scheme, the M1 scheme,
and the predictions of the leakage scheme for the value of the hydrodynamic variables obtained when evolving the system with the M1 scheme 
on Fig.~\ref{fig:lum}. The first thing to note is that, since the leakage scheme measures the instantaneous energy loss at each point of the domain while
the luminosities in the M1 scheme are measured through the neutrino fluxes at the outer boundary of the computational domain, there is naturally a
time shift between the two predictions. Even taking that time shift into account, however, it is quite clear that the electron neutrino luminosity
is larger in the leakage scheme, while the heavy-lepton neutrino luminosity is larger in the M1 scheme. The electron antineutrino luminosities
cannot be distinguished within the uncertainties due to the time shift in the measurements. Looking at the predictions of the leakage scheme
within the M1 simulation allows us to differentiate between the error due to the instantaneous estimate of the luminosity, and the error due to the diverging
evolution of the hydrodynamics variables. We see that by the end of the simulation, both sources of error are important.

Overall, these results appear consistent with the expected limitations of a leakage scheme. We also performed leakage simulations with two different methods to
compute the opacities, to check whether the differences observed between the leakage and M1 schemes could be due to differences in the estimated
neutrino opacities. But the differences between these two simulations were well below the estimated errors in the leakage scheme.

\subsubsection{Impact of the gray approximation}

By looking at M1 simulations using different energy-averaging approximations, we can also attempt to estimate errors
related to the use of a gray scheme. We use two different gray approximations which, in our test of neutrino-matter interactions
in a post-bounce supernova profile, bracketed the solution
obtained with an energy-dependent code (see Appendix~\ref{sec:tests}). Clearly, as an error estimate, this is a poor substitute for a comparison with an energy-dependent radiation transport code. Unfortunately, such a simulation remains too costly to
perform with the SpEC code at this point. We find that differences in the average properties of the disk, although measurable,
are much smaller than the differences between the M1 simulations and the leakage simulations. The average electron fraction varies by $\Delta Y_e < 0.01$, and the average temperature by $\Delta T < 0.1\,{\rm MeV}$. The total neutrino luminosities agree, for all species, within $\sim 20\%$. 

We also performed a simulation in which the gray opacities were computed assuming that the neutrinos obey a Fermi-Dirac distribution with 
temperature $T$ equal to the fluid temperature and equilibrium chemical potential $\mu_\nu$, a method
which significantly underestimates neutrino opacities: low density regions can have $T\sim 1-2\,{\rm MeV}$ while neutrinos are mostly emitted in regions with 
$T\gtrsim5\,{\rm MeV}$. The resulting differences are significantly smaller than in the test of the neutrino-matter coupling 
presented at the end of the appendices. In fact,
in the core of the disk, the errors are similar to the differences between the two M1 simulations using different prescription for the emissivities. Where
the two methods differ, unsurprisingly, is in the composition of the outflows: when assuming an equilibrium distribution at the fluid temperature
for the neutrinos, the electron fraction of the outflows is consistently $\Delta Y_e = 0.05$  lower than when using the more realistic neutrino energies
computed from the temperature in the emitting regions. This confirms our assumption that the electron fraction in the outflows is significantly affected
by neutrino absorption.

\subsubsection{Limitations of the M1 closure}

Finally, when analyzing our results, it is worth noting once more the limitations of the M1 closure. Crossing beams, caustics, and strongly focused beams
are known to be problematic when using the M1 closure. Some examples of these issues can be found in the tests presented in Appendix~\ref{sec:tests}.
Accordingly, the results of our simulations cannot be trusted close to the spin axis of the black hole, where neutrinos from all regions of the disk cross. Because of the relativistic beaming of the neutrinos, the energy density in that region is already low, but unphysical radiation shocks cause an even larger decrease. Unfortunately, the neutrino radiation in the polar region can have important consequences on the post-merger evolution of the system.
Indeed, neutrino-antineutrino annihilations into electron-positron pairs deposit energy in this low-density region, and affect the matter density there. 
The formation of a relativistic jet, desirable if black hole-neutron star mergers are to produce short gamma-ray bursts, is quite sensitive to baryon-loading
in the polar regions, but on the other hand could be helped by the energy deposition. The exact impact of the neutrinos in that context remains an open question, which can only be answered with an improved treatment of the neutrinos, and a better understanding of the jet forming mechanism.

\section{Conclusions}
\label{sec:concl}

We present a new module of the SpEC code~\cite{SpECwebsite}, allowing us to study the effects of neutrinos within a fully general relativistic - hydrodynamics code. The neutrinos are modeled using the M1 formalism, in which the first two moments of the neutrino distribution 
function are evolved (energy and fluxes). Although the formalism can in theory be energy-dependent, we limit ourselves to an energy-integrated version
of the code, due to the high cost of energy-dependent simulations. 
We offer here a detailed description of the implementation of the M1 algorithm in SpEC, as well as a series
of tests assessing our ability to study the evolution of neutrinos in flat and curved spacetimes, and their interaction with matter.

We also discuss the first simulation with a general relativistic code of the evolution of an accretion disk produced by a 
typical black hole-neutron star merger. We use as initial conditions a snapshot of an existing SpEC simulation using a simpler
treatment of the neutrinos (leakage scheme)~\cite{Foucart:2014nda}, right as the accretion disk begins to form and neutrino effects become important. 
This provides us with realistic initial conditions for the accretion disk, at least compared to the more commonly used equilibrium tori.
As we neglect magnetic fields and all material falling back on the disk on timescales longer than $20\,{\rm ms}$, we limit ourselves to a
relatively short evolution, up to $20\,{\rm ms}$ after merger. We find that the evolution of the forming accretion disk is initially dominated by the
circularization of the disk material. The disk expands and contracts in a cycle of about $6\,{\rm ms}$. These oscillations are the main
driver of the evolution of the average disk properties at early times. As the disk circularizes, however, these oscillations are rapidly damped.

Neutrinos cause a global cooling of the disk, but on a relatively long timescale of order $50\,{\rm ms}$. On the timescale of this simulation,
neutrinos have however other important effects. First, they drive the composition evolution of the core of the disk. The disk rapidly protonizes, reaching an average electron fraction $Y_e\sim 0.2$ (a larger value than predicted by the leakage scheme~\cite{Foucart:2014nda}). The electron fraction then more slowly decreases as the disk starts to cool down. Second, neutrinos cause changes in the electron fraction of the outer parts of the disk, and of the disk outflows. Although initially neutron rich, by the end of the simulations these regions have $Y_e>0.3$ ($Y_e\sim 0.4$ in the outflows). These low density regions above the disk would be the first to be unbound by disk winds and viscously-driven outflows~\cite{Lee:2009uc,Fernandez2013,Just2014,Fernandez:2014}, and their higher initial electron fraction could thus affect the composition of late time outflows, their nucleosynthesis output, and the light curve of associated electromagnetic transients. Third, energy transport by the neutrinos homogenize the temperature distribution in the disk, with cold regions absorbing higher energy neutrinos and hot regions radiating much more strongly. Finally, neutrinos will deposit energy and create electron-positron pairs in the low-density 
polar regions. This could plausibly affect, positively or negatively, the formation of relativistic jets in that region. Unfortunately, neutrino-antineutrino
annihilation in low density regions is not modeled within our formalism.
As opposed to what was observed in simulations of protoneutron stars~\cite{Guilet2014}, we also estimate that the impact of neutrino transport on the growth of the magnetorotational instability is likely to be small to inexistant in post-merger accretion disks.

We also find that, because the disk is very compact, and forms around a rapidly rotating black hole ($\chi_{\rm BH}=0.87$), relativistic effects are significant. Most of the radiation
comes from a region with mildly relativistic velocities ($v\sim 0.5c$). Accordingly, the neutrinos are beamed
tangentially to the disk, causing strong anisotropies in the neutrino luminosity. Light bending and gravitational redshift also naturally affect the luminosity
and neutrino spectrum. The luminosity is low in the equatorial plane, due to the disk shadow, and in the polar regions, due to beaming (although the evolution
of polar regions with the M1 closure is unreliable). We note that even though a black hole spin $\chi_{\rm BH}=0.87$ might seem large, this should be 
fairly typical of a black hole-neutron star merger in which an accretion disk forms. Indeed, for the most likely black hole masses, slowly spinning black holes
cannot disrupt the neutron star before it plunges into the black hole~\cite{Foucart2012}.

A number of simulations using approximate treatments of gravity have also considered the impact of neutrinos on disk evolutions, with methods generally
more advanced than contemporary general relativistic simulations (e.g.~\cite{Rosswog:2003rv,Setiawan2006,oechslin:07,Perego2014,Just2014}), 
and/or capable to evolve the system over longer timescales (e.g.~\cite{Lee:2005se,Fernandez2013,Fernandez:2014}). 
However, direct comparisons with our results are difficult. This is in part because of the importance of relativistic effects for accretion disks around
rapidly spinning black holes, and also because simulations which do not include general relativistic effects typically start from either an idealized equilibrium torus, or from the result of a merger in which the neutron star was disrupted by tidal effects modeled by a pseudo-Newtonian potential, whose realism close to a rapidly spinning black hole is difficult to assess. Our simulation shows that general relativistic effects significantly impact the neutrino radiation and the disk formation, and the forming accretion disk is more compact and hotter than in non-relativistic studies. An important application of our results would in fact be to provide better initial conditions for long-term disk evolutions, or for nucleosynthesis studies requiring a detailed knowledge of the disk structure and composition (e.g.~\cite{2013arXiv1312.1199S}).

Finally, we note that the joint effects of shocks during the disk circularization, instabilities at the disk/tail interface, and neutrino absorption unbinds a small
amount of material in the polar regions ($\sim 3\times 10^{-4}M_\odot$). This might seem negligible compared to the material ejected dynamically in the equatorial plane during the disruption of the neutron star ($\sim 0.06M_\odot$). However, this ejecta could be important because
it is unbound in a direction in which no material has been ejected so far, and could thus impact the formation of a relativistic jet. 
Additionally, over longer timescales, we expect neutrino-powered winds to become active 
and eject material in the polar regions, maybe of the order of $1\%$ of the mass of the disk. Later on, viscously-driven outflows
could eject a more significant amount of material (probably $5\%-25\%$ of the disk). But the outflows observed here will be the outermost
layer of ejected material in the polar region. Their opacity might affect the properties of electromagnetic transients for observers in directions close to the spin axis of the black hole. From our simulations, it appears that most of the matter in these outflows is too neutron rich and cold to avoid strong r-process nucleosynthesis and the formation of high-opacity lanthanides, and thus that these early outflows could obscure later disk winds. However, we caution
that the ejected mass and geometry of the outflows are likely to depend on the parameter of the binary, and thus that it would be dangerous to draw overly
generic conclusions from a single initial configuration.

With these results demonstrating the ability of the SpEC code to evolve neutrinos within the moment formalism, we are now in a position to improve on the
short simulations presented here. The gray scheme used in this work, which would be inadequate for the study of core collapse supernovae, appears 
at this point sufficient for the study of neutron star mergers.

\acknowledgments
The authors wish to thank Brett Deaton, Rodrigo Fernandez, 
and Dan Kasen for useful discussions over the course of this project,
and the members of the SXS collaboration for their suggestions and support.
F.F. gratefully acknowledges support from the
Vincent and Beatrice Tremaine Postdoctoral Fellowship.
Support for this work was provided
by NASA through Einstein Postdoctoral Fellowship
grants numbered PF4-150122 (F.F.) and PF3-140114 (L.R.) awarded 
by the Chandra X-ray Center, which is operated by the Smithsonian 
Astrophysical Observatory for NASA under contract NAS8-03060; 
and through Hubble Fellowship grant number 51344.001 awarded 
by the Space Telescope Science Institute, which is operated by the Association
 of Universities for Research in Astronomy, Inc., for NASA, under contract NAS 5-26555.
The authors at CITA gratefully acknowledge support from the NSERC
Canada. 
M.D.D. acknowledges support through NSF Grant PHY-1402916.
L.K. acknowledges support from NSF grants PHY-1306125 and AST-1333129 at
Cornell, while the authors at Caltech acknowledge support from NSF
Grants PHY-1068881, PHY-1404569, AST-1205732 and AST-1333520, and from NSF
CAREER Award PHY-1151197.
Authors at both Cornell and Caltech also thank the Sherman Fairchild Foundation
for their support. 
Computations were performed on the
supercomputer Briar\'ee from the Universit\'e de Montr\'eal, and Guillimin
from McGill University, both managed
by Calcul Qu\'ebec and Compute Canada. The operation
of these supercomputers is funded by the Canada Foundation
for Innovation (CFI), NanoQu\'ebec, RMGA and the Fonds de
recherche du Qu\'ebec - Nature et Technologie (FRQ-NT). Computations were
also performed on the Zwicky cluster at Caltech, supported by the Sherman
Fairchild Foundation and by NSF award PHY-0960291.
This work also used the Extreme Science and Engineering
Discovery Environment (XSEDE) through allocation No. TGPHY990007N,
supported by NSF Grant No. ACI-1053575.
\bibliography{References/References}

\appendix

\section{Closure relation}
\label{sec:closure}

The choice of a closure $P^{ij}(E,F_k)$ is the main  approximation used in the moment formalism. In this work,
we use the M1 closure, which relies on an interpolation between the expected closure relation in the optically thin and
optically thick regimes. Although we evolve $E$ and $F_i$, the closure relation is generally easier to express
as a function of the fluid frame quantities $J$ and $H^\mu$. Following~\cite{2011PThPh.125.1255S}, 
we define the parameter $\zeta$ as
\beq
\zeta^2 = \frac{H^\alpha H_\alpha}{J^2}\,\,,
\eeq 
such that $\zeta\sim 0$ in the optically thick limit and $\zeta\sim 1$ in the optically thin limit. The closure is then
\beq
P^{ij} = \frac{3\chi(\zeta)-1}{2} P^{ij}_{\rm thin} + \frac{3(1-\chi(\zeta))}{2} P^{ij}_{\rm thick}\,\,.
\eeq
By default, we use the Minerbo closure~\cite{Minerbo1978}
\beq
\chi(\zeta) = \frac{1}{3} + \zeta^2 \frac{6 - 2\zeta + 6 \zeta^2}{15}\,\,.
\eeq
In the optically thin limit, we then use
\beq
P^{ij}_{\rm thin} = \frac{F^i F^j}{F^k F_k} E\,\,.
\eeq
Although this expression is exact in the limit $E^2=F^iF_i$, Shibata {\it et al.}~\cite{2011PThPh.125.1255S}
have shown that it does not respect causality when $E^2>F^iF_i$. However, proposed alternatives have more serious
limitations~\cite{2011PThPh.125.1255S}, and $P^{ij}_{\rm thin}$ only matters when $E^2\sim F^iF_i$. 

For the optically thick limit, we choose
\beq
\label{opticallythickS_equation}
S^{\mu \nu}_{\rm thick} = h^{\mu \nu} \frac{J_{\rm thick}}{3}\,\,,
\eeq
which is equivalent to
\beqn
J_{\rm thick} &=& \frac{3}{2W^2+1}\left[(2W^2-1)E-2W^2 F^iv_i\right]\,\,,\\
\gamma^\alpha_\beta H^\beta_{\rm thick} &=& \frac{F^\alpha}{W}+\frac{Wv^\alpha}{2W^2+1} \left[(4W^2+1)v_iF^i-4W^2E\right]\,\,. \nonumber
\eeqn
$P^{ij}_{\rm thick}$ can then be computed from Eq.~(\ref{eq:Pij}).

In practice, because $P^{ij}$ is a function of $\zeta$, which itself depends on $J,H^\alpha$, which are themselves functions 
of $P^{ij}$, the equations that we just described can only be solved through the use of a root-finding algorithm. We thus define
\beq
R(\zeta;E,F_i) = \frac{\zeta^2 J^2 - H^\alpha H_\alpha}{E^2} \label{eq:clos:NR}\,\,,
\eeq
and solve for $R(\zeta;E,F_i)=0$ using a Newton-Raphson algorithm (with $\zeta$ initialized at its last computed value at the given point).
In Eq.~(\ref{eq:clos:NR}), $J$ and $H^\mu$ are computed from $E$, $F_i$ and $P_{ij}$, where $P_{ij}$ is now itself a known function of
$E$, $F_i$ and $\zeta$.

\section{Energy integrated source terms}
\label{sec:sources}

To evolve the energy-integrated moments $E$ and $F_i$
of the neutrino distribution function $f_{(\nu)}$, we need to define
the emissivity $\eta$ and opacities $(\kappa_a,\kappa_s)$
of the fluid. In theory, these source terms are functions
of $f_{(\nu)}$. However, as we only evolve energy-integrated
moments, we do not know the neutrino spectrum, and have instead to
rely on estimates of $f_{(\nu)}$. There are multiple ways to
do so, and we detail the various choices that we have implemented below.
We note that a number of those choices rely on information obtained
from a simpler leakage scheme for the treatment of neutrino cooling~\cite{Deaton2013}
(e.g. average energy of neutrinos, optical depth),
so that even when we evolve the neutrino radiation using the moment
formalism, we still leave the leakage scheme turned on --- but without
coupling it to the fluid equations.

We can first compute the source terms assuming that the neutrinos obey a Fermi-Dirac distribution 
with temperature $T$ and chemical potential $\mu_\nu$. In that case, we have
\beq
f_{(\nu}) = \frac{1}{1+\exp{[(\epsilon_\nu-\mu_\nu)/(k_BT)]}}\,\,,
\eeq
where $\epsilon_\nu$ is the neutrino energy, $\mu_\nu$ the
neutrino chemical potential, $k_B$ the Boltzmann constant, and $T$ the temperature of the fluid.

We now have to choose the value of the chemical potentials
 $\mu_\nu(\rho_0,T,Y_e)$. The value of the chemical potentials for neutrinos
 in equilibrium with the fluid, $\mu_\nu^{\rm eq}$, is taken
directly from the equation of state table. We consider two different choices for
$\mu_\nu$: either we set $\mu_\nu=\mu_\nu^{\rm eq}$ everywhere, or we make
the same choice as in many leakage codes,
\beq
\mu_\nu = \mu_\nu^{\rm eq} (1-e^{-\tau_\nu})\,\,,
\eeq
which is chosen so that $\mu_\nu \rightarrow 0$ in the optically thin region. 
We will refer to these two choices as `equilibrium' and `leakage' chemical potentials.
The choice of $\mu_\nu$ in the low optical depth regions does not matter much in
practice when using our default method for the computation of the source terms
(described below), but can have an effect in some of the alternative schemes
that we tested.

The energy integrated
source terms are then computed as in our leakage code~\cite{Deaton2013}
(which is itself based on the GR1D code~\cite{OConnor2010}, 
and previous work by Ruffert et al.~\cite{Ruffert1996}, 
Rosswog and Liebendorfer~\cite{Rosswog:2003rv} and 
Burrows et al.~\cite{Burrows2006b}):
we compute the absorption opacity due to electron and positron capture,
the free streaming emission due to $e^+e^-$ pair annihilation, plasmon decay
and nucleon-nucleon Bremsstrahlung, and the scattering opacities
due to elastic scattering on nucleons and heavy nuclei. 
We note however that, for the emissivity in optically thick regions, we use
the free streaming emissivity, while the leakage scheme replaces the
emissivity by the diffusion rate in those regions. 
Indeed, diffusion through the optically thick regions is handled naturally
when using the moment formalism.

In order to guarantee that
the neutrinos are in equilibrium with the fluid in the optically thick regions,
we then compute the free streaming emission due to charge current reactions
and the absorption due to pair processes through an energy-integrated
version of Kirchhoff's law. 
At a given energy $\epsilon_\nu$, Kirchhoff's law 
gives us a relation between the emissivity $\eta(\epsilon_\nu)$, the 
absorption opacity $\kappa_a(\epsilon_\nu)$, and the equilibrium spectrum of the
neutrino radiation $B_\nu(\epsilon_\nu)$
\footnote{Note that Eq.~\ref{eq:kir} is always true for charged-current reactions, but only
in the equilibrium limit for pair processes.}:
\beq
\eta(\epsilon_\nu) = \kappa_a(\epsilon_\nu) B_\nu(\epsilon_\nu)\,\,.
\label{eq:kir}
\eeq
The absorption opacity entering the evolution equations for $E$ and $F_i$,
on the other hand, is the energy averaged
\beq
\kappa_a = \frac{\int_0^\infty \kappa_a(\epsilon_\nu) I(\epsilon_\nu)d\epsilon_\nu}{\int_0^\infty I(\epsilon_\nu)d\epsilon_\nu} \label{eq:ka-avg}\,\,,
\eeq
where $I(\epsilon_\nu)$ is the specific intensity of the neutrino radiation.
In our code, when computing $\kappa_a$ from $\eta$ (for pair processes), or
$\eta$ from $\kappa_a$ (for charged-current reactions), we assume that $I_\nu = B_\nu$.
Then, we have
\beqn
\eta &=& \int_0^\infty \eta(\epsilon_\nu) d\epsilon_\nu = \int_0^\infty \kappa_a(\epsilon_\nu) B_\nu(\epsilon_\nu)\nonumber\\
&\approx& \kappa_a \int_0^\infty B_\nu(\epsilon_\nu) d\epsilon_\nu\,\,.\label{eq:Kirchhoff}
\eeqn
As $B_\nu$ is a known function of the fluid properties, this equation can easily be enforced.
In optically thick regions, this prescription will be accurate, as it maintains the equilibrium between
the fluid and the neutrino radiation. In optically thin regions, on the other hand, the neutrino
radiation can be far from equilibrium. Computing charged-current interactions from the energy-integrated
Kirchhoff's law assuming an equilibrium spectrum can affect the relative emission of electron neutrinos
and antineutrinos, and thus the evolution of the electron fraction in low-density regions.
We thus consider an alternative to the application of Kirchhoff's law when computing the emissivity
of charged-current reactions: we can smoothly interpolate between the value of the emissivity $\eta_K$ predicted by the 
application of Kirchhoff's law and the emissivity $\eta_{\rm free}$ predicted by Ruffert et al.~\cite{Ruffert1996}
for free emission in an optically thin region, using
\beq
\eta = \eta_K f(\tau_\nu) + \eta_{\rm free} [1-f(\tau_\nu)]\,\,,
\label{eq:etacorr}
\eeq
with
\beqn
f(\tau_\nu) &=& 1 \,\,\,{\rm if}\,\tau_\nu>2 \nonumber\\
 &=& 0 \,\,\,{\rm if}\,\tau_\nu<2/3 \nonumber\\
 &=& \frac{\tau_\nu-2/3}{4/3}\,\,\,{\rm otherwise}\,\,.
\eeqn
Note that even when using this corrected emissivity, the opacity due to charged-current reactions is not modified.
Accordingly, this corrected emissivity no longer satisfies Eq.~(\ref{eq:Kirchhoff}).
Although apparently more ad-hoc than the previous method, this prescription gives the best results
in our test of neutrino-matter interactions (see Appendix~\ref{sec:tests}). This is probably because the energy integrated emissivity $\eta_{\rm free}$
was specifically meant to approximate the emissivity in optically thin regions.
We note that the deviations from Kirchhoff's law for the energy integrated $\kappa_a$ and $\eta_{\rm free}$ come
from the approximations made in computing energy-averaged Pauli blocking factors, as well as from neglecting
the electron rest mass and the difference between the neutron and proton masses when integrating
over neutrino energies.

Finally, we can go one step further in improving our evolution scheme by noting that for most reactions considered
here, including the dominant charged-current reactions,
the absorption and scattering cross-sections scale like $\epsilon_\nu^2$. This means that in low-temperature regions,
where the fluid is much colder than the neutrinos (which, presumably, are emitted in higher temperature regions),
we largely underestimate $\kappa_{a,s}$. We can thus apply the correction
\beq
\kappa_{a,s} \rightarrow \kappa_{a,s} \left[\max{\left\{1,\left(\frac{\langle\epsilon^2_{\nu,leak}\rangle}{\langle\epsilon^2_{\nu,fluid}\rangle}\right)\right\}}\right]\,\,,
\label{eq:kcorr}
\eeq
where $\langle\epsilon^2_{\nu,leak}\rangle$ is the average of $\epsilon_\nu^2$ predicted by the leakage scheme (taken over the
entire grid) while
\beq
\epsilon^2_{\nu,fluid} = \frac{F_5(\eta_\nu)}{F_3(\eta_\nu)}T^2
\eeq
is the average of $\epsilon^2_\nu$ for neutrinos in equilibrium with the fluid. 
The average is weighted by the energy density of neutrinos
and not the number density of neutrinos, as we are interested in the energy-averaged $\kappa_{a,s}$ given by Eq.~(\ref{eq:ka-avg}).
In our tests (see Appendix~\ref{sec:tests}), we find that using this corrected
value of $\kappa_{a,s}$ significantly improves the agreement between the gray scheme used in this work
and energy-dependent radiation transport.

We thus leave open three choices when computing the source terms. The chemical potential can be obtained from
its equilibrium value or using the `leakage' prescription. The emissivity of charged-current reactions in the optically thin limit
can be obtained from the opacities of the fluid using the energy-integrated version of Kirchhoff's law or from a direct estimate 
of the emissivity. And finally, the opacities in low temperature regions can be obtained
by assuming equilibrium between the neutrinos and the fluid, or by applying a correction for the higher energy of the neutrinos.
Our default algorithm is to apply the energy correction to $\kappa_{a,s}$, to compute $\eta$ in optically thin regions
from a direct estimate of the emissivity, and to use the equilibrium chemical potential everywhere (although that latest choice
has nearly no impact when combined with our prescription for $\eta$). We should however note that, even though these choices
lead to significant differences in our test based on the evolution of the post-bounce configuration of a core collapse simulation, 
they are in much better agreement
in the case of binary mergers. 

\section{Computation of the fluxes at cell faces}
\label{sec:flux}

When computing the fluxes at cell faces for the evolution of $\tilde E$ and $\tilde F_i$,
we first use shock-capturing methods to estimate the value of $(E,F_i/E)$ at cell faces from their value
at cell centers. For the simulations presented in this paper, we use the 2nd order {\it minmod} reconstruction 
(see e.g.~\cite{1986AnRFM..18..337R,1990JCoPh..87..408N}),
although higher order reconstruction methods are available in SpEC (the fluid variables at cell faces, for example,
are reconstructed using the WENO5 algorithm~\cite{Liu1994200,Jiang1996202}). These reconstruction methods are used with both left- and right-biased
stencils, leaving us with two reconstructed values $U_L$ and $U_R$ for the evolved variable $U=(\tilde E,\tilde F_i)$
and $\bar F_L^i$ and $\bar F_R^i$ for the fluxes $\bar F^i=(\alpha \tilde F^i-\beta^i \tilde E,\alpha \tilde P^i_j - \beta^i \tilde F_j)$.
The fluxes at cell interfaces are then approximated by the HLL formula
\beq
\bar F = \frac{c_+ \bar F_L + c_- \bar F_R - c_+ c_- (U_R-U_L)}{c_+ + c_-}\,\,,
\eeq
where $c_+$ and $c_-$ are the absolute values of the largest right- and left-going characteristic speeds of the evolution
system (or zero if there is no left/right going characteristic speeds).  
For the characteristic speeds $\lambda_{1,2}$ at a cell interface along direction $d$, we use
\beqn
p &=&\alpha \frac{v^d}{W}\,\,,\\
r &=& \sqrt{\alpha^2 g^{dd}(2W^2+1)-2W^2p^2}\,\,,\\
\lambda_{1,\rm thin} &=& -\beta^d - \alpha \frac{|F^d|}{\sqrt{F^iF_i}}\,\,,\\
\lambda_{1,\rm  thick} &=& \min{(-\beta^d+\frac{2pW^2-r}{2W^2+1},-\beta^d+p)}\,\,,\\ 
\lambda_{2,\rm thin} &=& -\beta^d + \alpha \frac{|F^d|}{\sqrt{F^iF_i}}\,\,,\\
\lambda_{2,\rm  thick} &=& \min{(-\beta^d+\frac{2pW^2+r}{2W^2+1},-\beta^d+p)}\,\,,\\ 
\lambda_i &=& \frac{3\chi(\zeta)-1}{2} \lambda_{i,\rm thin} + \frac{3(1-\chi(\zeta))}{2} \lambda_{i,\rm thick}\,\,.
\eeqn
where the characteristic speeds in the optically thick and thin limits are taken from~\cite{2011PThPh.125.1255S}
and the interpolation between the two regimes uses the same formula as for the closure.

This prescription works well in optically thin regions. In the optically thick limit, however, these fluxes do not properly reproduce
the diffusion rate of the neutrinos through the fluid. We thus
apply corrections to the flux $\bar F_E$ of $\tilde E$
\cite{Jin1996}
\beq
\bar F_{E,\rm corr} = a\bar F_E+(1-a) \bar F_{E,\rm asym}\,\,,
\eeq
with
\beqn
a &=& \tanh{\frac{1}{\bar \kappa \Delta x^d}}\,\,,\\
\bar \kappa_{i+1/2} &=& \sqrt{(\kappa_a+\kappa_s)_i (\kappa_a+\kappa_s)_{i+1}} \label{eq:ktface}\,\,,
\eeqn
and where half-integer indices refer to values of the opacities at cell faces while integer indices refer
to value of the opacities at cell centers. Here, $d$ is the direction in which we are reconstructing,
$\Delta x^d=\sqrt{g_{dd} (\Delta x^d_{\rm grid})^2}$ is the proper distance between two grid points along that direction,
and $\Delta x^d_{\rm grid}$ the coordinate grid spacing along that direction.
The asymptotic flux in the fluid rest frame, which corresponds to
the flux in the diffusion limit, is \cite{1981MNRAS.194..439T} 
\beqn
H_\alpha = -\frac{1}{3 \kappa} \partial_\alpha J_{\rm thick}\,\,. 
\eeqn  
Using this and equation \ref{opticallythickS_equation} in
equation \ref{flux_equation} gives the asymptotic observer frame
flux 
\beqn
\bar F_{E,\rm asym} &=& \sqrt{\gamma}\left(\frac{4}{3} W^2 \alpha v^d J_{\rm thick} -\beta^d E \right) \nonumber\\
	&& - \alpha \frac{W}{3\bar
	\kappa}\left(\gamma^{di}+v^dv^i
	\right)\sqrt{\gamma}\frac{dJ_{\rm thick}}{dx^i}\,\,.
\eeqn
Numerically, this term can be fairly complex to evaluate. It requires derivatives of the neutrino energy density
in the fluid frame along all coordinate directions, estimated at cell faces. In practice, we compute
the first and second term of $\bar F_{E,\rm asym}$ separately. For the second term, which models the diffusion
of neutrinos, all quantities are estimated by averaging the values at neighboring cell centers, except for $\bar \kappa$,
which is given by Eq.~(\ref{eq:ktface}), and $dJ/dx^d$, for which we use
\beq
\left(\frac{dJ}{dx^d}\right)_{i+1/2} = \frac{J_{i+1}-J_i}{\Delta x^d}\,\,. 
\eeq
For the first term, which represents the advection of neutrinos with the fluid, we make separate estimates from
the left and right states $(U_L,U_R)$. For both states, we also compute the advection speed
\beq
c_{\rm adv} = -\beta^d + 4\alpha \frac{W^2}{2W^2+1} v^d\,\,.
\eeq
If both speeds are positive, we use the value computed from $U_L$. If both are negative, we use the value
computed from $U_R$. If their signs disagree, the advection term is set to zero.

A simpler correction is also applied to the flux $\bar F_F$ of
$\tilde F_i$, following \cite{Audit2002}
\beq
\bar F_{F,\rm corr} = A^2\bar F_F+(1-A^2) \frac{\bar F_{F,R} + \bar F_{F,L}}{2}\,\,,
\eeq
with
\beq
A = \frac{1}{\bar \kappa \Delta x^d}\,\,.
\eeq

\section{Implicit time stepping}
\label{sec:timestepping}

The collisional source terms in equations~(\ref{eq:Enu}-\ref{eq:Fnu}) can be very stiff in optically thick regions.
If we were to treat those source terms explicitly, we would need to use prohibitively small time steps. Accordingly,
we will split the evolution equations into implicit and explicit terms. The variables $(\tilde E_{n+1},\tilde F_{i,n+1})$
evolved from $(\tilde E_n, \tilde F_{i,n})$ by taking a time step $dt$ are given by
\beqn
\frac{\tilde E_{n+1}- \tilde E_n}{dt} &+& \partial_j(\alpha \tilde F^j_n - \beta^j \tilde E_n) \\
	&=& \alpha \left(\tilde P^{ij}_n K_{ij} - \tilde F^j_n \partial_j \ln \alpha - \tilde S^\alpha_{n+1} n_\alpha \right) \nonumber \,\,,\\
\frac{\tilde F_{n+1} - \tilde F_n}{dt} &+& \partial_j(\alpha \tilde P^j_{i,n} - \beta^j \tilde F_{i,n}) 
-\alpha \tilde S^\alpha_{n+1} \gamma_{i\alpha}\\
&=& 	\left( -\tilde E_n \partial_i \alpha + \tilde F_{k,n} \partial_i \beta^k + \frac{\alpha}{2} \tilde P^{jk}_n \partial_i \gamma_{jk}\right)\,\,. \nonumber
\eeqn
Solving this exactly would require the use of a four-dimensional non-linear root-finder algorithm to get $(\tilde E_{n+1},\tilde F_{n+1})$.
To limit the cost of each time step, we instead use a linear approximation to $\tilde S^\alpha_{n+1}$,
\beq
\tilde S^\alpha_{n+1} = \sqrt{\gamma} \eta u^\alpha + A^\alpha \tilde E_{n+1} + 
B^{\alpha j}  \tilde F_{j,n+1}\,\,,
\eeq
where the coefficients $A^\alpha$ and $B^{\alpha j}$ are computed assuming that the closure factor $\xi$ and the angle
between the neutrino flux $F_i$ and the fluid velocity $v^i$ are constant. The system then reduces to four linear equations
for $(\tilde E_{n+1},\tilde F_{i,n+1})$, which can be simply solved by the inversion of a $4 \times 4$ matrix at each point and
for each neutrino species. To ensure that this procedure is reasonable, however, we need an estimate of the time stepping
error -- which will help us choose the time step $dt$. We consider four different types of errors.
\begin{itemize}
\item Large fluxes may cause the explicit portion of the time stepping algorithm to be unstable. We thus require
that
\beq
dt < \frac{\alpha_F \tilde E + \alpha_{\rm Abs} \max{(\tilde E)}}{|\partial_j (\alpha \tilde F^j - \beta^j \tilde E)|} \label{eq:dtCFL}
\eeq
 at any point where the denominator is positive. For post-merger evolutions, we typically choose $\alpha_F=0.3$ and 
 $\alpha_{\rm Abs}=0.001$. The maximum is taken over the entire computational domain.
\item Strong coupling of the neutrino radiation to the fluid can cause large oscillations in the fluid quantities when the radiation
transport code is first turned on, causing the evolution to be unstable. These swings, when they occur, are particularly
noticeable in the evolution of the electron fraction $Y_e$. Accordingly, we define
\beq
\epsilon_{Y} = \frac{|\Delta Y_e|}{\alpha_{\rm Rel}}\frac{\rho_0}{\rho_0 + \alpha_{\rm Abs} \max{(\rho_0)}}\,\,,
\eeq
where $\Delta Y_e$ is the change in $Y_e$ over the last neutrino time step due to the coupling between the fluid and 
the neutrino radiation. For post-merger evolutions, we use $\alpha_{\rm Rel}=0.01$.
\item The implicit portion of the time step might be inaccurate, for example because of the linearization of $\tilde S^\alpha$.
Accordingly, we solve the implicit problem twice: once with a time step $dt$, and once using two time steps $dt/2$. The
explicit terms are kept identical for both time steps, but the linearization of $\tilde S^\alpha$ is recomputed at the intermediate
point $t+dt/2$. For any variable $U$, we thus have two estimates $U_1$ and $U_2$ for $U(t+dt)$, using respectively
one and two intermediate time steps. We can then obtain a second-order accurate estimate $U(t+dt)= 2U_2-U_1$ and
an estimate of the error, $\delta U = U_2-U_1$. We then define
\beqn
\epsilon_E &=& \frac{\delta \tilde E}{\alpha_{\rm Rel} \tilde E + \alpha_{\rm Abs} \max{(\tilde E)}}\,\,,\\
\epsilon_F &=& \frac{\gamma^{ij}\delta\tilde F_i \delta \tilde F_j}{\alpha_{\rm Rel} \tilde E + \alpha_{\rm Abs} \max{(\tilde E)}}\,\,,
\eeqn
as the errors from the implicit time step.
\item Errors in the evolution of $\tilde E$ and $\tilde F_i$ can cause $\tilde E$ to become negative, or the fluxes to violate
causality ($\gamma^{ij}\tilde F_i \tilde F_j > \tilde E^2$). To be safe, we additionally define
\beq
\epsilon'_E = \frac{|\tilde E|}{\alpha_{\rm Rel} \tilde E + \alpha_{\rm Abs} \max{(\tilde E)}}
\eeq
when $\tilde E<0$ and
\beq
\epsilon'_F = \frac{\gamma^{ij}\tilde F_i \tilde F_j-\tilde E^2}{\alpha_{\rm Rel} \tilde E + \alpha_{\rm Abs} \max{(\tilde E)}}
\eeq
when $\gamma^{ij}\tilde F_i \tilde F_j > \tilde E^2$. In practice, we find that these errors are generally smaller than $\epsilon_E,\epsilon_F$.
\end{itemize}
After taking a time step, we then define a global error $\epsilon$ as the largest error among all points, all neutrino species,
and all five errors $\epsilon_F,\epsilon_E,\epsilon_Y,\epsilon'_E,\epsilon'_F$. If the inequality~(\ref{eq:dtCFL}) is not satisfied
or if $\epsilon>10$, we reject the last time step and start again with a new time step given either by~(\ref{eq:dtCFL}) or
by
\beq
dt' = dt \sqrt{\frac{0.9}{\epsilon}}\,\,.
\eeq
Otherwise, we accept the time step and set the next time step for the evolution of neutrinos to be
\beq
dt' = dt \sqrt{\frac{0.9}{\max{(\epsilon,0.3)}}}\,\,.
\eeq
Finally, if $dt'$ is larger than the time step required for the neutrino evolution to catch up with the fluid evolution, we take a
time step bringing the two sets of equations to the same time. In practice, for most of the post-merger evolution, the code
ends up taking a single neutrino time step for each GR-Hydro time step, as can be expected for a quasi-equilibrium radiation
field. The adaptive time stepping algorithm is thus mostly useful to get stable evolutions when the neutrino radiation is first turned
on, as well as during the plunge of the neutron star into the black hole. Occasionally, taking 2-3 neutrino time steps per hydrodynamical
time step can also be necessary when time stepping is limited by the Courant condition, which is more restrictive for neutrinos than for the
fluid.

\section{Code Tests}
\label{sec:tests}

In order to test the ability of our code to perform neutrino transport simulations in compact binary mergers, we perform
a series of tests showing that the moment formalism has been properly implemented in SpEC, that it suffers from
the same known limitations as in other codes, and finally that the choices made when averaging emissivities and opacities
still allow us to reproduce reasonably well the output of an energy-dependent code in a spherically symmetric test evolving
a post-bounce configuration taken from a core collapse supernova simulation.

\subsection{Single beam in flat spacetime}

The simplest test that we perform is the propagation of a beam of radiation in vacuum and for a flat spacetime.
We use a 3-dimensional grid with spacing $\Delta x=0.01$ and bounds $0\leq x \leq 1$, $0 \leq y \leq 0.2$ 
and $0 \leq z \leq 0.2$. The field variables are frozen for $x<0.1$, with the condition $E=1$, $F_i = (0.9999999,0,0)$
inside of a ``beam'' confined to $0.05 \leq y \leq 0.15$, $0.05 \leq z \leq 0.15$, and are suppressed by a factor of
$10^{-10}$ in the region outside of the beam.

\begin{figure}
\flushleft
\includegraphics[width=1.\columnwidth]{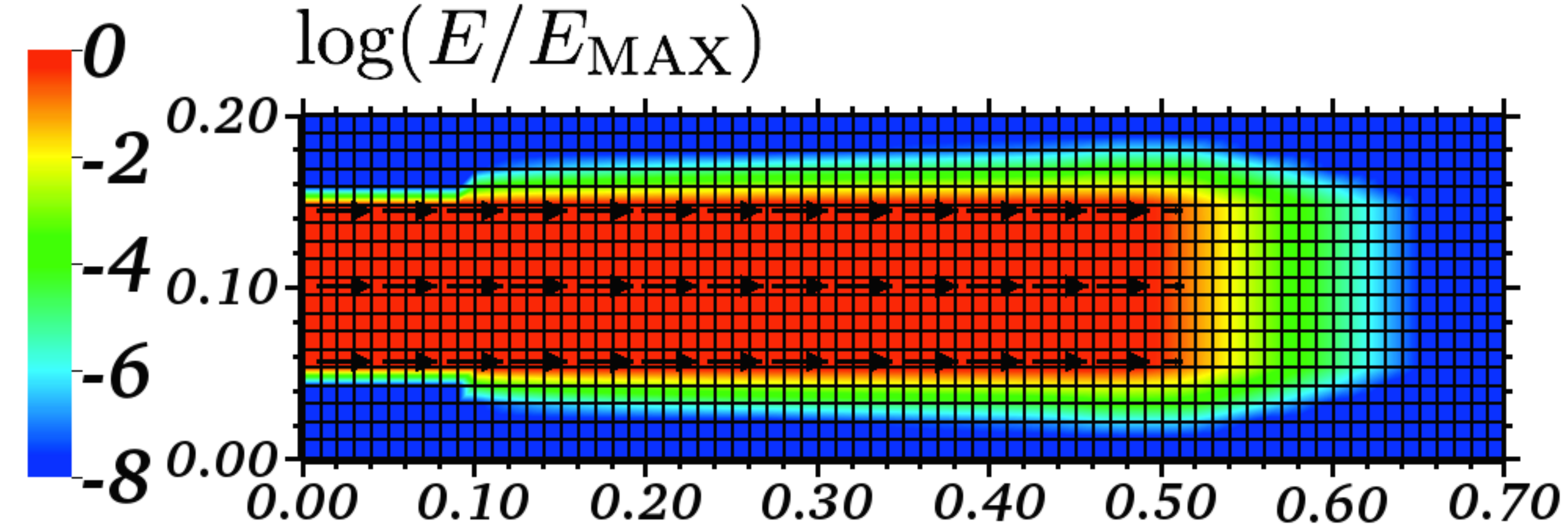}
\caption{Energy density for the single beam test, after $\Delta t = 0.4$. The black arrows show the flux $F_i$.}
\label{fig:SingleBeam}
\end{figure}

The results of this test are shown in Fig.~\ref{fig:SingleBeam}, at a time $t = t_0 +0.4$.
As expected, the beam propagates in a straight line and at the speed of light. Numerical errors
cause a slight widening of the beam, but by only one grid spacing at an energy density around $10^{-3}E_{\rm beam}$.
The beam front is also smoothed upstream of the beam, with an exponential decay of about $10^{-1/2}$ per grid point.
This pattern establishes itself quickly as the beam leaves the frozen region, and then propagates at the speed of light with
the beam.

\begin{figure}
\flushleft
\includegraphics[width=1\columnwidth]{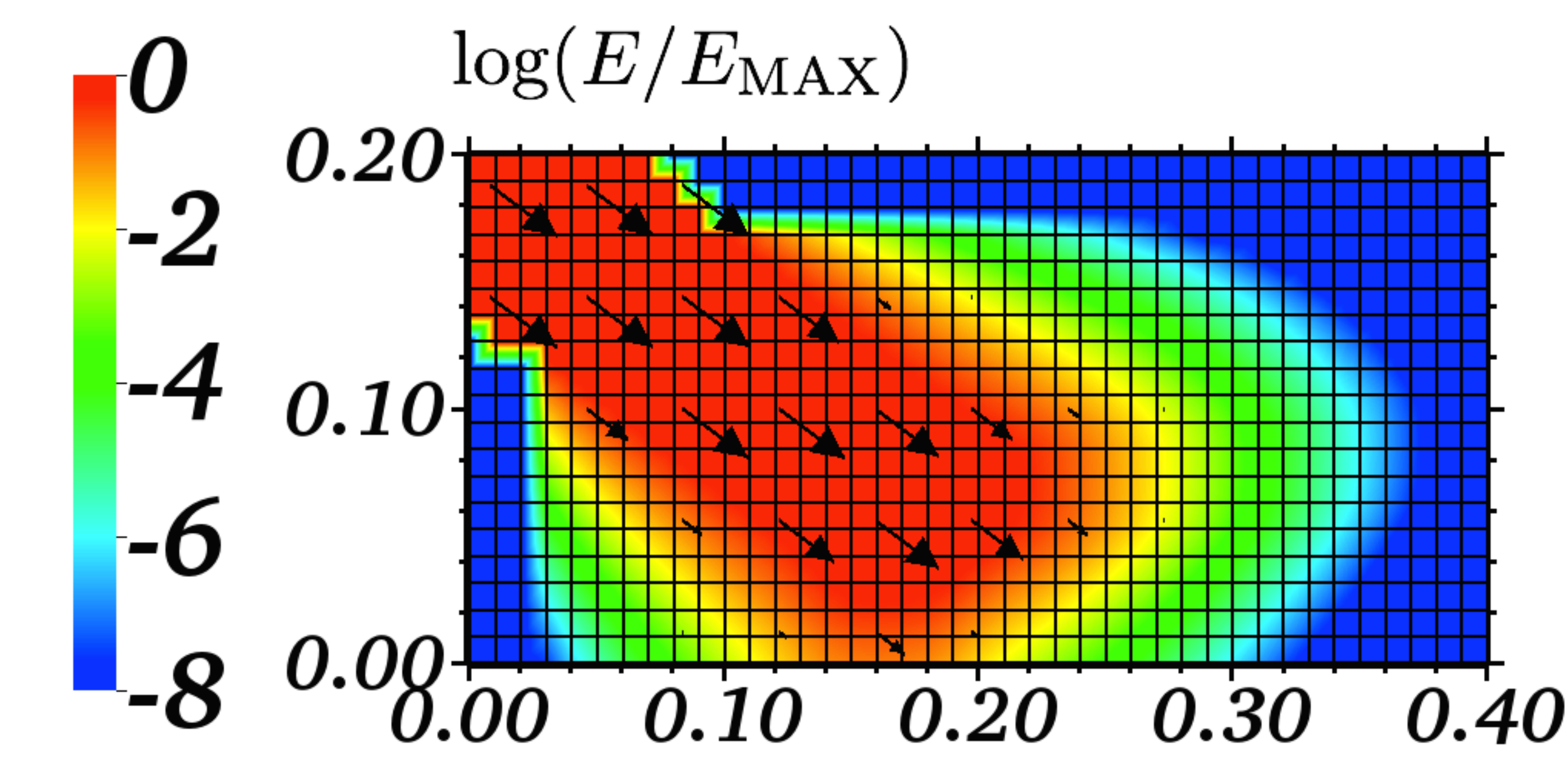}
\caption{Energy density for the oblique beam test, after $\Delta t = 0.2$. The arrows show the flux $F_i$.}
\label{fig:ObliqueBeam}
\end{figure}

We then performed the same test, but for a beam no longer propagating along a coordinate direction. 
In this case, we freeze the region with $x<0.03$ or $y>0.17$, and set $F_i=(0.8,-0.6,0)$ in the beam region.
The results are shown in Fig.~\ref{fig:ObliqueBeam}. The main difference with the aligned beam is that the edges
of the beam are not as sharp (while the front of the beam is slightly sharper). The exponential decay of the energy
away from the beam is now about the same in all directions. 

\subsection{Shadow}

\begin{figure}
\flushleft
\includegraphics[width=1.\columnwidth]{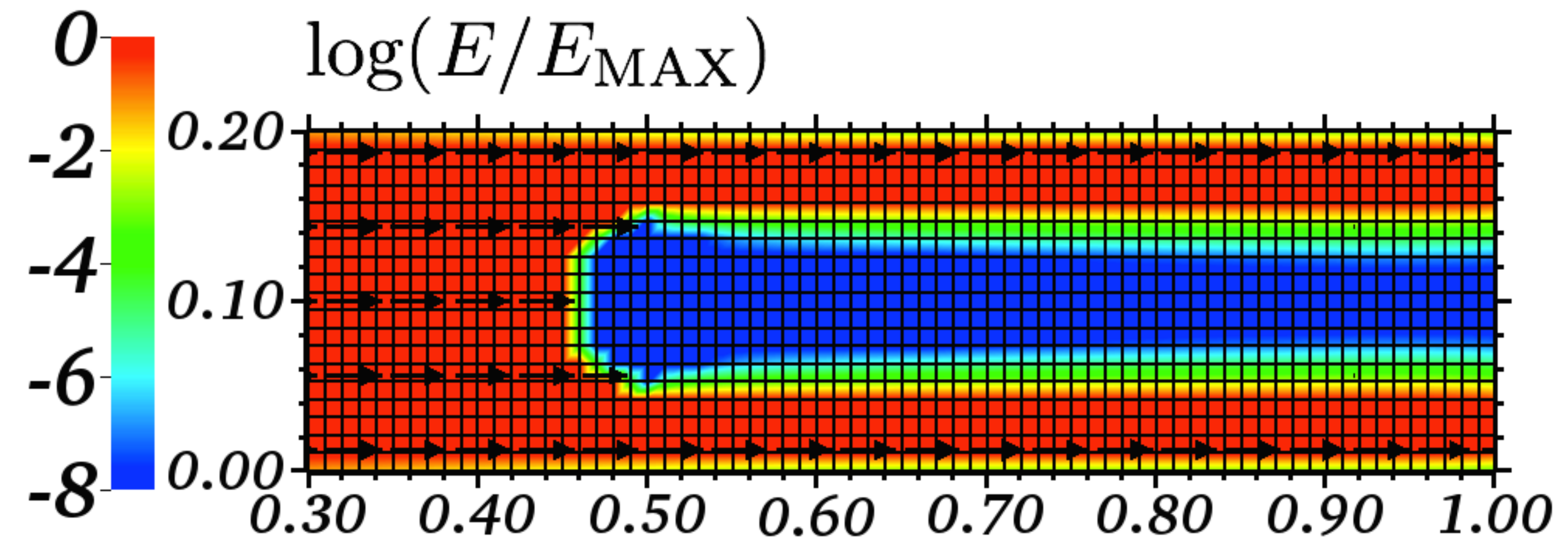}
\caption{Energy density for the shadow test after $\Delta t = 1$. The arrows show the flux $F_i$.}
\label{fig:Shadow}
\end{figure}

We now consider a uniform radiation field with $E=1$ and $F_i=(0.9999999,0,0)$, hitting a sphere of optically thick material.
In this test, we set the absorption opacity $\kappa_a=10^6$ within a sphere of radius $r_S=0.05$ and centered on
${\bf c_S}=(0.5,0.1,0.1)$. The grid is identical to the one used for the beam tests. 
The results of the evolution are shown in Fig.~\ref{fig:Shadow}.

The M1 closure is known to perform well for such a test, in which we expect to see the shadow of the optically thick sphere
in the fields for $x>0.5$ (see e.g.~\cite{2014MNRAS.441.3177M}). 
Indeed, we observe a clear shadow behind the sphere, with the formation of two independent beams each having
properties similar to those observed in the previous tests. Our implementation of the M1 formalism thus appears to
properly project the shadow of optically thick objects.

\subsection{Single beam in curved spacetime}

As our simulations require the evolution of the neutrinos close to a black hole, we want to determine how well the M1 formalism performs in
a strongly curved spacetime. To do so, we perform another set of tests on beams propagating in vacuum, but now in a black hole
spacetime. These tests are largely inspired from those presented in McKinney et al.~\cite{2014MNRAS.441.3177M} (on their Fig.\ 13
and Sec.\ 5.9). The tests in~\cite{2014MNRAS.441.3177M} are performed in 2D in spherical polar coordinates, however, while the SpEC
code always uses 3D cartesian grids for radiation hydrodynamics. For the sake of comparison, we thus first consider
a ``Kerr-like'' spacetime in which the metric is set by $g_{\mu \nu} (t,x,y,z) = g_{\mu\nu}^{Kerr}(0,x,0,z)$, with $g_{\mu\nu}^{Kerr}$
the Kerr metric for a non-spinning black hole, in Kerr-Schild coordinates. 
This metric is unphysical, but allows us to consider effectively 2D beams without having to develop an axisymmetric
version of our code. In all tests, the beams are emitted from a region of width $\Delta z = M$ in the $x=0$ plane, in which $E=1$
and the flux is chosen so that $F_i F^i = 0.998 E^2$ and $\alpha F^i - \beta^i E$ is along the $x$-axis. We use a grid
spacing $\Delta x=0.2M$ (i.e. the beam is initially 5 grid points wide), where $M$ is the black hole mass.

\begin{figure}
\flushleft
\includegraphics[width=1.\columnwidth]{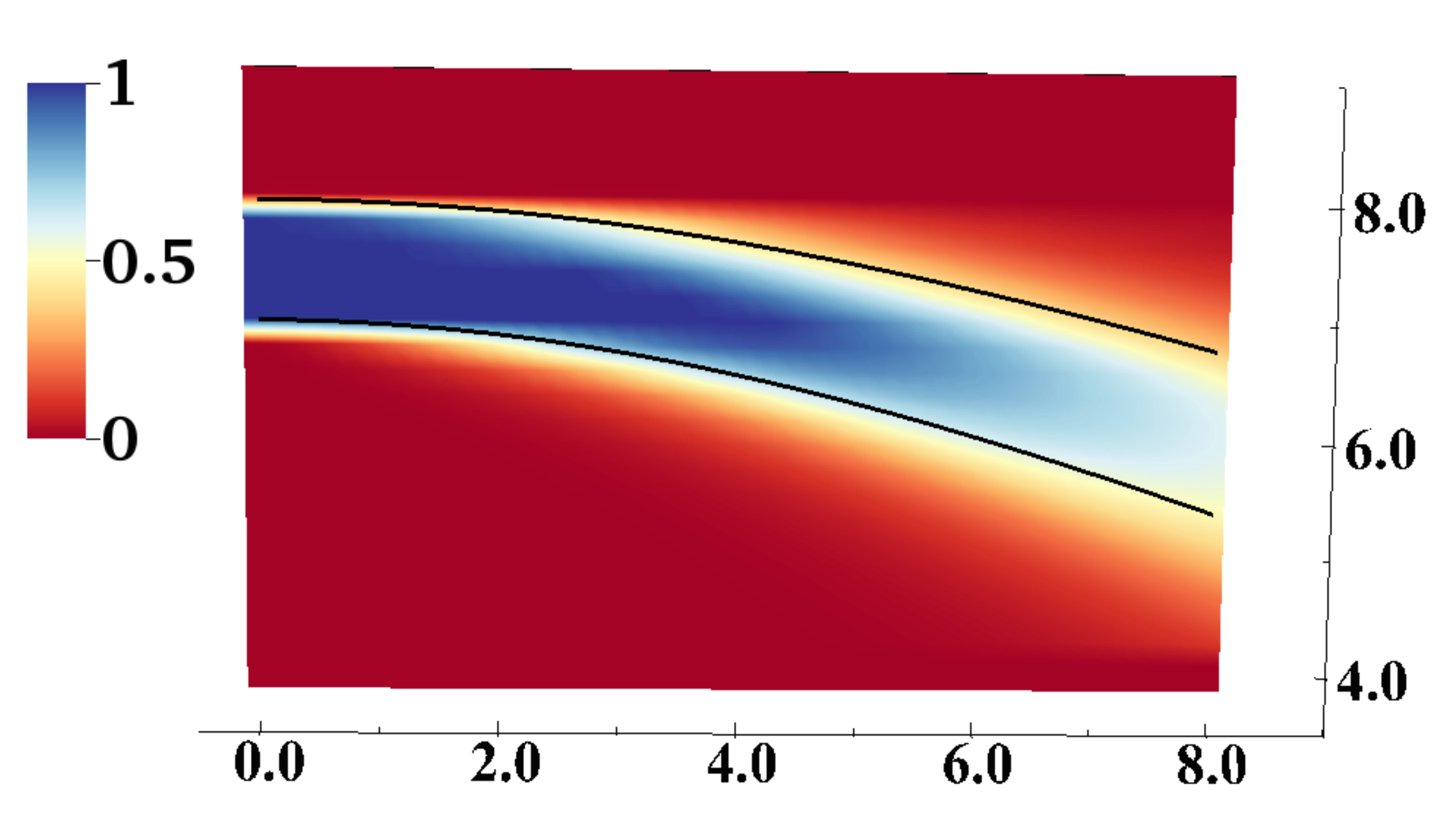}
\caption{Energy density in steady-state for a 2D beam in a Kerr-like spacetime,
with the beam emitted from $r=7M-8M$ and $M$ the mass of the non-spinning
central black hole. The solid lines show null geodesics bounding the expected
location of the beam. The simulation is performed in full 3D, but the metric and beam are independent 
of $z$.}
\label{fig:CurvedBeam2DLargeR}
\end{figure}

We first consider a beam emitted from a relatively large distance, $7M<z<8M$, shown in Fig.~\ref{fig:CurvedBeam2DLargeR}.
The energy density in the beam decreases as the neutrinos get farther from the black hole, in part due to the gravitational redshift
(about a $10\%$ effect over the length of the beam on the grid), and in part due to the spreading of the beam caused by
the gravitational bending of the radiation. Within the accuracy shown in the previous tests, the beam remains otherwise 
well contained within the region delimited by the null geodesics defined by ${\bf x}(t=0)=(0,0,8M)$, $dr/dt(t=0)=0$ and 
${\bf x}(t=0)=(0,0,7M)$, $dr/dt(t=0)=0$ (shown as solid black lines on the figure).

\begin{figure}
\flushleft
\includegraphics[width=1.\columnwidth]{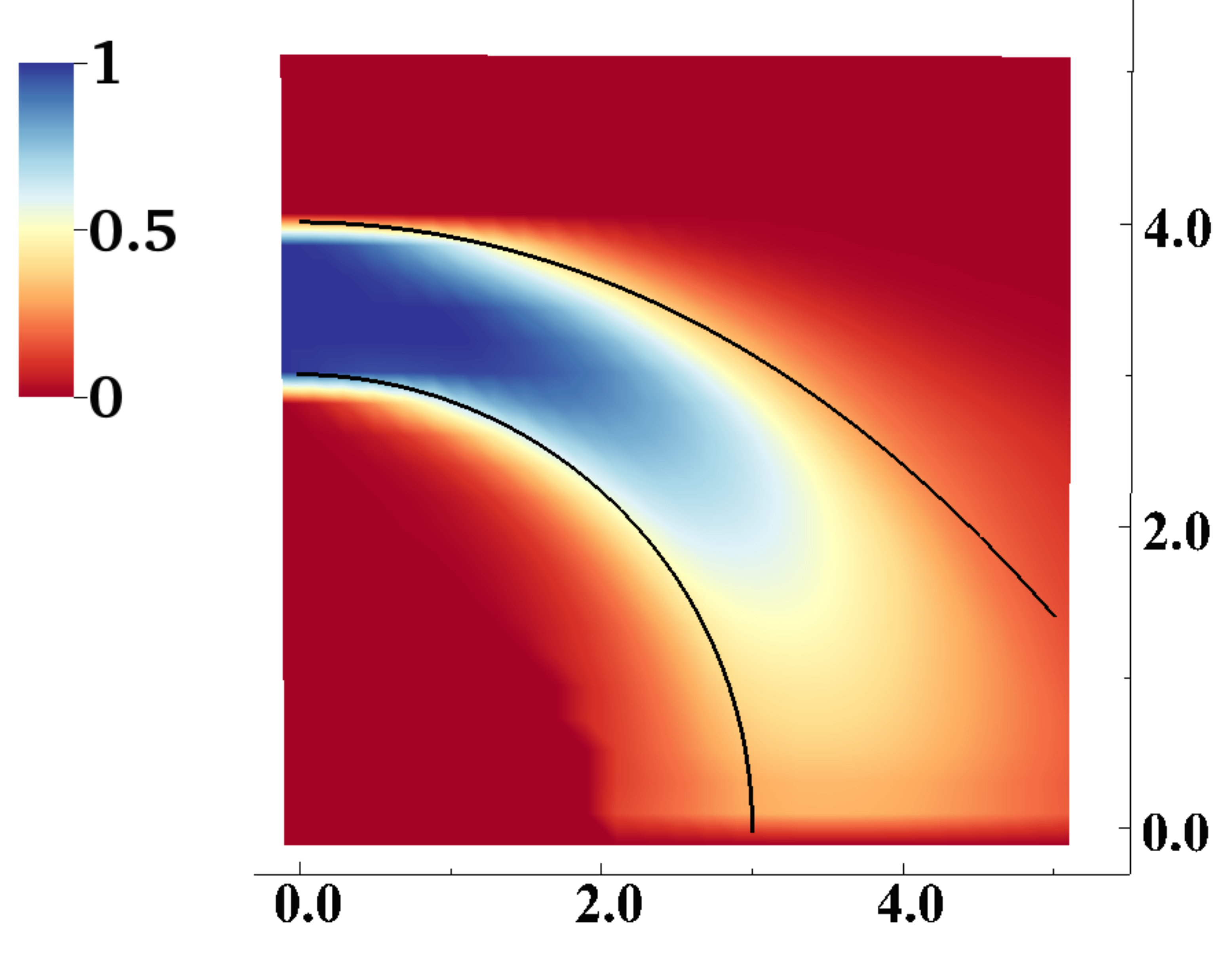}
\caption{Same as in Fig.~\ref{fig:CurvedBeam2DLargeR}, but
with the beam emitted from $r=3M-4M$. As before, the solid lines trace null-geodesics.}
\label{fig:CurvedBeam2D}
\end{figure}

We now move on to a beam initially closer to the black hole, with $3M<z<4M$. The inner edge of the beam then lies on
the unstable circular photon orbit. Both the gravitational redshift effect and the divergence of the null geodesics are much
stronger in this case, so that the beam widens and its energy density decreases quickly as the beam orbits the black hole.
The results of the simulation are shown in Fig.~\ref{fig:CurvedBeam2D}. As before, we can check that the beam remains
mostly within the region predicted for free-streaming massless particles.

\begin{figure}
\flushleft
\includegraphics[width=1.\columnwidth]{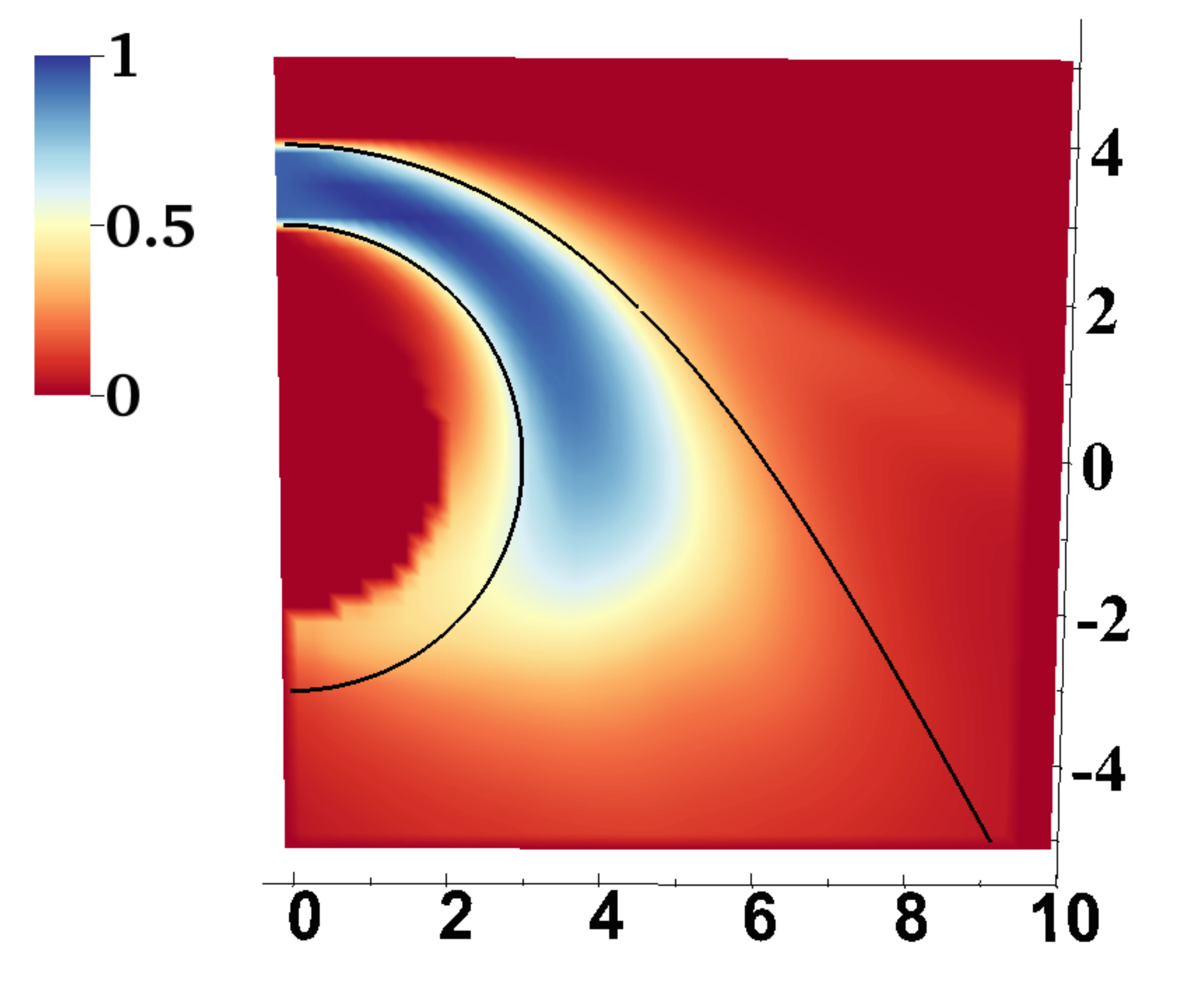}
\caption{Same as Fig.~\ref{fig:CurvedBeam2D}, but using the full 3-dimensional Kerr-metric and a beam of 
finite transverse size.}
\label{fig:CurvedBeam3D}
\end{figure}

Finally, we consider the same configuration but in full 3D. The background metric is now that of a non-spinning 
black hole in Kerr-Schild coordinates. In 3D, the null geodesics followed by our beam converge towards the equatorial plane, 
and should all cross on the $z=0,y=0$ axis. The numerical results are shown in Fig.~\ref{fig:CurvedBeam3D}.
As in the 2D test, the widening of the beam is consistent with the predictions obtained by tracing null geodesics.
The decrease in energy density due to the widening of the beam in the $xy$-plane and to the gravitational
redshift is however compensated by the collapse of the beam along the $z$-axis, at least until the beam 
crosses the $x$-axis. In the $z<0$ plane, the beam should rapidly widen again along the $y$-axis.
In practice, however, the beam remains slightly more collimated than expected in that region.
Although the qualitative differences are fairly mild in this case, it is a first indication of the problems 
intrinsic to the use of the M1 formalism, which we will made clearer in the next test.

\subsection{Crossing beams}

\begin{figure}
\flushleft
\includegraphics[width=1.\columnwidth]{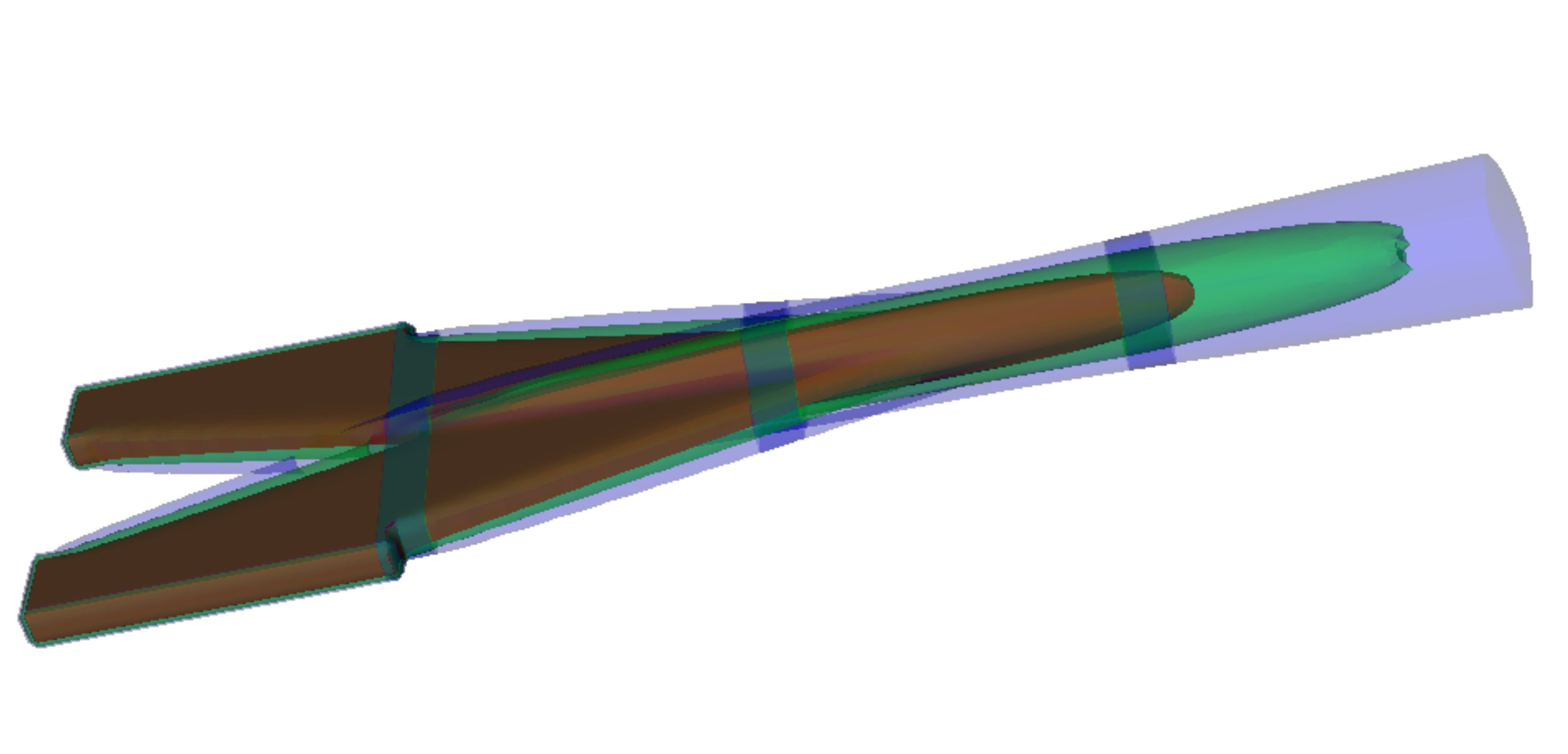}
\caption{Contours of constant energy density for the crossing beams test. The contours are for
10\%, 1\% and 0.1\% of the original energy density of the beams.}
\label{fig:CrossingBeams}
\end{figure}

The problems with converging beams, which we alluded to in the previous section, can be more easily observed if we consider
a much simpler setup: two crossing beams in flat spacetime. As we do not include any interaction between neutrinos,
except indirectly through the effective opacity due to the inverse reactions of thermal processes, the two beams should
simply pass through each other. However, this does not occur when using the moment formalism with the M1 closure. 
We show in Fig.~\ref{fig:CrossingBeams} what happens when such a system is evolved: the beams collide and form a wider,
single beam along the average direction of propagation of the crossing beams. This is, effectively, due to the difference
between the approximate form of the second moment of the distribution function in the analytical M1 closure, and its true
second moment.

This inability of the M1 formalism to deal with crossing beams, and the inaccuracies existing in the presence of converging beams, 
means that the evolution of the moments
of the neutrino distribution function in the region along the polar axis of the black hole, as well as
close to the inner edge of the disk, is not entirely reliable. The M1 formalism performs very well for diverging free-streaming neutrinos
(which is what we see in most of the regions surrounding the disk), and in optically thick regions (as in the core of the disk,
see next test). But its performance in regions in which higher moments of the distribution function are required to properly
model the neutrino pressure is, by definition, quite poor (see also~\cite{2014MNRAS.441.3177M}).

\subsection{Optically thick radiative sphere}

The tests presented above are mainly intended to determine how well our code can evolve the neutrino
moments for various geometric configurations in the free-streaming limit. In order to evolve neutron stars and the
optically thick accretion disks created as a result of neutron star mergers, we also need to determine whether
we can properly model optically thick regions and, more importantly maybe, how closely we can reproduce
the transition between optically thick and optically thin regions. Indeed, these will determine how well
we can predict the neutrino luminosity from neutron stars and accretion disks.

\begin{figure}
\flushleft
\includegraphics[width=1.\columnwidth]{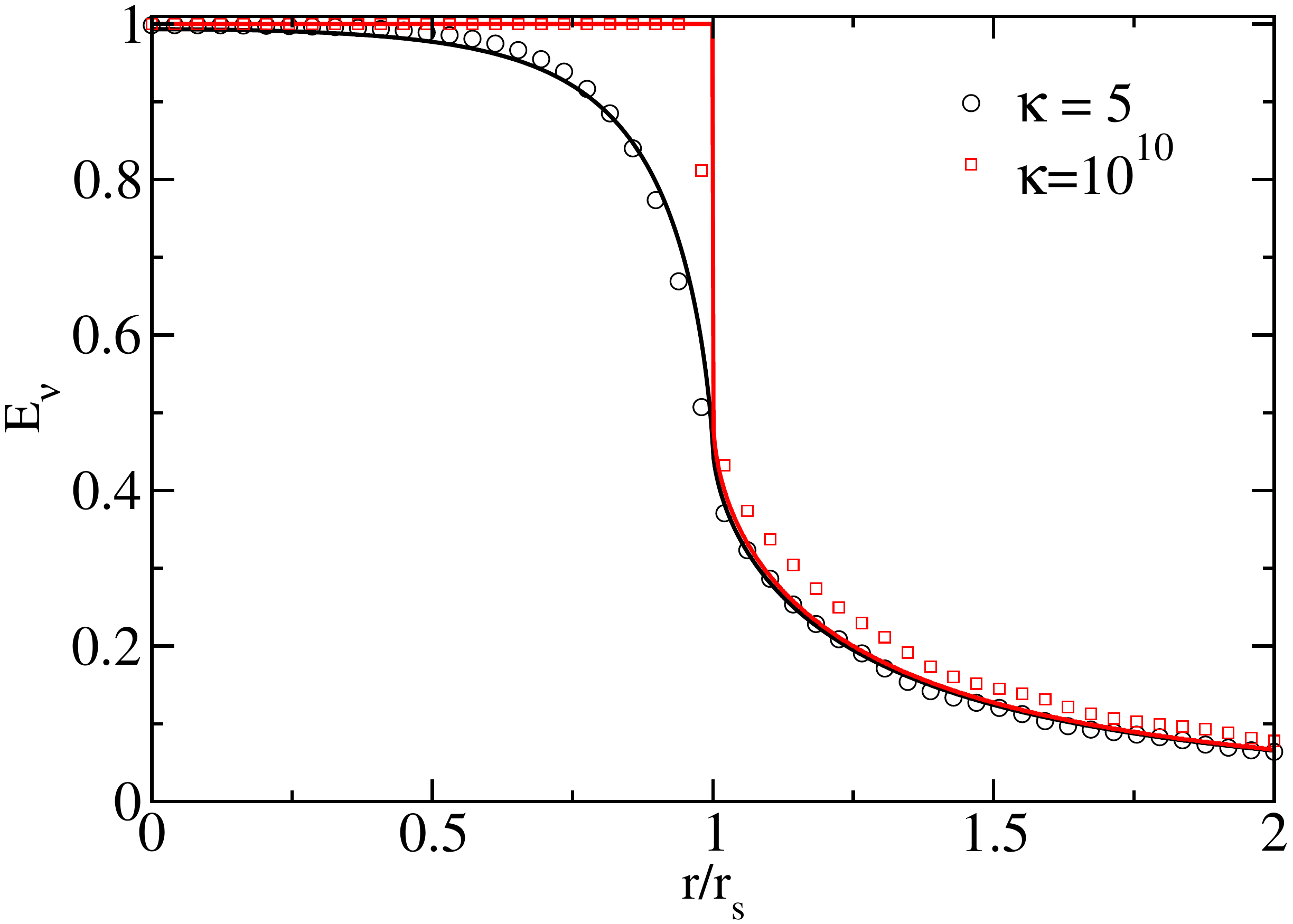}
\caption{Energy density of neutrinos at $t=2$ for the emitting sphere tests with $\kappa=5$ and $\kappa=10^{10}$.
The numerical results are shown by circles/squares while the solid lines are the analytical predictions.}
\label{fig:RadSphere}
\end{figure}

We first consider optically thick radiative spheres, for which we have an analytical solution for the 
distribution function~\cite{Smit:1997}:
\beq
f(r,\mu) = b[1-\exp(-\kappa s(r,\mu))]
\eeq
for a sphere of radius $r_s$ and absorption opacity $\kappa$, and for $\mu = \cos(\theta)$ with $\theta$ the angle between
the neutrino momentum and the radial direction. The function $s(r,\mu)$ is given by
\beqn
s(r,\mu) &=& r\mu + r_s g(r,\mu) \,\,\, [r<r_s; \,  -1<\mu<1]\,\,,\\
&=& 2r_sg(r,\mu) \,\,\,\,\,\,\,\,\,\,\,\,\,\,\, [r\geq r_s; \, \sqrt{1-\left(\frac{r_s}{r}\right)^2}<\mu<1] \nonumber \,\,,\\
&=& 0 \,\,\,\,\,\,\,\,\,\,\,\,\,\,\,\,\,\,\,\,\,\,\,\,\,\,\,\,\,\,\,\,\,\,\, {\rm otherwise} \nonumber\,\,,
\eeqn
with
\beq
g(r,\mu) = \left(1-\left(\frac{r}{r_s}\right)^2 (1-\mu^2)\right)\,\,.
\eeq
We can then obtain the exact solution by numerical integration of $f(r,\mu)$ over $\mu$. In this test, we consider
two different regimes. First we use a sphere of moderate optical depth ($r_s=1$, $\kappa=5$), which is fairly typical
of the conditions in post-merger accretion disks. Then, we use a sphere of extremely high optical depth ($r_s=1$, $\kappa=10^{10}$),
which provides a sharp neutrinosphere similar to what may be found at the surface of a neutron star. In both cases, we use a 3-dimensional
Cartesian grid with grid spacing $\Delta x=0.04$. The results of the numerical
evolutions are compared with the exact solution in Fig.~\ref{fig:RadSphere}. We find good agreement between the two solutions.
For the most optically thick case, this is mostly a consequence of the corrections to high optical depth regions 
described in Appendix~\ref{sec:flux}. The configuration
with $\kappa=5$, on the other hand, has a very small correction to the fluxes, as $\kappa \Delta x = 0.2 < 1$.

\subsection{Neutrino emission and coupling with matter in a post-bounce configuration}
\label{sec:testcollapse}

As a last test of our code, we consider a 1D profile constructed as a spherical average from a 2D, Newtonian, multi-neutrino energy, multi-neutrino angle neutrino transport simulation at 160\,ms after core bounce \cite{Ott2008}. Since we spherically average the 2D simulation, do not use precisely the same nuclear equation of state and neutrino opacities, and freeze the hydrodynamics, the initial profile is slightly out of equilibrium.  Therefore, this provides a good test of the lepton and energy coupling and a venue for comparing with energy-dependent transport.  Similar tests were carried on with this profile for testing Monte Carlo transport \cite{Abdikamalov2012}. We compare the SpEC results with the output of the GR1D code~\cite{OConnor2010}. 
For this test, GR1D uses an energy-dependent M1 scheme with 12 energy groups, and was itself shown to agree well with the results of full transport codes \cite{OConnor2014}.
During this test, the matter density is fixed and the fluid is assumed to be at rest. But the internal energy and electron fraction of the fluid are coupled to
the neutrino evolution. We perform this test with various choices of gray schemes: with the standard SpEC methods described in Appendix~\ref{sec:sources} (SpEC-Std),
with the average energy of neutrinos always obtained by assuming equilibrium between the neutrinos and the fluid (i.e. ignoring the correction given
by Eq.~\ref{eq:kcorr}, SpEC-$T_{fl}$), and without the correction~\ref{eq:etacorr} to the charged-current emissivities in low-opacity regions (SpEC-$\eta_K$).
The SpEC simulations, which are performed with the full 3D code assuming octant symmetry, have a fairly coarse resolution of $6\,{\rm km}$ at the lowest resolution,
$3\,{\rm km}$ at the medium resolution, and $1.5\,{\rm km}$ at the highest resolution.

\begin{figure}
\flushleft
\includegraphics[width=1.\columnwidth]{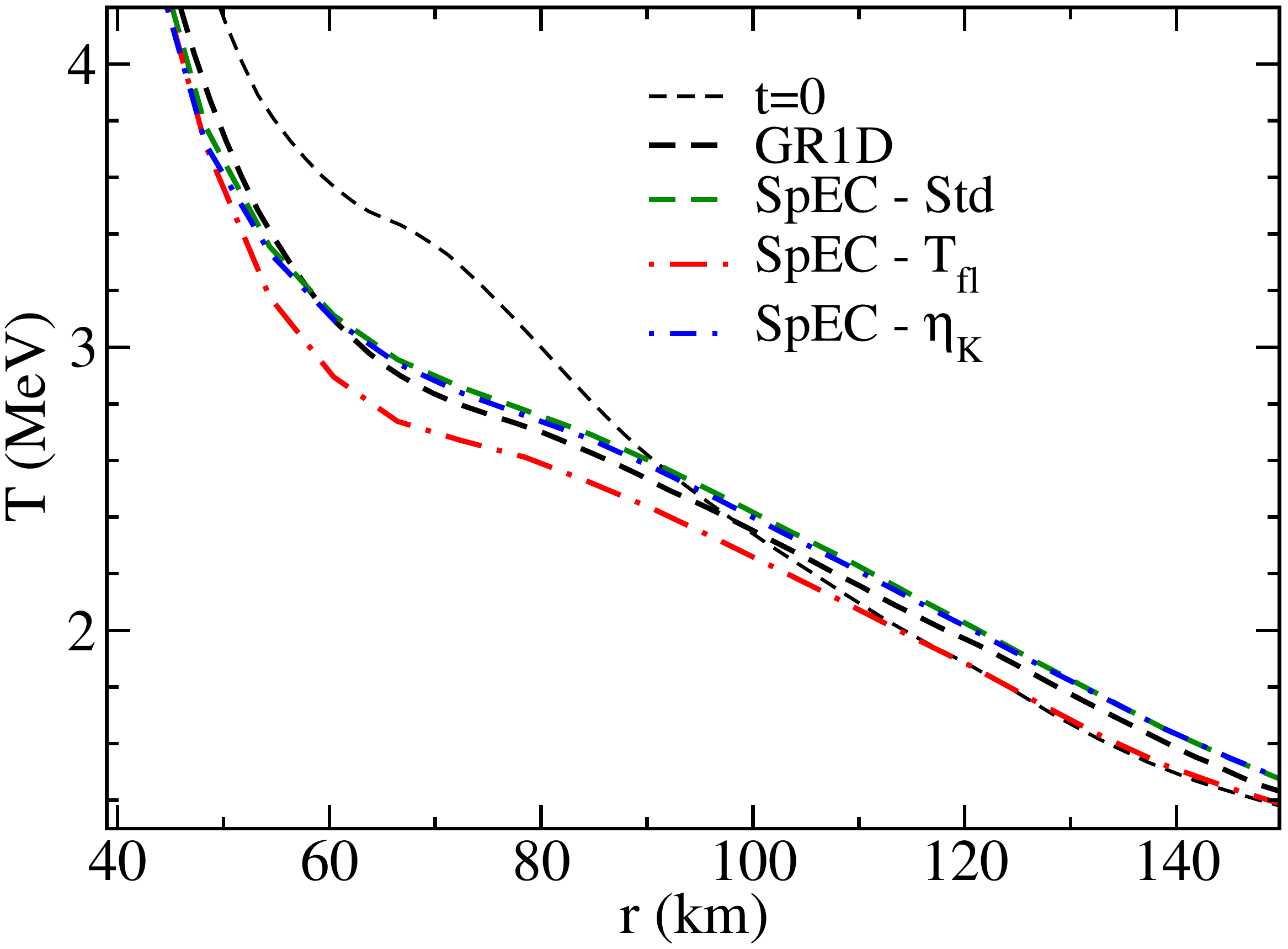}
\caption{Temperature after $8\,{\rm ms}$ of evolution for the evolution of a post-bounce configuration, in the region
in which the temperature evolution is most affected by the gray approximation.}
\label{fig:SCTemp}
\end{figure}

\begin{figure}
\flushleft
\includegraphics[width=1.\columnwidth]{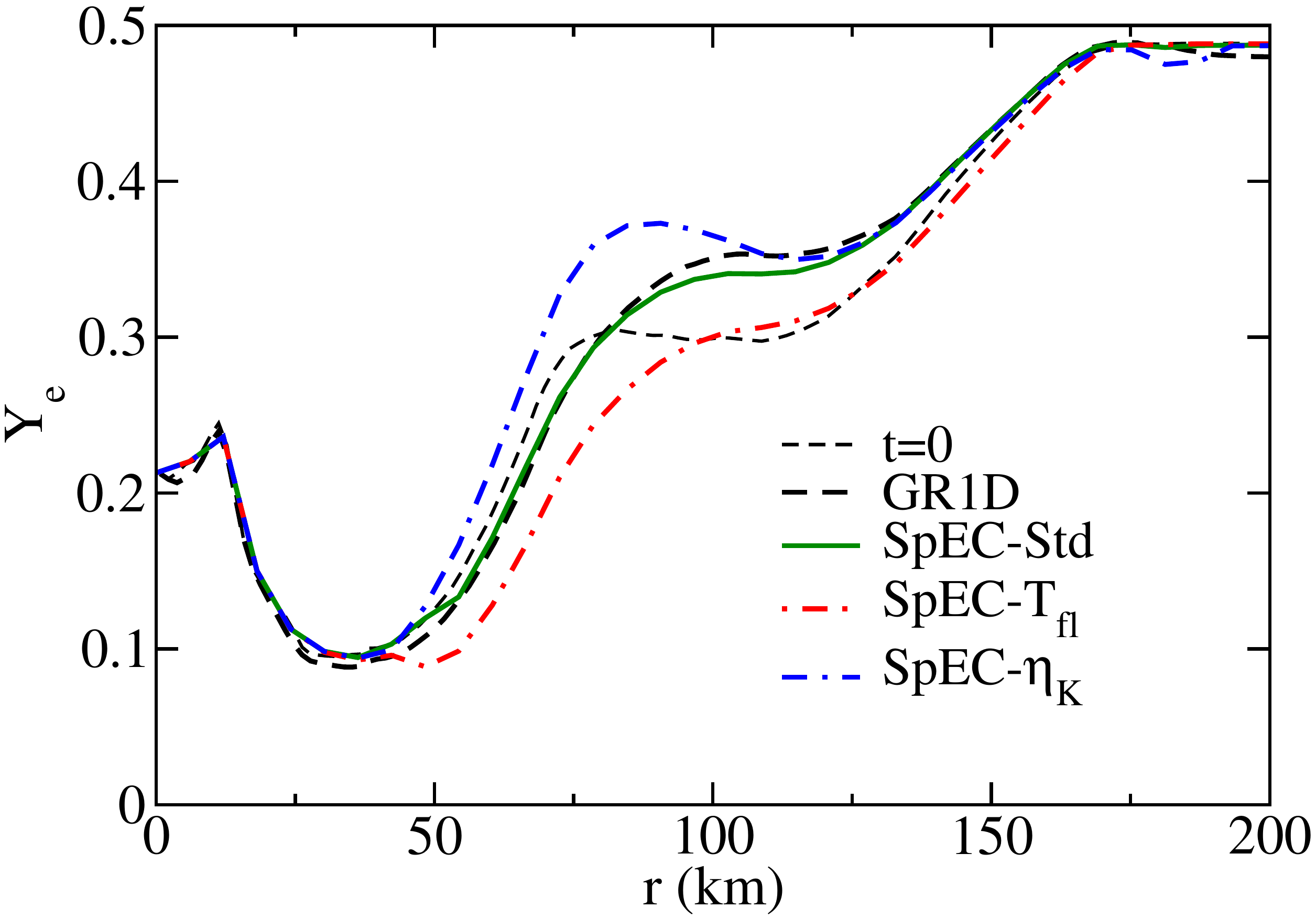}
\caption{Electron fraction $Y_e$ after $8\,{\rm ms}$ of evolution for the evolution of a post-bounce configuration.}
\label{fig:SCYe}
\end{figure}

We first compare radial profiles of the fluid variables $T$ and $Y_e$ after $8\,{\rm ms}$ of evolution, shown in Figs.~\ref{fig:SCTemp}-\ref{fig:SCYe}.
The temperature profiles agree well with the GR1D results, as long as we correct the average energy of neutrinos according to Eq.~\ref{eq:kcorr}.
If we do not, the absorption of neutrinos in the low-density regions is widely underestimated -- and in particular, the code completely misses the existence
of a gain region at $r>100\,{\rm km}$. With correction~\ref{eq:kcorr}, the heating in the gain region is reproduced better, but now occurs faster than expected. 
This is due to an overestimate of the average energy of neutrinos. Whether or not we modify the emissivities according to equation~\ref{eq:etacorr} does not appear to affect the temperature evolution. 

The evolution of the electron fraction $Y_e$ is more sensitive to the choices made when energy-integrating.
With our standard choices, we find results close to GR1D, while the composition is widely inaccurate when neglecting
correction~\ref{eq:kcorr}, and the equilibrium composition can in some regions be wrong by $\Delta Y_e \sim 0.05$ when neglecting
correction~\ref{eq:etacorr}. The grid resolution does not significantly affect these results, except close to $r=0$.

Finally, we observe over $20\,{\rm ms}$ the neutrino luminosity extracted on a sphere of radius $r=250\,{\rm km}$, and 
compare the average energy-squared $\langle\epsilon^2_{\rm leak}\rangle$ predicted by the leakage 
(and used to compute opacities in colder regions) and the same quantity measured by the energy-dependent M1 code. 
We find that the luminosities are accurate to $\sim 10\%$ for $\nu_e$ and $\nu_x$, and to $20\%-30\%$ for $\bar\nu_e$.
For comparison, the luminosities predicted by the leakage scheme are off by factors
of 2-3 for $\nu_e$ and $\bar\nu_e$ (but are very good for $\nu_x$), while the errors due to the finite grid resolution are $\sim 5\%-10\%$. 
The energies $\sqrt{\langle\epsilon^2\rangle}$ are typically overestimated
by the leakage scheme, by up to $50\%$. This causes the larger-than-expected heating rate observed in the gain region ($r\gtrsim 90\,{\rm km}$). 
The reason
for these large errors is that the neutrino energies in the hot, highly degenerate matter present at high optical depth
are very large and significantly affect the average energy (due to weighting of the electron number emission rate by the square
of the neutrino energies). In practice, however, these high energy neutrinos are thermalized as they propagate through the
optically thick regions. Significant improvements in $\sqrt{\langle\epsilon^2\rangle}$ can be obtained if we define
\beq
\langle\epsilon^2\rangle = \frac{\int R_\nu({\bf x}) \langle\epsilon_\nu^2({\bf x})\rangle \min\left(1,\exp[\tau_{\rm threshold}-\tau]\right)dV}{\int R_\nu({\bf x}) \min\left(1,\exp[\tau_{\rm threshold}-\tau]\right)dV}\,\,,
\eeq
where $R_\nu$ is the neutrino number emission predicted by the leakage scheme, and $\tau$ is the optical depth.
Any value of the threshold $\tau_{\rm threshold}\sim 1-10$ allows us to recover $\sqrt{\langle\epsilon^2\rangle}$ to within
$20\%$ for $\nu_e$ and $\bar{\nu_e}$, and within $30\%$ for $\nu_x$. In all cases, the finite resolution error is negligible compared
with the error due to the gray approximation. We have checked that these variations are indeed due to the gray approximation,
and not to the specific implementation of the M1 scheme in SpEC: an energy-dependent version of the M1 code in SpEC, which can easily be used in this
case due to the relatively low computational cost of the test problem and the lack of gravitational redshift or velocity gradients, agrees with the GR1D results.

These results are already much better than if we got the neutrino energies by assuming equilibrium with the fluid: in the gain region, the equilibrium 
average neutrino energies are $\sqrt{\langle\epsilon^2\rangle} \sim 5-10\,{\rm MeV}$ while the actual average neutrino energies are 
$\sqrt{\langle\epsilon^2\rangle} \sim 10-25\,{\rm MeV}$. And indeed, with this corrected $\langle\epsilon^2\rangle$, heating in the gain region agrees
well with the results of the GR1D simulations. The correction also causes changes to the evolution of $Y_e$, with results
with and without correction~\ref{eq:etacorr} to the emissivities now bracketing the GR1D results (with typical errors $\Delta Y_e \sim 0.02$).
The simulation without correction~\ref{eq:etacorr} still slightly overestimates $Y_e$ around $r\sim 100\,{\rm km}$, while with correction~\ref{eq:etacorr}
the simulation now slightly underestimates $Y_e$ in that same region.

Such errors would be very significant in the context of core collapse supernova simulations, where properly
modeling neutrinos is critical to the explosion mechanism. In compact binary mergers, however, the impact of neutrinos
is more modest. The computation of $\langle\epsilon^2\rangle$ from the leakage scheme is also more accurate in post-merger
accretion disks: the temperature does not vary as much within post-merger accretion disks as in the test presented here, 
the disks are only moderately optically thick ($\tau \leq 10$), and matter within the disks reaches densities of at most
$\rho_0\sim 10^{12}\,{\rm g/cm^3}$, at which only electrons are degenerate. Using emissivities with and without 
correction~\ref{eq:etacorr}, we can get an estimate of the errors due to the use of a gray scheme. These errors, discussed in
the main text of the paper, are much smaller than the errors observed in this test. 
Using the M1 formalism instead of our leakage scheme then significantly improves 
our ability to determine the cooling time and composition evolution of the disk, and gives us a reasonable first estimate of
the energy deposition in the corona (except along the polar axis of the black hole, where the M1 approximation is unreliable).

\section{Accuracy of the post-merger evolution}
\label{sec:accuracy}

Most of the discussion of our simulations of the formation of an accretion disk after a neutron star-black hole merger (Sec.~\ref{sec:disk})
focuses on the qualitative features of the system, and on the differences between various algorithms for the evolution of neutrinos.
In this section, we will discuss the accuracy of these results, and argue that the features of the system discussed in Sec.~\ref{sec:disk}
are appropriately resolved by our simulations. We identify four main sources of error. The first is due to the finite resolution of our
numerical grid during the simulations. The second is the numerical error in the simulation of the neutron star-black hole merger
before we turn on the neutrino transport code, which causes errors in the initial conditions used for our simulations. The third is the
approximate treatment of the neutrinos, and in particular the use of a gray scheme and of the M1 closure. And the fourth
comes from only turning on the neutrino transport and changing the treatment of low-density material about $6.1\,{\rm ms}$
after merger. The first two can be estimated through the use of simulations at lower and/or higher resolution. The last two 
are more difficult to assess, and will only be rigorously measured through the use of more advanced (and more costly) simulations.
Until such simulations are available, we have to rely on simpler estimates of these errors.

To determine the importance of the first source of error, we performed simulations of a post-merger accretion disk using $60^3$ and $140^3$
points for each level of refinement, instead of the $100^3$ points used in our "standard" configuration. The higher resolution simulation was only evolved
for $1.1{\rm ms}$, to verify that the solution was converging with resolution. Even the simulation using the coarsest grid shows surprisingly good agreement with our standard runs. The average temperature in the disk agrees to $\Delta \langle T \rangle \sim 0.2\,{\rm MeV}$ and the electron fraction to 
$\Delta \langle Y_e \rangle \sim 0.01$. The neutrino luminosities agree to better than $20\%$, and the average neutrino energies within $0.5\,{\rm MeV}$. 
This is comparable to the differences observed between various choices of gray approximations, and much smaller than the differences between the leakage
and M1 simulations. Additionally, the difference between the standard and high resolutions is, for all observed quantities, more than a factor of two smaller than the difference between the low and standard resolutions. Most of those differences arise immediately after the neutrino transport is turned on, while evolution at later times is very similar for all numerical resolutions. Accordingly, we do not expect numerical resolution during the post-merger evolution
to be a significant source of error.

The numerical error in the simulations before we turn on neutrino transport can easily be determined from the lower resolution simulations
of the same system performed in Foucart {\it et al.} 2014~\cite{Foucart:2014nda}. The largest error in these simulations was in the determination
of the properties of the dynamical ejecta, which does not concern us here as that material is allowed to escape the numerical grid. At the time
at which we begin the simulations with neutrino transport, we otherwise find relative errors of less than $10\%$ in the total mass outside of the black hole,
average temperature and average electron fraction in the disk. Considering that the grid spacing used in~\cite{Foucart:2014nda} was about a factor
of two coarser than the grid spacing used in the standard simulations of this paper, these are clearly overestimates of the numerical error. The only
numerical error which could affect our results is thus the initial temperature of the disk ,which generally decreases with resolution. The disk could be, on average, about $0.4\,{\rm MeV}$ cooler at infinite resolution. This is only slightly smaller than the difference between leakage and M1 simulations, but does
not affect the conclusion of this paper: both the leakage and M1 simulations presented here use the same imperfect initial data. We should also note
that even our imperfect initial data remains a much better starting point than the commonly used alternative, an equilibrium torus of constant entropy. 

The effect of an approximate treatment of the neutrinos is discussed in Sec.~\ref{sec:nutreat}, to which we refer the interested reader. Here we will simply
note that our rough estimate of the error due to the use of a gray scheme is comparable to our estimate of the error due to numerical resolution, and much
smaller than the differences between simulations using the M1 scheme and the leakage scheme. However, a true measure of the error would require
a simulation using an energy-dependent transport scheme without the weaknesses of the M1 closure. Such a simulation is currently out of reach for
our code.

Finally, we have to consider the impact of suddenly turning on the neutrino transport code $6.1\,{\rm ms}$ after merger, and of the use of a higher
threshold for the application of atmosphere corrections at earlier times. The only way to rigorously measure this would be to perform a simulation using
the M1 code and a lower atmosphere threshold, starting before the disruption of the neutron star. Now that we have a well-tested M1 code and good estimates of the atmosphere thresholds which should be used after merger, we intend to perform such simulations. However, the disruption of the neutron star
and accretion of the neutron star core onto the black hole is the most computationally intensive part of the evolution, while we believe
that the simulations performed here capture the most important properties of the merger and post-merger evolution (except for the absence of magnetic
fields). Indeed, we begin the simulation around the time at which the neutrino luminosity and the electron fraction begin to significantly increase due to the heating of the disk and the emission of neutrinos. We also find that the disk outflows are not immediately launched when we decrease the atmosphere
threshold. At the beginning of the simulation, those outflows are still contained by material from the tidal tail falling onto the forming disk. It is only about
$2\,{\rm ms}$ later that outflows begin to form above the disk. A potentially more important source of error is the transient occurring when the transport
code is turned on. To study its effects, we consider two possible initializations of the moments of the neutrino distribution function. First, we initialize them
assuming that the neutrinos are in equilibrium with the fluid, in which case the neutrino energy density is overestimated, and the initial transient consists
in a transfer of energy from the neutrinos to the fluid and a slight decrease of the average electron fraction. Then, we initialize the neutrino energy density to a negligible value, in which case the transient has the opposite effect. We verify that after about $2\,{\rm ms}$ the two solutions are in good agreement (at early times, this transient is of course the largest source of error in the simulation). 
\end{document}